# Coloring Panchromatic Nighttime Satellite Images: Comparing the Performance of Several Machine Learning Methods

Nataliya Rybnikova, Boris A. Portnov, Evgeny M. Mirkes, Andrei Zinovyev, Anna Brook, and Alexander N. Gorban

*Abstract*— **Artificial light-at-night (ALAN), emitted from the ground and visible from space, marks human presence on Earth. Since the launch of the Suomi National Polar Partnership satellite with the Visible Infrared Imaging Radiometer Suite Day/Night Band (VIIRS/DNB) onboard, global nighttime images have significantly improved; however, they remained panchromatic. Although multispectral images are also available, they are either commercial or free of charge, but sporadic. In this paper, we use several machine learning techniques, such as linear, kernel, random forest regressions, and elastic map approach, to transform panchromatic VIIRS/DBN into Red Green Blue (RGB) images. To validate the proposed approach, we analyze RGB images for eight urban areas worldwide. We link RGB values, obtained from ISS photographs, to panchromatic ALAN intensities, their pixel-wise differences, and several land-use type proxies. Each dataset is used for model training, while other datasets are used for the model validation. The analysis shows that model-estimated RGB images demonstrate a high degree of correspondence with the original RGB images from the ISS database. Yet, estimates, based on linear, kernel and random forest regressions, provide better correlations, contrast similarity and lower WMSEs levels, while RGB images, generated using elastic map approach, provide higher consistency of predictions.**

*Index Terms*—**Artificial Light-at-Night (ALAN), Day-Night Band of the Visible Infrared Imaging Radiometer Suite (VIIRS/DNB), elastic map approach, International Space Station (ISS), multiple linear regression, non-linear kernel regression, panchromatic nighttime imagery, RGB nighttime imagery, validation.**

## I. INTRODUCTION

ARTIFICIAL Light-at-Night (ALAN), emitted from streetlights, residential areas, places of entertainment, industrial zones, and captured by satellites' nighttime sensors, has been used in previous studies for remote identification of different Earth phenomena, such as stellar visibility [1]–[3]; ecosystem events [4], [5]; monitoring urban development and population concentrations [6]–[11]; assessing the economic performance of countries and regions [12]–[17], and in health geography research [18]–[20].

Compared to traditional techniques, which national statistical offices use to monitor the concentrations of human activities (such as, e.g., monitoring the level of urbanization, production density, *etc.*), using ALAN as a remote sensing tool has several advantages (see [21] for a recent review). *First and foremost*, satellite-generated ALAN data are available seamlessly all over the world, providing researchers and decision-makers with an opportunity to generate data even for countries and regions with extremely poor reporting behavior. *Second*, ALAN data are mutually comparable for different geographic regions, which minimizes the problem of comparability between socio-economic activity estimates, potentially originating from differences in national reporting procedures. *Third*, data on remotely sensed ALAN intensities are now available worldwide on a daily basis [22], which enables researchers and public decision-makers to obtain prompt estimates of ongoing changes in the geographic spread of different human activities and their temporal dynamics. The latter is especially important for

This paper was submitted for review on September, 3, 2020. Work of N. Rybnikova was supported by the Council for Higher Education of Israel. Work of N. Rybnikova, E. M. Mirkes, and A. N. Gorban was supported by the University of Leicester. Work of A. Zinovyev was supported by Agence Nationale de la Recherche in the program Investissements d'Avenir (Project No. ANR-19-P3IA-0001; PRAIRIE 3IA Institute). Work of E. M. Mirkes, A. Zinovyev, and A. N. Gorban was supported by the Ministry of Science and Higher Education of the Russian Federation (Project 075-15-2020-808).

N. Rybnikova is with Dept. of Mathematics, University of Leicester, Leicester LE1 7RH, United Kingdom; Dept. of Natural Resources and Environmental Management, and Dept. of Geography and Environmental Studies, University of Haifa, Haifa 3498838, Israel (e-mail: nataliya.rybnikova@gmail.com).

B. A. Portnov is with Dept. of Natural Resources and Environmental Management, University of Haifa, Haifa 3498838, Israel (e-mail: portnov@research.haifa.ac.il).

E. M. Mirkes is with Dept. of Mathematics, University of Leicester, Leicester LE1 7RH, United Kingdom; Lobachevsky University, Nizhny Novgorod 603105, Russia (e-mail: em322@le.ac.uk).

A. Zinovyev is with Institut Curie, PSL Research University, Paris 75248, France; Institut National de la Santé et de la Recherche Médicale, U900, Paris 75013, France; MINES ParisTech, CBIO-Centre for Computational Biology, PSL Research University, Paris 77305, France; Lobachevsky University, Nizhny Novgorod 603105, Russia (e-mail: Andrei.Zinovyev@curie.fr).

A. Brook is with Dept. of Geography and Environmental Studies, University of Haifa, Haifa 3498838, Israel (e-mail: abrook@geo.haifa.ac.il).

A. N. Gorban is with Dept. of Mathematics, University of Leicester, Leicester LE1 7RH, United Kingdom; Lobachevsky University, Nizhny Novgorod 603105, Russia (e-mail: ag153@le.ac.uk).

socioeconomic activities, for which estimates based on traditional techniques, are unavailable with a desired frequency or time-consuming to generate.

Several sources of *global* nighttime imagery exist today. Between 1992 and 2013, nighttime satellite imagery was provided by the U.S. Defense Meteorological Satellite Program (DMSP/OLS) on an annual basis, with the spatial resolution of about 2.7 km per pixel [23]. From April 2012 on, nighttime images, generated by the Day-Night Band of the Visible Infrared Imaging Radiometer Suite (VIIRS/DNB) instrument of the Suomi National Polar Partnership (SNPP) satellite, have become available. The satellite moves through a sun-synchronous polar orbit at the altitude of about 824 km, and captures ALAN emissions at about 1:30 am local time [23]. The VIIRS/DNB program routinely provides panchromatic global imagery in the 500-900 ηm range at about 742 m per pixel spatial resolution, on annual and monthly bases. From the first quarter of 2019 on, ALAN data are available daily from the NASA Black Marble night-time lights product suite, or VNP46 [22], the Distributed Active Archive Center [24].

In comparison to DMSP/OLS images, VIIRS/DNB data have a better spatial resolution and lower light detection limits (2E-11 Watts/cm$^2$/sr vs. 5E-10 Watts/cm$^2$/sr in US-DMSP), which is especially important for analyzing *dimly lit* areas. VIIRS/DNB data also do not exhibit bright light saturation [23], which is essential for the analysis of *brightly lit* areas, such as major cities and their environs.

However, despite the above-mentioned improvements in the ALAN image quality and resolution, the main drawback of *global* ALAN data, available today, is that they remain *panchromatic*, reporting the *summarized* intensity of light in the 500-900 ηm diapason [23]. This limitation makes it difficult to use such data to differentiate *between* specific economic activities, which are characterized by varying spectral signatures [25], because they use light sources of different spectral properties, to fit their resources and needs [26]. As a recent study [27] shows, night-time multispectral ALAN imagery also helps to study and understand better urban land use types.

Panchromatic ALAN data also do *not* make it possible to investigate health effects, associated with ALAN exposures to different sub-spectra, such as e.g., hormone-dependent cancers, known to be strongly related to ALAN exposure in the short-wavelength (blue) light spectra [28], [29].

In addition, the 500-900ηm sensitivity diapason, reported by global VIIRS/DNB images, omits some important intervals of the visible light spectrum (see Fig. A1 in Appendix). In particular, it omits the emission peaks of the incandescent and quartz halogen lamps that are at about 1000 ηm, and a large share of ALAN emissions from the Light Emitting Diodes (LED), which occur in the 450-460 ηm range [30]. This means that the reported *summarized ALAN intensities* are essentially *biased*, and this bias, potentially introduced by local lighting standards and/or cultural preferences, is *not random* but may vary systematically across different geographical areas, depending e.g., on the level of propagation of specific light sources, such as LEDs, which light emission is outside the captured ALAN range. In this respect, the ongoing rapid propagation of LEDs is of particular concern, as it might gradually diminish the capability of presently available global ALAN images to serve as a reliable proxy for monitoring the human footprint, and may thus impede research progress on estimating various side effects of light pollution.

RGB nighttime images of better spectral resolution, provided by the habitable International Space Station (ISS) [31], is also available. However, the use of ISS data for a global analysis is often problematic. The matter is that these night-time images are photographs, captured sporadically by varying cameras, which need to be geo-referenced and calibrated, to produce a continuous image from a mosaic of fragmented local pictures, taken by different cameras and different astronauts [32]. In addition, the ISS images in question are not available on a regular basis.

Considering these limitations of the globally available polychromatic ALAN data, the present study aims to demonstrate a possibility that the spectral resolution of global panchromatic VIIRS/DBN night-time imagery can be enhanced, by transforming such panchromatic data into RGB images. To achieve this goal, we use machine learning techniques to build and cross-validate the models associating light intensities of red, green, or blue sub-spectra with panchromatic ALAN data, pixel-wise neighborhood difference measures and several land-use proxies. As the study demonstrates, using regression tools and the elastic map approach, originating from the manifold learning field, helps to produce reasonably accurate RGB estimates from panchromatic data. The importance of this result is that it may help to generate more informative and freely available remote proxies for a human presence on Earth.

The rest of the paper is organized as follows. We start by outlining our study design and describe the datasets used for model training and validation. Next, we itemize criteria used for model validation, report the obtained results, and discuss controversial issues raised by the analysis and limitations that should be addressed in future studies.

## II. METHODS

### A. Research hypothesis and study design

According to [33], each type of land use is characterized by a certain combination of different luminaires. As a result, different land uses differ in terms of both aggregated light flux, spectral power distribution (SPD), and the primary emission peak diapason [25]). In addition, some types of light emission are *spatially localized* (such as e.g., blue-light emissions from commercial and industrial hubs), while other light emissions are more geographically uniform, such as e.g., long-wavelength light emissions from homogeneous low-density residential areas. Therefore, we *hypothesize that information on different ALAN sub-spectra (red, green, and blue) can be extracted from a combination of panchromatic ALAN data, pixel-wide neighborhood differences, and built-up area characteristics.*

To test this hypothesis, we link the intensity of each ALAN sub-spectra (Red-Green-Blue) with the intensity of panchromatic ALAN, pixel-wise neighborhood ALAN difference measures and characteristics of built-up areas available for several major metropolitan areas worldwide (see section II-B).

The former group of neighborhood controls includes differences between the panchromatic ALAN intensity in a given pixel and either average or the most extreme ALAN intensity in its neighborhood. The potential importance of such differences is expected to be due to the fact that substantial differences in neighboring ALAN intensities may occur, if, for instance, a brightly lit commercial facility, often characterized by blue luminaries, stands out against nearby dimly lit areas or if such a facility is separated from its surrounding by a dimly-lit buffer zone. In contrast, similar light emissions in the pixel's neighborhood may result from the pixel's location in a homogenously lit residential area, where long wavelength luminaries (such as incandescent or vapor lamps) are often used. Concurrently, the above-mentioned built-up area characteristics include the percent of built-up area and its spatial homogeneity, considering that each type of land use has its spatial configuration and land cover [34].

We examine four types of machine learning models. The first one is the elastic map approach [35], originating from the manifold learning field, and three standard methods, represented by multiple linear, non-linear kernel, and random forest regressions (see section II-D). Using these methods, the models are first estimated for training sets and then are validated against testing sets (see section II-D). In each case, the models' performance is assessed by mutually comparing the model-estimated and original RGB data. To perform assessments, different similarity measures are used – Pearson's correlation coefficients, weighted mean squared error (WMSE), and contrast similarity. In addition, we control for the consistency of these measures by comparing the results obtained for training and testing datasets (see section II-E).

### B. Data Sources

For *each* metropolitan area under analysis, we built a dataset that includes three separate images of the RGB sub-spectra (red, green and blue), a panchromatic image of ALAN intensity, a layer of neighborhood differences, calculated for the panchromatic ALAN layer, and a land-use layer (see section II-A).

As RGB ALAN data source, we use local night-time images provided by the International Space Station (ISS) and available from the Astronaut Photography Search Photo service [31]. Concurrently, panchromatic ALAN images are obtained from the VIIRS/DNB image database, maintained by the Earth Observation Group site [36], while land-use characteristics of *built-up area* are computed from the global raster layer of human built-up area and settlement extent (HBASE) database available at the NASA Socioeconomic Data and Application Centre site [37]. The HBASE dataset is a 30-meter resolution global map derived from the Global Land Survey Landsat dataset for the year-2010. In the present analysis, we use the HBASE layer that reports the pixel-wise probability of the built-up area in the range from 0 to 100%. The HBASE is a companion for the Global Manmade Impervious Surfaces (GMIS) dataset, which addresses GMIS's commission errors arising from over-prediction of impervious cover in areas which are full of dry soil, sands, rocks, *etc*. [38].

It should be noted that ISS images report ALAN levels in digital numbers, which are camera-specific [39]. Therefore, to ensure the comparability of RGB levels, reported for different localities, we selected from the ISS database *only* images taken by the same – a Nikon D4 Electronic Still – camera. In addition, to enable the comparability of ISS images with panchromatic ALAN images, we selected the ISS images taken at the time close to the VIIRS/DNB image acquisition, that is, at about 01:30 a.m., local time.

The ISS images were matched with spatially referenced layers using the Geo-referencing tool of the ArcGISv10.x software by matching key points in the raster photos with corresponding points in the Streets Basemaps obtained from the ArcGIS online archive [40]. Next, the ISS images were paired with corresponding monthly VIIRS/DNB composites, and clipped to the extent of the corresponding RGB image. In particular, the following pairs of images were used:

(i) For the Atlanta region, the USA, the ISS image (ID ISS047-E-26897, taken on March 29, 2016), was matched with the VIIRS/DNB image taken in March 2016 Tile 1 (75N/180W) composite;

(ii) For the Beijing region, China, the ISS image (ID ISS047-E-11998, taken on March 20, 2016), was matched with the VIIRS/DNB image taken in March 2016 Tile 3 (75N/060E) composite;

(iii) For the Haifa region, Israel, the ISS image (ID ISS045-E-148262, taken on November 29, 2015), was matched with the VIIRS/DNB image taken in November 2015 Tile 2 (75N/060W) composite;

(iv) For the Khabarovsk region, Russia, the ISS image (ID ISS047-E-12012, taken on March 20, 2016), was matched with the VIIRS/DNB image taken in March 2016 Tile 3 (75N/060EW) composite;

(v) For the London region, the UK, the ISS image (ID ISS045-E-32242, taken on September 27, 2015), was matched with the VIIRS/DNB image taken in September 2015 Tile 2 (75N/060W) composite;

(vi) For the Naples region, Italy, the ISS image (ID ISS050-E-37024, taken on January 30, 2017), was matched with the VIIRS/DNB image taken in January 2017 Tile 2 (75N/060W) composite;

(vii) For the Nashville region, the USA, the ISS image (ID ISS045-E-162944, taken on December 6, 2015), was matched with the VIIRS/DNB image taken in December 2015 Tile 1 (75N/180W) composite;

(viii) For the Tianjing region, China, the ISS image (ID ISS047-E-12004, taken on March 20, 2016) was matched with the VIIRS/DNB image taken in March 2016 Tile 3 (75N/060EW) composite.

Fig. 1 reports examples of images used for the Greater Haifa metropolitan area in Israel. [Images for other areas under analysis are not reported here, for brevity's sake, and can be obtained from the authors upon request]. We should note that daily nighttime VIIRS images have recently become available as the VNP46A2 product, reporting moonlight and atmosphere corrected nighttime lights [22]. However, in such images, poor-quality pixels, caused either by outliers, cloud contamination, *etc.*, might be present [41]. For example, in the Khabarovsk region of Russia, used in the paper as one of the test sites, the daily image for the required date comprises about 20% of pixels, flagged as poor-quality ones. Therefore, in the present analysis, we opted to use cloud-free monthly composites, considering that future studies may consider using daily nighttime images for coloring, while employing the data modelling method we propose.

### C. Image Processing

The data for the analysis were processed in several stages. First, we reduced the high-resolution of ISS RGB images (~10 meters per pixel), by averaging neighboring pixel values, to match the resolution of corresponding VIIRS/DNB images (~750 meters per pixel) and then converted the resized images into point layers, using the *Raster-to-Point* conversion tool in ArcGIS v.10.x software. Next, to each point in the layer (i.e., reference points), we assigned the corresponding values of the red, green, and blue light sub-spectra from the corresponding ISS RGB image. The task was performed using the *Extract MultiValues to Points* tool in ArcGIS v.10.x software. *Next,* after VIIRS/DNB images were converted into points, each point was assigned with the following information: 1) panchromatic ALAN flux; 2) average difference between ALAN intensity in the point and ALAN intensities in its eight neighboring points, and 3) maximum difference between the ALAN intensity in a given point and ALAN values in eight neighboring points in the point's immediate neighborhood. *Lastly,* after the HBASE image was converted into points, its pixel averages and standard deviations (SDs) were calculated and assigned to the reference points as well.

During data processing, all the points located outside the study area (for instance, points falling into water bodies) or classified as outliers in each dataset (see Outliers Analysis Box in Appendix) were excluded from the analysis. Table AI reports the number of observations for each geographic site, and other relevant information, while descriptive statistics for research variables are reported in Table AII in Appendix.

### D. Data modelling

To estimate the models linking ALAN intensities of red, green and blue sub-spectra with the set of explanatory variables (see section II-A), we used, as previously mentioned, four alternative modeling approaches: the elastic map approach, originating from the manifold learning field [42], and three standard supervised multivariate modeling methods, that is, ordinary multiple linear regression, non-linear kernel regression (see *inter alia* [43]), and random forest approach [44].

All the approaches belong to the field of supervised machine learning, as they model the relations between variables based on some training data and use the revealed relationships to make predictions for others – that is, testing – data. This generates a so-called bias-variance dilemma [45]. The better a model fits the training data, the worse it is expected to fit the test data. As a result, while linear regression's performance may be relatively poor for the training dataset, it may generate reasonably good predictions for test datasets. By contrast, non-linear kernel regression or random forest regression might fit training data perfectly but may fare poorly, when applied to new datasets. In this context, elastic maps with *varying bending regimes* can be viewed as an approach for optimizing such a bias-variance trade-off.

Each of the aforementioned models was first estimated separately for the red, green, blue light intensities for *each* of the eight cities covered by analysis – i.e., Atlanta, Beijing, Haifa, Khabarovsk, London, Naples, Nashville, and Tianjing (see section II-B). During the model estimation, *all pixels* belonging to a city were included into the training set, while the testing sets were formed by seven other cities, which were not used for training. Each estimated model was next applied to the other metropolitan areas, to validate its performance. In the sections below, we describe, in brief, each modeling approach used in the analysis.

#### 1) Elastic map approach

Elastic map approach implies constructing a non-linear principal manifold approximation, represented by nodes, edges (connecting pairs of nodes) and ribs (connecting triples of nodes), by minimizing the squared distances from the dataset points to the nodes, while penalizing for stretching of the edges and bending of the ribs [38]. Elastic map, eventually presented by multidimensional surface, built of piece-wise linear simplex fragments, might be considered as a non-linear 1D, 2D, or 3D screen, on which the multidimensional data point vectors are projected. It is built, on the one hand, to fit the data, and, on the other hand, not to be too stretched and too bent.

The general algorithm of the elastic map follows the standard splitting approach. Elastic map is initialized as a regular net, characterized by nodes, edges, connecting two closest nodes, and ribs, connecting two adjacent continuing edges. This net is embedded into a space of multidimensional data, and the node embeddings are optimized to achieve the smooth and regular data approximation. The optimization is done in iterations, similarly to the k-means clustering algorithm. At the first step of each iteration, the data point cloud is partitioned accordingly to the closest elastic net's node embedment. At the second step, the total energy of the net (U) is minimized, and the node embedment is updated. After this, a new iteration starts, and this process continues till a maximum number of iterations is achieved or the changes in the node positions in the multidimensional data space become sufficiently small at each iteration. For detailed formal methodology description, see [41] and [42]).

The energy of the elastic map is represented by the following three components: summarized energy of nodes (U(Y)), calculated as the averaged squared distance between the node and the corresponding subset of data points closest to it; summarized energy of edges (U(E)), which is the analog to the energy of elastic stretching and is proportional – via a certain penalty – to the sum of squared distances between edge-connected nodes; and summarized energy of ribs (U(R)), which might be considered as the analog to the of elastic deformation of the net and is calculated as proportional – via a certain

penalty – to the sum of squared distances between the utmost and center nodes of the ribs. Fig. 2 provides the reader with a simplified explanation for elastic maps approach, summarizing the above-mentioned components of elastic map and their energies: Each node is connected by elastic bonds to the closest data points and simultaneously to the adjacent nodes.

It is important to note that, unlike standard supervised methods, such as linear or kernel regression models, elastic map is, by its nature, a non-supervised manifold learning method which does not treat any variable as dependent one; it is designed to explain – under pre-defined penalties for stretching and bending – *total* variance of the data. However, similarly to Principal Component Analysis (see, for example, [46]), the elastic map data approximations can be used for predicting the values of some of the variables (e.g., those which are considered to be dependent) through imputing them. The imputing approach, in this case, consists in fitting the elastic map using the part of the dataset containing no missing values and then projecting the data vectors containing a missing value for the dependent variable. The imputed (or, predicted) value is the value of the variable in the point of its projection onto the elastic map.

By construction, elastic map, represented by a *sufficient* number of nodes, and given the *low penalties* for stretching of edges and bending of ribs, would fit input data perfectly. Theoretically, when the number of elastic map nodes approaches the number of points in the input dataset, and under zero penalties, the fraction of the total unexplained variance of input point cloud by corresponding elastic map would equal to zero. At that, this elastic map's ability to generalize to another dataset or predict one of its variables levels is expected to be low. Increasing the elastic penalty is expected to increase the generalization power of the approach while respecting the non-linearities in the variable dependences. In the limit of very stiff elastic penalty, the performance of elastic maps is expected to match those of linear methods. However, the optimal performance can be found in between these two extremes (absolute flexibility vs absolute rigidity).

The present elastic map analysis was conducted in MATLAB v.R2020x software [47]. We utilized a two-dimensional net with a rectangular grid with nodes, which were brought into actual data subspace spanned by the first three principal components. Due to the outlier analysis performed, we settled the stretching penalty at a zero level. To prevent overfitting, the number of nodes was also fixed at a level of 144 (12x12), which is about 5-50 times smaller than the number of points of input datasets. We experimented with the bending coefficient only. In the attempt to optimize bias-variance trade-off, we tested elastic maps built under nine varying bending penalties. (Fig. A2 in Appendix, reporting corresponding models for blue light association with the set of predictors for Haifa dataset, gives an idea of how these maps look like. As one can see from the figure, representing the general tendency for either red, green or blue lights containing datasets, map smoothness gradually grows with an increasing penalty for bending, while the level of a fraction of total variance (that is, a fraction of variance by all six variables in the dataset) unexplained (FVU) by smoother map further decreases.)

### 2) Multiple linear regression

The general idea behind the multiple least-squares linear regression is fitting the observations (each represented by a point in N-dimensional space with (N-1) number of predictors and one dependent variable) by a linear relationship, represented by an (N-1)-dimensional linear surface, or hyperplane, by minimizing the sum of squared errors between the actual and estimated over this hyperplane levels of the dependent variable. In the current analysis, for each geographic site dataset, the following multiple ordinary least squares (OLS) regression model was estimated:

$$CL_{ij} = b_0 + \sum_k (b_k \times P_{ki}) + \varepsilon_i \ , \tag{1}$$

where $CL_i$ = observation $i$ of ALAN intensity in color band $j$ (either red, green or blue sub-spectra); $b_0$ = model intercept; $b_k$ = regression coefficient for the $k^{th}$ predictor; $P$ = vector of model predictors, represented by pixel-specific panchromatic ALAN intensity, reported by VIIRS/DNB (i); the difference between the pixel-specific panchromatic ALAN intensity and average panchromatic ALAN intensities of eight neighboring pixels (ii); the maximum difference between the panchromatic ALAN flux from a pixel and panchromatic ALAN fluxes from eight neighboring pixels (iii); average percent and standard deviation of land coverage, calculated from HBASE, and $\varepsilon$ = random error term.

The multiple regression analysis of the factors associated with RGB ALAN intensities was performed in the IBM SPSS v.25 software [48].

### 3) Non-linear kernel regression

Non-linear kernel regression is a non-parametric technique, fitting the observations into a hypersurface. The method uses a sliding window, with a dataset being divided into smaller subsets. Within each data subset, each data point is treated as a 'focal point', and its value along the dependent variable axis is re-estimated from a hyperplane (or hypersurface), built to minimize the errors, weighted for the distance to the focal point along independent variables axes and for the difference between estimated and actual levels of the dependent variable [49].

Under this estimation technique, many parameters are a matter of choice. First, the size of the sliding window may vary from several points to significant amounts of the whole dataset, providing correspondingly less or flatter hypersurface. Second, the modelled association between a dependent variable and its predictors might be either linear, parabolic, exponential, *etc*. Third, the errors between estimated and actual levels might be either minimized or not allowed to exceed a certain value. Fourth, the 'weights' function might vary, implying paying more or less attention to more distant data points. Finally, the number of iterations on re-estimating dependent variable actual levels might also be increased, so the resulting hypersurface would be flatter.

In the present analysis, we used a standard realization of the Gaussian kernel regression built-in MATLAB v.R2020x software under the chosen automatic option for the kernel regression parameters optimization [50]. The latter implies the optimization of the kernel regression parameters *by using* five-fold cross-validation based on mean squared errors.

*4) Random forest regression*

We also tested the random forest approach [44], which implies building an ensemble of decision trees, each 'voting' for a certain class or level of the dependent variable, with subsequent averaging of the estimates across all the decision trees. In the present analysis, we implemented a standard realization of the random forest regression (the *TreeBagger* module) in the MATLAB v.R2020x software [51]. During the estimation procedure, the following two parameters were a matter of choice – the number of independent variables used for the individual decision tree construction and the number of decision trees comprising the "forest." Following [52], all the predictors available were used for the decision trees' construction and the total number of decision trees was set to 32.

*E. Criteria for the models' comparison*

To compare models estimated using the above-discussed statistical techniques, we used the following indicators:

(i) Pearson correlation coefficients were calculated to determine the strength of association between the actual and predicted levels of RGB sub-spectra. This metric assesses the model's ability to produce RGB estimates, which – in their *relative tendency*, – correspond well with the actually observed RGB levels;

(ii) Weighted mean squared errors (WMSE) between the actual and predicted levels of ALAN emissions in the red, green, and blue sub-spectra. This metric is calculated as mean squared difference between the model-estimated and actually observed RGB levels, divided by the actually observed value; the metric helps to assess differences between the estimated and actual RGB levels on an *absolute scale*;

(iii) Contrast similarity index between the original and model-predicted RGB images. This measure generates a pairwise comparison of local standard deviations of the signals from the original and model-generated images [53]. In our analysis, this indicator was used to compare the *spatial patterns* of differences between light intensities of a variety of restored RGB images and corresponding RGB originals. The calculations of the index were performed in MATLAB v.R2020x software using its structural similarity computing module [54], while setting the exponents of two other terms, that is, luminance and structural terms, to *zero*.

(iv) Consistency of the estimated obtained using the aforementioned metrics – Pearson correlation, WMSE, and the contrast similarity index, – was estimated as the geometric mean of the ratio between the *average* value and *standard deviations* of a given measure, assessed for the training and testing sets, respectively. The consistency was considered a measure of *universality* of the modeling approach.

## III. RESULTS

*A. General comparison of the models' performance*

Fig. 3 & Figs. A3-A9 report results of the analysis, in which different models are estimated for one metropolitan area (Haifa) and then applied to either this area (Fig. 3) or to *seven* other metropolitan areas under analysis (Figs. A3-A9 in Appendix). (For the reader's convenience, we also report day-time images of all study areas in Fig.1 (a) and Fig. A10 in Appendix.) In particular, each of the Figs. 3&A3-A9 report the original ISS RGB image, resized to the spatial resolution of the corresponding panchromatic VIIRS/DNB image, and, next to it, RGB images generated from panchromatic ALAN VIIRS/DNB images and HBASE maps. The figures also report several assessment criteria – Person correlation, WMSE, and contrast similarity. Although we performed similar assessments for *all other* metropolitan areas, by applying the models estimated for one of them to all the "counterpart" geographical areas, in the following discussion, we report only general statistics of such assessments (see Fig. 4 and Table I), while the RGB images generated thereby are not reported in the following discussion, for brevity's sake, and can be obtained from the authors upon request.

As Figs. 3&A3-A9 show, the model-generated RGB maps are, in all cases, visually similar to the original ISS RGB data. In addition, the models' performance measures show a close correspondence between original and model-generated RGB images, with Pearson correlation coefficients, both for training and testing sets, ranging between 0.719 and 0.963, WMSE varying from 0.029 to 4.223 and contrast similarity ranging from 0.931 to 0.993 (see Table AV; For corresponding statistics for other case studies covered by the analysis, see Tables AIII-AX in Appendix).

Fig. 4, which mutually compares the performance of linear regressions, kernel regressions, random forest regression, and elastic maps built under different bending penalties, for training and testing sets, also shows that models-generated RGB estimates demonstrate a high degree of correspondence with the original ISS RGB data. In particular, as Fig. 4 shows, Pearson correlation coefficients exceed in all cases, for both testing and training sets, 0.62, WMSE are smaller than 2.03, and contrast similarity is greater than 0.91 (91%), indicating a high level of correspondence with the original ISS data.

As Fig. 4 further shows, in terms of Pearson correlation coefficients and WMSE, random forest approach and kernel regressions perform somewhat better for *training sets* (with *r*=0.93-0.96 and *WMSE*=0.05-0.11 for random forest regressions and *r*=0.80-0.89 and *WMSE*=0.10-0.26 for kernel regressions vs. *r*=0.77-0.87 and *WMSE*=0.14-0.37 for linear regressions and *r*=0.69-0.85 and *WMSE*=0.12-0.59 for elastic map models). However, for *testing sets*, in terms of Pearson correlations, linear regression *outperforms* other modeling methods (*r*=0.75-0.85 vs. *r*=0.70-0.84 for random forest regressions, *r*=0.68-0.85 for kernel regressions, and *r*=0.62-0.82 for elastic map models). Concurrently, in terms of WMSE, linear regressions also perform better for the *blue* light band (*WMSE*=0.81 vs. *WMSE*=1.05 for random forest regression, *WMSE*=0.91 for kernel regression, and *WMSE*=0.97-1.17 for elastic map models), while random forest regressions perform better for the *red* and *green* light sub-spectra (*WMSE*=1.04-1.16, compared to *WMSE*=1.18-1.12 for kernel regressions,

*WMSE*=1.44-1.70 for linear regressions, and *WMSE*=1.17-2.03 for elastic map models). In terms of contrast similarity (*C_sim*), random forest models demonstrate better performance, for both training and testing sets (*C_sim*=0.931-0.989 *vs. C_sim*=0.913-0.966 for linear regressions, *C_sim*=0.922-0.973 for kernel regressions, and *C_sim*=0.917-0.979 for elastic maps models).

Table I reports consistency assessment of the models' performance across training and testing sets. As the table shows, in most cases, elastic map models outperform both linear, kernel, and random forest regressions, except for Pearson's correlation coefficients' consistency, assessed for green light datasets, for which linear regression outperforms other methods (*r*=0.979 *vs. r*=0.976 for elastic map models, *r*=0.804 for kernel regressions, and *r*=0.500 for random forest regressions).

### B. Factors affecting light flux in different RGB bands

As hypothesized in Section IIA, different types of land-use tend to emit nighttime lights, being different in terms of light intensity and spectra. This fact potentially enables a successful extraction of RGB information from panchromatic ALAN images. To verify this hypothesis, we ran multiple regression models, linking the set of predictors, described in Section IIA, with light intensities in different spectra – either red, green, or blue. We estimated the models for all eight study-datasets *together*, to identify a general trend.

Table II reports the results of this analysis and confirms the above hypothesis overall. In particular, as the models' pairwise comparison shows, differences between regression coefficients estimated for different RGB models are statistically significant for all the variables under analysis (*P*<0.01).

The table also indicates that, in line with our initial research hypothesis, *different RGB intensities are associated with different strength with different features in the panchromatic image and different land-use attributes. First*, panchromatic *ALAN intensities* contributes more to the Red and Green light emissions than to the Blue ones (**M1**: *t*=171.61; *P*<0.01 vs. **M2**: *t*=197.63; *P*<0.01 vs. **M3**: *t*=158.12; *P*<0.01). *Second*, Blue spectrum intensities appear to be strongly and negatively associated with the average percent of built-up area (**M3**: *t*=-4.90; *P*<0.01), while, in contrast, ALAN emissions in the Red and Green sub-spectra exhibit positive associations with built area percent (**M1**: *t*=32.77; *P*<0.01 and **M2**: *t*=16.68; *P*<0.01). *Third*, ALAN–Mean Diff. appears to be significantly and negatively associated with ALAN emissions in the Red and Green spectra, while its association with the Blue spectrum emissions is much weaker (**M1**: *t*=-20.99; *P*<0.01 vs. **M2**: *t*=-16.80; *P*<0.01 vs. **M3**: *t*=-3.80; *P*<0.01). *Lastly*, the ALAN–Max Diff. variable is positively and highly significantly associated with the Red and Green sub-spectra, while this variable is *insignificant* in the model, estimated for the Blue sub-spectrum (**M1**: *t*=26.78; *P*<0.01 vs. **M2**: *t*=20.32; *P*<0.01 vs. **M3**: *t*=1.41; *P*>0.1).

### C. Factor contribution test

Since none of the kernel, random forest regression, and elastic map models provide explicit estimates of the explanatory variables' coefficients, which multiple regression analysis enables (see Table II), we implemented a different strategy for a *cross-model* comparison. In particular, we explored the relative roles of different predictors by excluding them from the models one by one, and assessing the change in the models' performance attributed to such exclusions. As Fig. 5 shows, the factor ranking appears to be similar in all types of the models, with ALAN contributing most to the *r*-change (Δ*r*=0.197-0.251 for linear regressions, Δ*r*=0.131-0.159 for kernel regressions, Δ*r*=0.180-0.203 for elastic map models, Δ*r*=0.043-0.048 for random forest regressions). In compare to this major contribution, the relative contribution of the HBASE-based predictors, such as HBASE mean and standard deviation, is smaller (reaching Δ*r*=0.13, depending on the model). Yet, this contribution is not negligible, and varies by the ALAN band, which may be crucial for some applications (see an example reported in Fig. A11). The inter-pixel ALAN differences emerge third (Δ*r*<0.010 for all model types).

## IV. DISCUSSION

One of the most important findings of the study is that different predictors have different loadings on the explained variance of the Red, Green, and Blue ALAN emissions. In particular, as the multiple regression analysis shows, the association between panchromatic ALAN intensities appears to be stronger for the Red and Green sub-spectra, in compare to the Blue sub-spectrum. This difference may be explained by a smaller overlapping diapason of relative spectral sensitivities of the Blue channel (in comparison to the Red and Green diapasons) provided by the DSLR cameras, used by ISS astronauts, and that of the VIIRS/DNB sensor (see Fig. A1 in Appendix).

The regression coefficients for the mean and max ALAN-diff. indices also emerged with different strengths in different RGB bands, being stronger in the Red and Green band models than in the Blue band models. To understand these differences, we should keep in mind that ALAN-diff. can be *negative* for mean-difference, and *positive* for ALAN-max differences only if the following conditions are met: (i) a pixel is, *on the average, dimmer* than the adjacent pixels, but is (ii) *brighter* than, at least, one of the neighboring pixels. Such a situation might happen if a pixel in question is located at the edge of a lit area. As a result, it may not stand out against its surroundings. Since Red and Green lights are more associated with moderately lit residential areas (unlike industrial and commercial facilities often lit by Blue lights), we assume the aforementioned effect is more pronounced in the Red and Green light models.

In addition, percent of the built-up areas emerged *positive* in the Red and Green lights models, and *negative* in the Blue light model. Built-up area SD also emerged negative, being weaker in the Blue-light model than in the Red and Green light models. This phenomenon may be attributed to the fact that Red and Green lights are associated with densely and homogenously lit residential areas, while Blue lights may be more common in industrial and commercial areas, which are characterized by more sparse and heterogeneous illumination patterns. Kernel-based, random forest, and elastic maps models generally confirm these associations.

Another important finding of the study is that different models differ in performance, when used to convert panchromatic ALAN images into RGB. In particular, as the study reveals, random forest and non-linear kernel regression models generally perform well in terms of Pearson correlation, WMSE, and contrast similarity index for *training* sets, while multiple linear regressions outperform, in most cases, other methods for *testing* sets. As we suggest, this difference is due to the flexibility of random forest and kernel regressions, which helps to fit the training data more precisely, while linear regressions fare better in capturing *trends*. Concurrently, in terms of consistency of the models' performance estimated for training and testing sets, elastic map models, built under predominantly medium bending penalty, fared better than other model types. Given medium bending penalties, elastic map models also show better performance, compared to less and more bent counterparts, thus indicating diminishing benefits of under- and over-smoothing.

To the best of our knowledge, this study is the first that attempts to extract RGB information from panchromatic nighttime imagery, which determines its novelty. We should emphasize that this task is different from the gray-scale image or movie colorization task, which is based on the analysis of semantically similar images with matching the luminance and texture information and selecting the best color among a set of color candidates (see for example [55]–[59]). The present task, though, might be considered as similar to the day-time satellite or aerial image colorization task, which is used to enrich past images, to make them comparable with the present-day images or to obtain color images at the spatial resolution of their panchromatic counterparts (see [52], [60], [61]). Night-time satellite imagery, though, is way worse in terms of spatial resolution; thus, use the features' peculiarities, such as texture, is not beneficial. To overcome this difficulty, we use auxiliary HBASE data, to compensate this drawback.

The importance of the proposed approach is due to a possibility of obtaining *seamless RGB data coverage from panchromatic ALAN images*, which are widely available today globally with various temporal frequencies. In its turn, generating RGB information from freely available or easy-to-compute information from panchromatic nighttime imagery and built-up area data might contribute to research advances in different fields, by enabling more accurate analysis of various human economic activities and by opening more opportunities for ecological research. In particular, the panchromatic-to-RGB image conversion may enable studies of different health effects, associated with ALAN exposures to different sub-spectra, such e.g., breast and prostate cancers. The conversion in question may also help to correct a bias in the light pollution estimates, obtained from panchromatic VIIRS/DNB ALAN imagery by widening their spectrum sensitivity diapason.

One important question needs to be answered: Would colorized VIIRS images actually help empirical studies to obtain more robust effect estimates? To address the issue, we used breast cancer (BC) data, reported in [62], and compared the strength of association between BC rates and uncolored VIIRS image, and, then, between BC rates and the Blue, Green, and Red bands of the colored image generated for the Haifa metropolitan area using the random forest modelling approach. We run both the models incorporating the full set of predictors and also a truncated set, from which the HBASE-based predictors were excluded. Figure A11 reports the results of such comparison. As figure A11 *(a)* shows, for most BC rate cut-off thresholds, the association between the observed BC rates is consistently higher for blue lights than either for panchromatic or green- and red-band lights. This result is fully consistent with existing empirical evidence about most efficient melatonin suppression by short wavelength (blue) illumination ([28], [63], [64]), potentially associated with elevated risk of hormone-dependent cancers [65]. Importantly, the RGB estimates, obtained from the models without HBASE-based predictors (Fig. A11 *(b)*) show weaker and more similar one to another correlations with BC DKD rates, which evidences that retaining the HBASE predictors in the models is beneficial.

Another concern might arise about spatial autocorrelation between observations used in the modelling. If testing and training sets are geographically close, then this problem is crucially important and would require special methods [66]. In our analysis, however, we used observations from different cities to form the training and testing sets. In particular, if a city was used for training, it was excluded from testing and vice versa. Then, another city was used for training but was included from the testing set, and so on. By way of this, training and testing sets always referred to different geographical regions, without any spatial overlap between them. Therefore, no spatial autocorrelation between training and testing sets was present.

Several *limitations of the study* are yet to be mentioned. *First and foremost*, VIIRS/DNB reports panchromatic ALAN intensities in physical units ($nW/cm^2/sr$), while ISS-provided imagery reports raw data in digital numbers (DN) and, therefore, a direct comparison between the two might be problematic. However, since we do not mutually compare red, green, and blue light levels, but only compare each of them separately with panchromatic ALAN intensities, this consideration is less critical, and should not affect the results of our analysis substantially. Furthermore, as conversion of digital numbers into physical quantities should conform linear (or near-linear) transformation, our results are unlikely to be distorted by such a conversion. *Second*, it should be acknowledged that a time lag between the year-2010 HBASE-based predictors and years-2015/17 nighttime VIIRS data exists. However, this lag is not expected to influence the study's results crucially, since the built-up coverage of major metropolitan areas tends to stabilize in recent years [67]. Considering its 30-m resolution, the HBASE database is sufficient for the study, in which the observations are aggregated into the 750x750 m grids. However, to address the time gap between ALAN and HBASE datasets, future studies may consider other data sources for urban grey estimation, based on either Sentinel-2, Landsat 8, or Tan-DEM-X imagery [68].

*Third*, our analysis revealed some peculiar cases which demonstrate relatively poor low applicability of our models to some test datasets. One example is the application of the models estimated for Haifa and Naples to Atlanta, for which high WMSE levels of red and green light levels for testing sets were reported (see e.g., Tables AV and AVIII in Appendix). This suggests that the proposed approach should be further refined. It would be tenable to expect that mixing the observations from training and testing sets would result in a better performance of the models. We checked this assumption pooling the observations of all the eight cities together. This pooled dataset was 10 times randomly split into training and testing sets at the 90/10 ratio, and multiple linear regressions were used to estimate Pearson correlations between the estimated and actual RGB levels. The analysis, however, indicated no substantial improvement compared to models' performance in our main experiment ($r$=0.756-0.839 in the new experiment *vs.* $r$=0.745-0.871 in the previously reported experiment). Additionally, we suggest, to improve the performance of the models, in future studies, other combinations of predictors can be tested, and outlier analysis can be improved, by using alternative procedures for data normalization, and experimenting with elastic maps' pre-defined parameters. As we expect, these procedures will make it possible to obtain more robust results and thus to improve generic and area-specific algorithms used for predicting polychromatic ALAN intensities.

## V. Conclusions

The present analysis tests the possibility of generating RGB information from panchromatic ALAN images, combined with freely available, or, easy-to-compute, land-use proxies. As we hypothesized from the outset of the analysis, since different land-use types emit night-time light of different intensity and spectrum, it might be possible to extract RGB information from panchromatic ALAN-images, coupled with built-up-area-based predictors. To verify this possibility, we use ISS nighttime RGB images available for eight major metropolitan areas worldwide – Atlanta, Beijing, Haifa, Khabarovsk, London, Naples, Nashville, and Tianjing. In the analysis, four different data modeling approaches are used and their performance mutually compared – multiple linear regressions, non-linear kernel regressions, random forest regressions, and elastic map models. During the analysis, the dataset for each geographical site is used, once at a time, as a training set, while other datasets – as testing sets. To assess the models' performance, we use different measures of correspondence between the observed and model-estimated RGB data: Pearson correlation, WMSE, contrast similarity, and consistency of the models' performance for training and testing sets. The analysis supports our research hypothesis about the feasibility of extracting RGB information from panchromatic ALAN images coupled with built-up-area-based predictors, pointing, however, that linear, kernel, and random forest regressions produce better estimates in terms of Pearson's correlation, WMSE, and contrast similarity, while elastic maps models perform better in terms of consistency of these indicators upon training and testing sets. The proposed approach confirms that panchromatic ALAN data, which are currently freely available globally on a daily basis, might be colorized into RGB images, to serve as a better proxy for the human presence on Earth.

## References

[1] P. Cinzano, F. Falchi, C. D. Elvidge, and K. E. Baugh, "The artificial night sky brightness mapped from DMSP satellite Operational Linescan System measurements," *Mon. Not. R. Astron. Soc.*, vol. 318, no. 3, pp. 641–657, Nov. 2000, doi: 10.1046/j.1365-8711.2000.03562.x.

[2] F. Falchi *et al.*, "The new world atlas of artificial night sky brightness," *Sci. Adv.*, vol. 2, no. 6, p. e1600377, Jun. 2016, doi: 10.1126/sciadv.1600377.

[3] F. Falchi *et al.*, "Light pollution in USA and Europe: The good, the bad and the ugly," *J. Environ. Manage.*, vol. 248, p. 109227, Oct. 2019, doi: 10.1016/j.jenvman.2019.06.128.

[4] J. Bennie, J. Duffy, T. Davies, M. Correa-Cano, and K. Gaston, "Global Trends in Exposure to Light Pollution in Natural Terrestrial Ecosystems," *Remote Sens.*, vol. 7, no. 3, pp. 2715–2730, Mar. 2015, doi: 10.3390/rs70302715.

[5] Z. Hu, H. Hu, and Y. Huang, "Association between nighttime artificial light pollution and sea turtle nest density along Florida coast: A geospatial study using VIIRS remote sensing data," *Environ. Pollut.*, vol. 239, pp. 30–42, Aug. 2018, doi: 10.1016/j.envpol.2018.04.021.

[6] C. D. Elvidge, K. E. Baugh, E. A. Kihn, H. W. Kroehl, E. R. Davis, and C. W. Davis, "Relation between satellite observed visible-near infrared emissions, population, economic activity and electric power consumption," *Int. J. Remote Sens.*, vol. 18, no. 6, pp. 1373–1379, 1997, doi: 10.1080/014311697218485.

[7] P. Sutton, D. Roberts, C. Elvidge, and K. Baugh, "Census from Heaven: An estimate of the global human population using night-time satellite imagery," *Int. J. Remote Sens.*, vol. 22, no. 16, pp. 3061–3076, Nov. 2001, doi: 10.1080/01431160010007015.

[8] S. Amaral, A. M. V. Monteiro, G. Camara, and J. A. Quintanilha, "DMSP/OLS night-time light imagery for urban population estimates in the Brazilian Amazon," *Int. J. Remote Sens.*, vol. 27, no. 5, pp. 855–870, Mar. 2006, doi: 10.1080/01431160500181861.

[9] L. Zhuo, T. Ichinose, J. Zheng, J. Chen, P. J. Shi, and X. Li, "Modelling the population density of China at the pixel level based on DMSP/OLS non-radiance-calibrated night-time light images," *Int. J. Remote Sens.*, vol. 30, no. 4, pp. 1003–1018, Feb. 2009, doi: 10.1080/01431160802430693.

[10] S. J. Anderson, B. T. Tuttle, R. L. Powell, and P. C. Sutton, "Characterizing relationships between population density and nighttime imagery for Denver, Colorado: Issues of scale and representation," *Int. J. Remote Sens.*, vol. 31, no. 21, pp. 5733–5746, 2010, doi: 10.1080/01431161.2010.496798.

[11] G. R. Hopkins, K. J. Gaston, M. E. Visser, M. A. Elgar, and T. M. Jones, "Artificial light at night as a driver of evolution across urban-rural landscapes," *Front. Ecol. Environ.*, vol. 16, no. 8, pp. 472–479, Oct. 2018, doi: 10.1002/fee.1828.

[12] C. H. Doll, J.-P. Muller, and C. D. Elvidge, "Night-time Imagery as a Tool for Global Mapping of Socioeconomic Parameters and Greenhouse Gas Emissions," *AMBIO A J. Hum. Environ.*, vol. 29, no. 3, pp. 157–162, May 2000, doi: 10.1579/0044-7447-29.3.157.

[13] S. Ebener, C. Murray, A. Tandon, and C. C. Elvidge, "From wealth to health: Modelling the distribution of income per capita at the sub-national level using night-time light imagery," *Int. J. Health Geogr.*, vol. 4, Feb. 2005, doi: 10.1186/1476-072X-4-5.

[14] T. Ghosh, R. L Powell, C. D Elvidge, K. E Baugh, P. C Sutton, and S. Anderson, "Shedding light on the global distribution of economic activity," *Open Geogr. J.*, vol. 3, no. 1, 2010.

[15] J. V. Henderson, A. Storeygard, and D. N. Weil, "Measuring economic growth from outer space," *Am. Econ. Rev.*, vol. 102, no. 2, pp. 994–1028, 2012.

[16] C. Mellander, J. Lobo, K. Stolarick, and Z. Matheson, "Night-time light data: A good proxy measure for economic activity?," *PLoS One*, vol. 10, no. 10, 2015.

[17] R. Wu, D. Yang, J. Dong, L. Zhang, and F. Xia, "Regional Inequality in China Based on NPP-VIIRS Night-Time Light Imagery," *Remote Sens.*, vol. 10, no. 2, p. 240, Feb. 2018, doi: 10.3390/rs10020240.

[18] I. Kloog, A. Haim, R. G. Stevens, and B. A. Portnov, "Global co-distribution of light at night (LAN) and cancers of prostate, colon,

[19] I. Kloog, R. G. Stevens, A. Haim, and B. A. Portnov, "Nighttime light level co-distributes with breast cancer incidence worldwide," *Cancer Causes Control*, vol. 21, no. 12, pp. 2059–2068, 2010, doi: 10.1007/s10552-010-9624-4.

[20] N. A. Rybnikova, A. Haim, and B. A. Portnov, "Does artificial light-at-night exposure contribute to the worldwide obesity pandemic?," *Int. J. Obes.*, vol. 40, no. 5, pp. 815–823, 2016, doi: 10.1038/ijo.2015.255.

[21] N. Levin *et al.*, "Remote sensing of night lights: A review and an outlook for the future," *Remote Sens. Environ.*, vol. 237, p. 111443, Feb. 2020, doi: 10.1016/j.rse.2019.111443.

[22] M. O. Román *et al.*, "NASA's Black Marble nighttime lights product suite," *Remote Sens. Environ.*, vol. 210, pp. 113–143, Jun. 2018, doi: 10.1016/j.rse.2018.03.017.

[23] C. D. Elvidge, K. E. Baugh, M. Zhizhin, and F.-C. Hsu, "Why VIIRS data are superior to DMSP for mapping nighttime lights," *Proc. Asia-Pacific Adv. Netw.*, vol. 35, no. 62, 2013.

[24] "LP DAAC - Homepage." https://lpdaac.usgs.gov/ (accessed Apr. 07, 2020).

[25] N. A. Rybnikova and B. A. Portnov, "Remote identification of research and educational activities using spectral properties of nighttime light," *ISPRS J. Photogramm. Remote Sens.*, vol. 128, pp. 212–222, 2017, doi: 10.1016/j.isprsjprs.2017.03.021.

[26] J. Veitch, G. Newsham, P. Boyce, and C. Jones, "Lighting appraisal, well-being and performance in open-plan offices: A linked mechanisms approach," *Light. Res. Technol.*, vol. 40, no. 2, pp. 133–151, Jun. 2008, doi: 10.1177/1477153507086279.

[27] E. Guk and N. Levin, "Analyzing spatial variability in night-time lights using a high spatial resolution color Jilin-1 image – Jerusalem as a case study," *ISPRS J. Photogramm. Remote Sens.*, vol. 163, pp. 121–136, May 2020, doi: 10.1016/j.isprsjprs.2020.02.016.

[28] C. Cajochen *et al.*, "High Sensitivity of Human Melatonin, Alertness, Thermoregulation, and Heart Rate to Short Wavelength Light," *J. Clin. Endocrinol. Metab.*, vol. 90, no. 3, pp. 1311–1316, Mar. 2005, doi: 10.1210/jc.2004-0957.

[29] C. A. Czeisler, "Perspective: Casting light on sleep deficiency," *Nature*, vol. 497, no. 7450, p. S13, May 2013, doi: 10.1038/497S13a.

[30] C. D. Elvidge, D. M. Keith, B. T. Tuttle, and K. E. Baugh, "Spectral identification of lighting type and character," *Sensors*, vol. 10, no. 4, pp. 3961–3988, Apr. 2010, doi: 10.3390/s100403961.

[31] "Search Photos." https://eol.jsc.nasa.gov/SearchPhotos/ (accessed Apr. 07, 2020).

[32] "Cities at night – mapping the world at night." https://citiesatnight.org/ (accessed Apr. 07, 2020).

[33] J. D. Hale, G. Davies, A. J. Fairbrass, T. J. Matthews, C. D. F. Rogers, and J. P. Sadler, "Mapping Lightscapes: Spatial Patterning of Artificial Lighting in an Urban Landscape," *PLoS One*, vol. 8, no. 5, May 2013, doi: 10.1371/journal.pone.0061460.

[34] M. Herold, X. H. Liu, and K. C. Clarke, "Spatial metrics and image texture for mapping urban land use," *Photogrammetric Engineering and Remote Sensing*, vol. 69, no. 9. American Society for Photogrammetry and Remote Sensing, pp. 991–1001, Sep. 01, 2003, doi: 10.14358/PERS.69.9.991.

[35] A. Gorban and A. Zinovyev, "Elastic principal graphs and manifolds and their practical applications," *Computing*, vol. 75, no. 4, pp. 359–379, 2005.

[36] "Earth Observation Group." https://ngdc.noaa.gov/eog/download.html (accessed Mar. 17, 2020).

[37] "HBASE Dataset From Landsat." https://sedac.ciesin.columbia.edu/data/set/ulandsat-hbase-v1/data-download (accessed Apr. 09, 2020).

[38] P. Wang, C. Huang, E. C. Brown de Colstoun, J. C. Tilton, and B. Tan, "Global Human Built-up And Settlement Extent (HBASE) Dataset From Landsat. Palisades, NY: NASA Socioeconomic Data and Applications Center (SEDAC)." 2017.

[39] "How Digital Cameras Work." https://www.astropix.com/html/i_astrop/how.html (accessed Jul. 26, 2020).

[40] "ArcGIS Online | Cloud-Based GIS Mapping Software." https://www.esri.com/en-us/arcgis/products/arcgis-online/overview (accessed Jan. 14, 2021).

[41] M. O. Román, Z. Wang, R. Shrestha, T. Yao, and V. Kalb, "Black

[42] Marble User Guide Version 1.0," 2019.

[42] A. N. Gorban and A. Zinovyev, "Principal manifolds and graphs in practice: From molecular biology to dynamical systems," *Int. J. Neural Syst.*, vol. 20, no. 3, pp. 219–232, Jun. 2010, doi: 10.1142/S0129065710002383.

[43] T. Hastie, R. Tibshirani, and J. Friedman, "The Elements of Statistical Learning: Data Mining, Inference, and Prediction ... - Trevor Hastie, Robert Tibshirani, Jerome Friedman - Google Книги," *Springer*, 2017. https://books.google.co.il/books?hl=ru&lr=&id=tVIjmNS3Ob8C&oi=fnd&pg=PR13&dq=Tibshirani+Nonparametric+Regression+Statistical+Machine+Learning&ots=ENKbMaD2Z1&sig=WQcXNywDZyHTk3mGlfSW1UZE67Y&redir_esc=y#v=onepage&q=Tibshirani Nonparametric Regression Statis (accessed Jan. 22, 2020).

[44] L. Breiman, "Random forests," *Mach. Learn.*, vol. 45, no. 1, pp. 5–32, 2001.

[45] U. von Luxburg and B. Schölkopf, *Statistical Learning Theory: Models, Concepts, and Results*, vol. 10. North-Holland, 2011.

[46] B. Grung and R. Manne, "Missing values in principal component analysis," *Chemom. Intell. Lab. Syst.*, vol. 42, no. 1–2, pp. 125–139, Aug. 1998, doi: 10.1016/S0169-7439(98)00031-8.

[47] "GitHub - Elastic map." https://github.com/Mirkes/ElMap (accessed Apr. 11, 2020).

[48] "SPSS Software | IBM." https://www.ibm.com/analytics/spss-statistics-software (accessed Mar. 17, 2020).

[49] M. P. Wand and M. C. Jones, *Kernel Smoothing - M.P. Wand, M.C. Jones - Google Книги*. .

[50] "Fit Gaussian kernel regression model using random feature expansion - MATLAB fitrkernel." https://www.mathworks.com/help/stats/fitrkernel.html (accessed May 21, 2020).

[51] "Create bag of decision trees - MATLAB." https://www.mathworks.com/help/stats/treebagger.html (accessed Jan. 05, 2021).

[52] D. Seo, Y. Kim, Y. Eo, and W. Park, "Learning-Based Colorization of Grayscale Aerial Images Using Random Forest Regression," *Appl. Sci.*, vol. 8, no. 8, p. 1269, Jul. 2018, doi: 10.3390/app8081269.

[53] Z. Wang, A. C. Bovik, H. R. Sheikh, and E. P. Simoncelli, "Image quality assessment: From error visibility to structural similarity," *IEEE Trans. Image Process.*, vol. 13, no. 4, pp. 600–612, Apr. 2004, doi: 10.1109/TIP.2003.819861.

[54] "Structural similarity (SSIM) index for measuring image quality - MATLAB ssim." https://www.mathworks.com/help/images/ref/ssim.html (accessed May 26, 2020).

[55] T. Welsh, M. Ashikhmin, and K. Mueller, "Transferring color to greyscale images," in *Proceedings of the 29th Annual Conference on Computer Graphics and Interactive Techniques, SIGGRAPH '02*, 2002, pp. 277–280, doi: 10.1145/566570.566576.

[56] R. K. Gupta, A. Y. S. Chia, D. Rajan, E. S. Ng, and H. Zhiyong, "Image colorization using similar images," in *MM 2012 - Proceedings of the 20th ACM International Conference on Multimedia*, 2012, pp. 369–378, doi: 10.1145/2393347.2393402.

[57] A. Bugeau, V. T. Ta, and N. Papadakis, "Variational exemplar-based image colorization," *IEEE Trans. Image Process.*, vol. 23, no. 1, pp. 298–307, Jan. 2014, doi: 10.1109/TIP.2013.2288929.

[58] R. Zhang, P. Isola, and A. A. Efros, "Colorful image colorization," in *Lecture Notes in Computer Science (including subseries Lecture Notes in Artificial Intelligence and Lecture Notes in Bioinformatics)*, 2016, vol. 9907 LNCS, pp. 649–666, doi: 10.1007/978-3-319-46487-9_40.

[59] G. Larsson, M. Maire, and G. Shakhnarovich, "Learning representations for automatic colorization," in *Lecture Notes in Computer Science (including subseries Lecture Notes in Artificial Intelligence and Lecture Notes in Bioinformatics)*, 2016, vol. 9908 LNCS, pp. 577–593, doi: 10.1007/978-3-319-46493-0_35.

[60] H. Liu, Z. Fu, J. Han, L. Shao, and H. Liu, "Single satellite imagery simultaneous super-resolution and colorization using multi-task deep neural networks," *J. Vis. Commun. Image Represent.*, vol. 53, pp. 20–30, May 2018, doi: 10.1016/j.jvcir.2018.02.016.

[61] M. Gravey, L. G. Rasera, and G. Mariethoz, "Analogue-based colorization of remote sensing images using textural information," *ISPRS J. Photogramm. Remote Sens.*, vol. 147, pp. 242–254, Jan. 2019, doi: 10.1016/j.isprsjprs.2018.11.003.

[62] N. A. Rybnikova and B. A. Portnov, "Outdoor light and breast cancer incidence: a comparative analysis of DMSP and VIIRS-DNB satellite data," *Int. J. Remote Sens.*, vol. 38, no. 21, pp. 5952–5961, Nov. 2017, doi: 10.1080/01431161.2016.1246778.

[63] G. C. Brainard *et al.*, "Action spectrum for melatonin regulation in humans: Evidence for a novel circadian photoreceptor," *J. Neurosci.*, vol. 21, no. 16, pp. 6405–6412, Aug. 2001, doi: 10.1523/jneurosci.21-16-06405.2001.

[64] H. R. Wright, L. C. Lack, and D. J. Kennaway, "Differential effects of light wavelength in phase advancing the melatonin rhythm," *J. Pineal Res.*, vol. 36, no. 2, pp. 140–144, Feb. 2004, doi: 10.1046/j.1600-079X.2003.00108.x.

[65] A. Haim and B. A. Portnov, *Light pollution as a new risk factor for human breast and prostate cancers*. Springer Netherlands, 2013.

[66] M. Wurm, T. Stark, X. X. Zhu, M. Weigand, and H. Taubenböck, "Semantic segmentation of slums in satellite images using transfer learning on fully convolutional neural networks," *ISPRS J. Photogramm. Remote Sens.*, vol. 150, pp. 59–69, Apr. 2019, doi: 10.1016/j.isprsjprs.2019.02.006.

[67] M. Kasanko *et al.*, "Are European cities becoming dispersed?. A comparative analysis of 15 European urban areas," *Landsc. Urban Plan.*, vol. 77, no. 1–2, pp. 111–130, Jun. 2006, doi: 10.1016/j.landurbplan.2005.02.003.

[68] "Global and Continental Urban Extent / Settlement Layers: Summary Characteristics | POPGRID." https://www.popgrid.org/data-docs-table4 (accessed Jan. 05, 2021).

[69] P. A. N. Gorban and D. A. Y. Zinovyev, "Visualization of Data by Method of Elastic Maps and Its Applications in Genomics, Economics and Sociology," 2001.

FIGURES AND TABLES

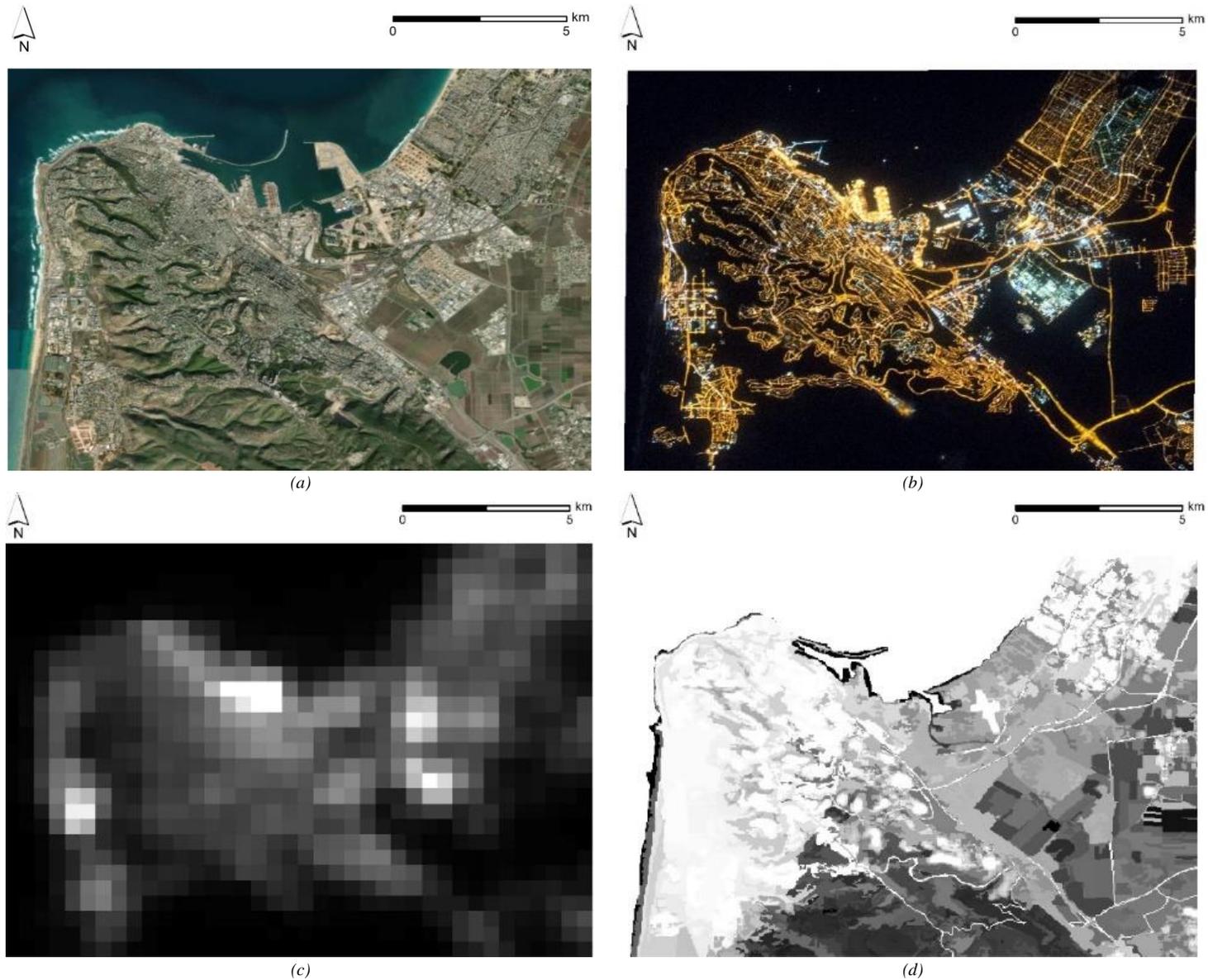

*(a)*

*(b)*

*(c)*

*(d)*

Fig. 1. Satellite images of the Haifa metropolitan area: *(a)* day-time image ([40]), *(b)* ~10-meter resolution RGB image with the range of values of 0-255 dn for each band; *(c)* ~750-meter resolution panchromatic image with the values in the range of 1-293 nW/cm$^2$/sr, and *(d)* ~30-meter resolution HBASE image with the values in the 0-100 % range.

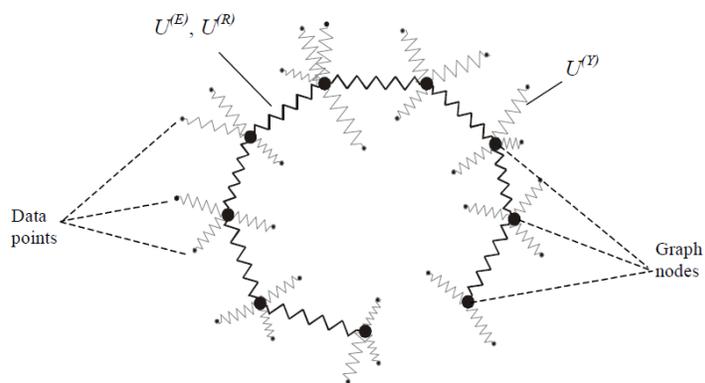

Fig. 2. Energies of elastic map (principal manifold approximation). The principal manifold is represented by a regular grid of nodes (large black circles) connected by attractive springs (shown by thick zigzag lines and representing the stretching energy). In addition, the triples of nodes in the grid are assigned the bending energy (not represented here). The data points shown by small circles are assigned to the closest node of the grid similarly to the $k$-means clustering. Then the data approximation term (Mean Squared Error) can be represented as the total elastic energy of springs connecting the data points and the grid nodes (thin zigzag lines here).
(*Source*: [69])

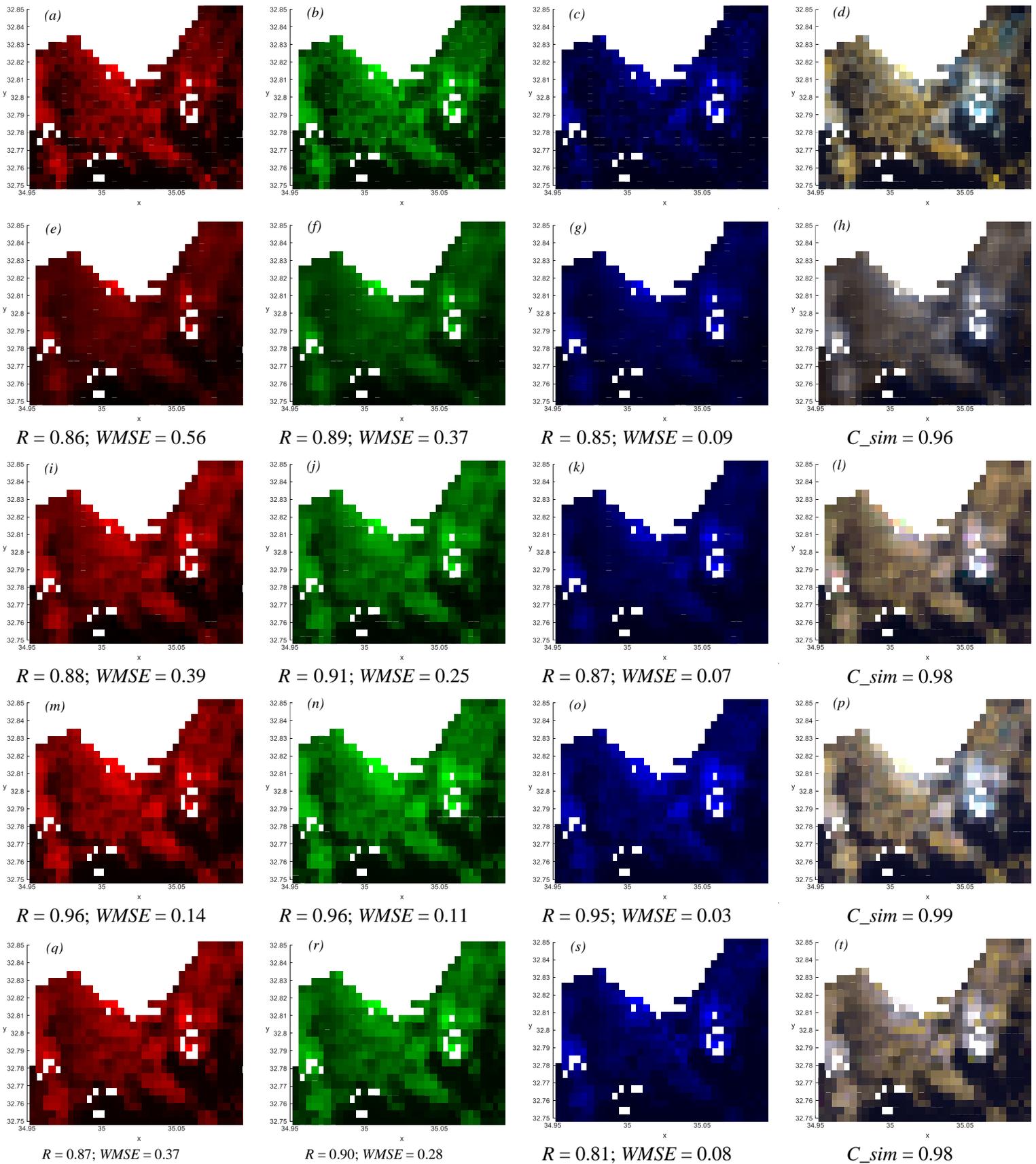

**Fig. 3.** *Haifa metropolitan area (Israel)*: Red (the first column), Green (the second column), Blue (the third column)) bands, and RGB images (the fourth column); ISS-provided, resampled to the spatial resolution of VIIRS imagery (the first row), and outputs of four models trained on *Haifa datasets*: linear multiple regressions (the second row), non-linear kernel regressions (the third row), random forest regressions (the fourth row), and elastic map models (the fifth row).

*Notes:* Output generated by elastic maps, built under the 0.05 bending penalty, is reported. *R* and *WMSE* denote correspondingly for Pearson's correlation and weighted mean squared error of the red, green, and blue lights' estimates, *C_sim* – for contrast similarity between restored and original RGB images. White points in the city area correspond to outliers.

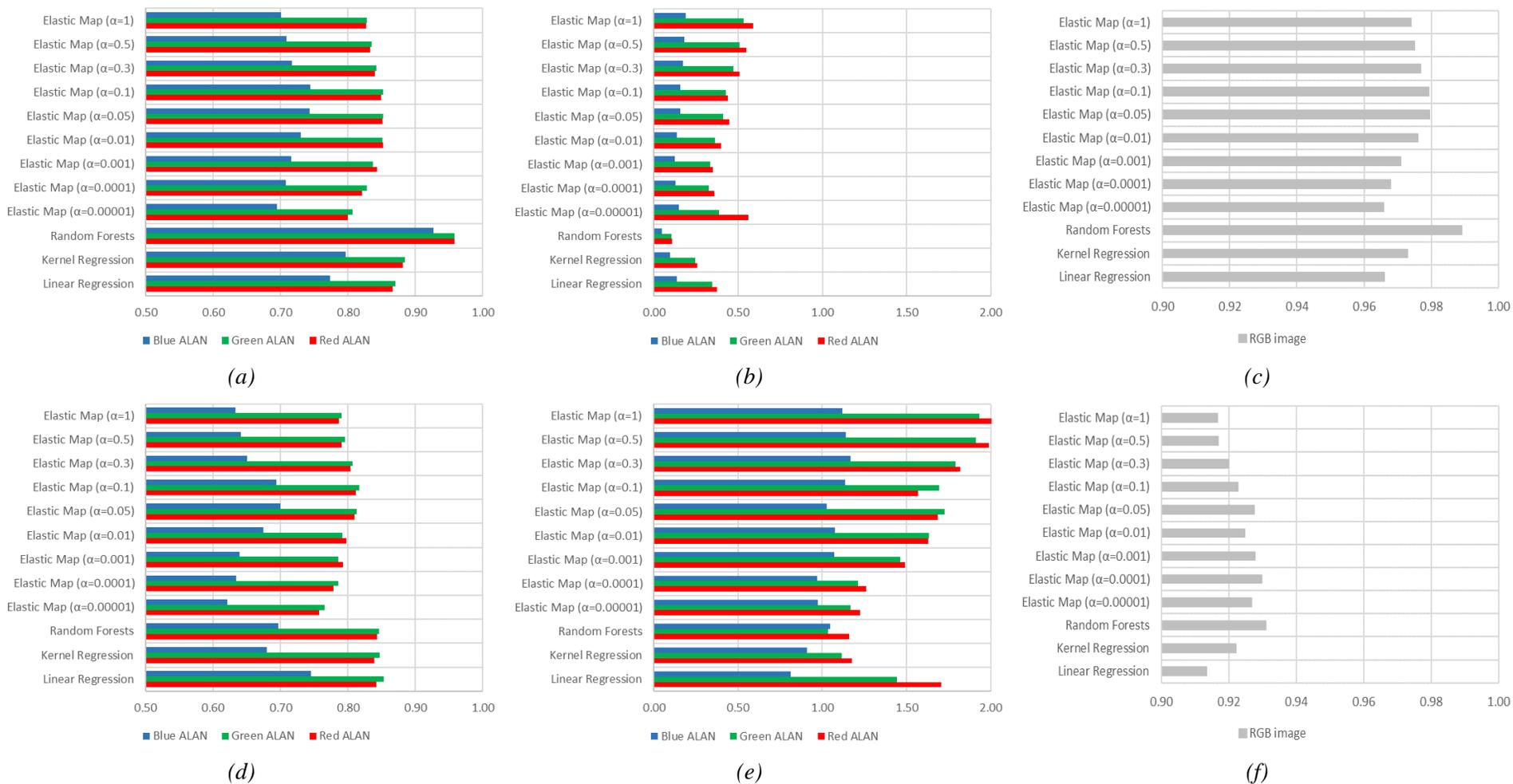

*(a)*          *(b)*          *(c)*

*(d)*          *(e)*          *(f)*

Fig. 4. Mutual comparison of linear, kernel, random forest, and elastic map models for the training (top row) and testing (bottom row) datasets, in terms of *averaged* Pearson correlation coefficients (*(a)* & *(d)*), WMSE (*(b)* & *(e)*), and contrast similarity (*(c)* & *(f)*).

*Notes:* In case of Pearson's correlation (*(a)* & *(d)*) and contrast similarity (*(c)* & *(f)*), greater means better; In case of WMSE (*(b)* & *(e)*), lower means better.

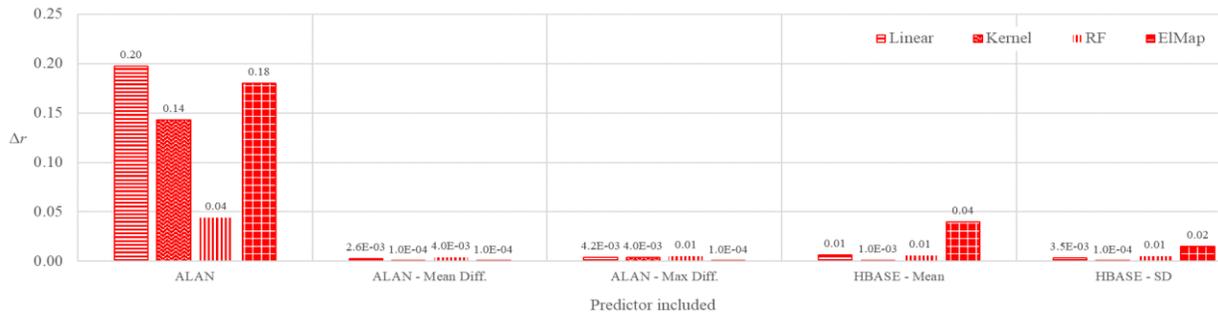

*(a)*

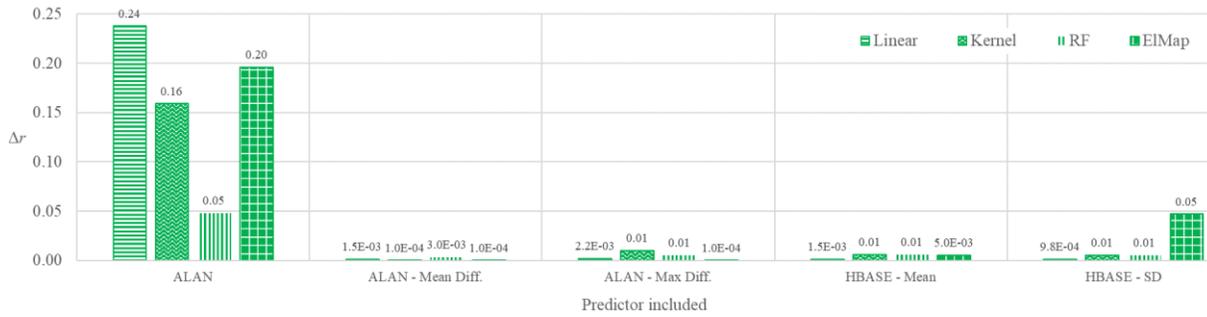

*(b)*

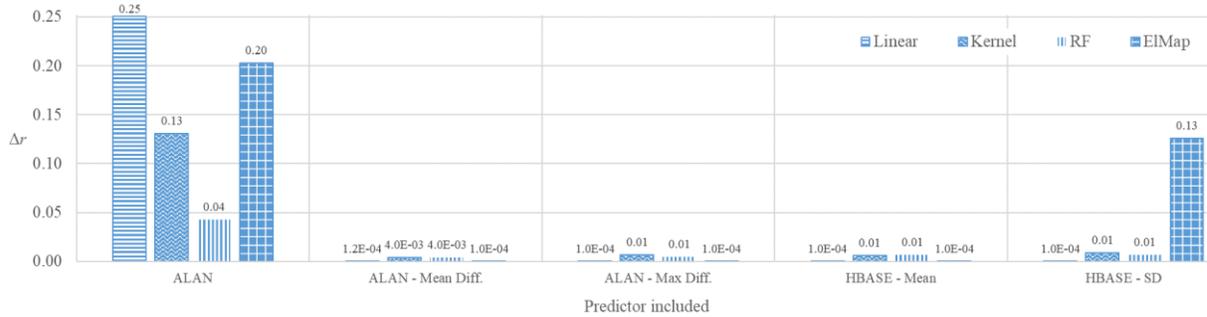

*(c)*

Fig. 5. Changes in the models' performance (Δ*r*), attributed to the exclusion of particular variables from the set of predictors, estimated separately for different model types (Study dataset: all metropolitan areas under analysis; N. of pixels/obs. = 33,846); the models are estimated separately for the Red *(a)*, Green *(b)*, and Blue *(c)* spectra

TABLE I
Mutual comparison of linear, kernel, random forest, and elastic map models in terms of estimate consistency for training and testing datasets

| | Model performance measure | | | | | | |
| Model type | Pearson correlation coefficient | | | WMSE | | | Contrast similarity |
| | R | G | B | R | G | B | RGB |
| Linear regression | 0.938 | 0.979 | 0.880 | 0.124 | 0.147 | 0.096 | 0.541 |
| Kernel regression | 0.680 | 0.804 | 0.572 | 0.135 | 0.148 | 0.059 | 0.514 |
| Random Forest regression | 0.472 | 0.500 | 0.400 | 0.061 | 0.074 | 0.028 | 0.357 |
| Elastic map model [1] | 0.970 [1a] | 0.976 | 0.975 [1b] | 0.371 [1c] | 0.257 [1c] | 0.108 [1c] | 0.598 [1d] |

The results of the best-performing model are reported with: [1a] $\alpha$=0.0001; [1b] $\alpha$=0.05; [1c] $\alpha$=0.00001; [1d] $\alpha$=0.001.
The grey cell backgrounds mark the best-performed model for specific measures.

TABLE II
The association between ALAN intensities in different RGB bands and predictors from the VIIRS and HBASE datasets (Study area – all geographical sites together (N. of pixels/obs. = 33,846); method – ordinary least square regression (OLS); dependent variables – ALAN intensities in different parts of the RGB spectra) and significance of differences in the regression coefficients

| | Models | | | | | | | Models' comparison | | | | | | | | |
| Predictors | **M1:** Dependent variable – ALAN intensity in the Red spectrum – dn | | **M2:** Dependent variable – ALAN intensity in the Green spectrum – dn | | **M3:** Dependent variable – ALAN intensity in the Blue spectrum – dn | | *VIF* | M1 vs. M2 | | | M1 vs. M3 | | | M2 vs. M3 | | |
| | $B$ | $t$ | $B$ | $t$ | $B$ | $t$ | | $\Delta B$ | SE | Sig. | $\Delta B$ | SE | Sig. | $\Delta B$ | SE | Sig. |
| (Constant) | 3.34 | (8.93)*** | 2.09 | (7.77)*** | 5.49 | (27.35)*** | - | - | - | - | - | - | - | - | - | - |
| ALAN | 0.98 | (171.61)*** | 0.81 | (197.63)*** | 0.49 | (158.12)*** | 1.90 | 0.17 | 0.003 | 0.00E0 | 0.50 | 0.005 | 0.00E0 | 0.33 | 0.003 | 0.00E0 |
| ALAN – Mean Diff. | -0.67 | (-20.99)*** | -0.39 | (-16.80)*** | -0.07 | (-3.80)*** | 4.02 | -0.28 | 0.015 | 4.98E-75 | -0.61 | 0.030 | 6.27E-92 | -0.32 | 0.016 | 2.64E-93 |
| ALAN – Max Diff. | 0.24 | (26.78)*** | 0.13 | (20.32)*** | 0.01 | (1.41) | 3.31 | 0.11 | 0.004 | 7.06E-138 | 0.23 | 0.008 | 8.79E-171 | 0.12 | 0.004 | 3.39E-175 |
| HBASE – mean | 0.15 | (32.77)*** | 0.05 | (16.68)*** | -0.01 | (-4.90)*** | 1.49 | 0.09 | 0.002 | 0.00E0 | 0.16 | 0.004 | 0.00E0 | 0.07 | 0.002 | 3.313E-194 |
| HBASE –SD | -0.32 | (-24.24)*** | -0.13 | (-13.71)*** | -0.01 | (-1.06) | 1.15 | -0.19 | 0.006 | 1.67E-191 | -0.31 | 0.012 | 6.55E-142 | -0.12 | 0.006 | 3.64E-80 |
| $R^2$ | 0.67 | | 0.70 | | 0.57 | | | F = (3487.79)*** | | | F = (5540.62)*** | | | F = (7074.78)*** | | |

B = unstandardized regression coefficients; *t* = *t*-statistics; *VIF* = variance of inflation; *, ** and *** indicate correspondingly 0.1, 0.05 and 0.01 significance levels

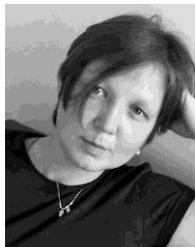

**Nataliya Rybnikova** was born in Lugansk, Ukraine, in 1981. She received her M.S. and Ph.D. degrees in economics from the Volodymyr Dahl Eastern Ukrainian National University, Ukraine, in 2004 and 2011. She received her second Ph.D. degree in remote sensing from the University of Haifa, Israel, in 2018.

From 2005 to 2010, she was a Researcher with the Cadastre of Natural Resources Laboratory, Ukraine. From 2011 to 2013, she was a Lecturer with the Dept. of Economics, Volodymyr Dahl Eastern-Ukrainian National University, Ukraine. From 2018 to 2019, she was a Postdoctoral Fellow with the Remote Sensing Laboratory, University of Haifa, Israel. Since 2019, she is a Postdoctoral Fellow with the Dept. of Mathematics, University of Leicester, UK, and School of Environmental Studies, University of Haifa, Israel. She is the author of more than 30 articles and book chapters. Her research interests include processing and using artificial night-time light imagery as a proxy for human presence on Earth.

Dr. Rybnikova is a member of the European Regional Science Association and the International Society for Photogrammetry and Remote Sensing. She was a recipient of the Israeli Ministry of Science, Technology and Space the Ilan Ramon Scholarship from 2015 to 2018; of the Israeli Regional Science Association Best Young Scientist's Paper Prize in 2018. Since 2019, she is a recipient of the Council for Higher Education of Israel Postdoctoral Excellence Scholarship.

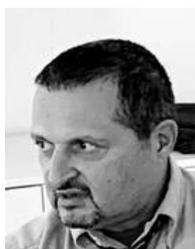

**Boris A. Portnov** was born in Odessa, former USSR. He received the M.A. degree in Architecture from the Poltava Civil Engineering Institute, Ukraine, in 1982; the Ph.D. degree in Urban Planning from the Central Scientific & Project Institute of Town-Building, Moscow, Russia, in 1987, and D.Sc. (2nd Russian Doctoral Degree) in urban planning from the Moscow Architectural Institute, Russia, in 1994.

From 1988 to 1995, he served as a Senior Lecturer, an Associate Professor, and a Full Professor in the Dept. of Architecture and Urban Planning, Krasnoyarsk Civil Engineering Institute – Siberian Federal University, USSR. From 1995 to 2002, he was a Senior Researcher, and then a Research Professor in the Dept. of Man in the Desert, Ben-Gurion University of the Negev, Israel. Since 2002, he was a Senior Research Fellow, an Associate Professor, and, in 2012, he was appointed a Full Professor in the Dept. of Natural Resources and Environmental Management, University of Haifa, Israel. Prof. Portnov authored or edited seven books, and more than 180 articles and book chapters. He is a member of the editorial board of several journals. His research interests include geographic information systems, urban planning, population geography, and real estate valuation and management.

Prof. Portnov was a recipient of the Ukraine State Committee of Town-Building Honor Diploma, Ukraine, in 1986; All-Union Architectural Competition 2nd Award, Russia, in 1989; Gelsenkirchen Municipality Redesigning of Gelsenkirchen Industrial district Summer School Award, Germany, in 1993; Russian Ministry of Higher Education, Increasing Land Use Effectiveness under Transition to a Market-Oriented Society Federal Text-Book Writing Award, Russia, in 1994; Deichmann Scholarship in Desert Studies in 1999-2001; 2000 PLEA Conference on 'Architecture and City Environment' Best Paper Award, UK, in 2000; Israel Central Bureau of Statistics, Award for Monograph Writing in 1998-2001; Gilladi Scholarship in 2001-2003; Sheikh Zayed Prize in Environmental Studies in 2006; Excellent Thesis Supervisor Award, Israel, in 2006; Ministry of Foreign Affairs, Republic of China - Taiwan Research Fellowship in 2011; the 1st International Conference on Sustainable Lighting and Light Pollution Best Paper Award, South Korea, in 2014. He is a member of Professional Union of Russian Architects, European Network for Housing Research, Association of Computer-Aided Design in Architecture, Israel Regional Science Association, Professional Union of Israeli Architects and Engineers, Association of Israeli Geographers, International Geographical Union, American Real Estate Society, and Network on European Communications and Transport Activities Research.

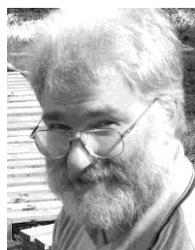

**Evgeny M. Mirkes** was born in Krasnoyarsk, Russia, in 1964. He received the M.S., Ph.D., and D.Sc. (2nd Russian Doctoral Degree) degrees in mathematics from the Krasnoyarsk State University, Russia, in 1985, 1990, and 2001, correspondingly.

From 2002 to 2012 he was a Professor with Dept. of Computer Science, Siberian Federal University, Russia. From 1993 to 2002 he was a Lecturer with Dept. of Computer Science, Krasnoyarsk State University, Russia. Since 2012, he is a Researcher with the School of Mathematics and Actuarial Science, University of Leicester, UK. He was a supervisor of 6 Ph.D. theses. He is the author of 6 books, more than 80 articles and book chapters. His research interests include biomathematics, data mining and software engineering, neural network and artificial intelligence.

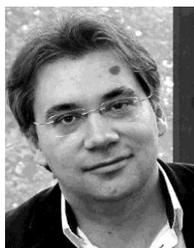

**Andrei Zinovyev** was born in Krasnoyarsk, Russia, in 1974. He received the M.S. degree in physics from Krasnoyarsk State University, Russia, in 1997 and Ph.D. degree in computer science from Institute of Computational Modeling of Russian Academy of Science in 2001.

From 2001 to 2004, he was a Postdoctoral Fellow at the Institut des Hautes Etudes Scientifiques, France. Since 2005, he is a Senior Researcher and the Leader of the Computational Systems Biology of Cancer group at Institut Curie, France. Since 2019, he is an Interdisciplinary Chair with the Paris Artificial Intelligence Research Institute, France. He is the author of three books, more than 120 articles and book chapters. His research interests include application of machine learning to biological and healthcare problems, learning latent spaces and structures in big data point clouds, cancer systems biology, modeling and dimension reduction in complex systems.

Dr. Zinovyev was a recipient of Student Soros Prize in 1994 and 1995 and the Agilent Thought Leader Award in 2014 as a part of the group.

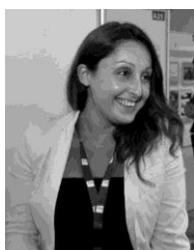

**Anna Brook** Anna Brook was born in Tbilisi, Georgia, in 1981. She received her Ph.D. degree in remote sensing and spectroscopy from Tel-Aviv University, Israel, in 2010.

From 2011 to 2013, she was a Researcher with the CISS Department at the Royal Military Academy of Belgium. Since 2014, she is a Lecturer with Dept. of geography and environmental studies, the University of Haifa, Israel. She is the author of 3 books, more than 50 articles, and book chapters. Her research interests include the development and implementation of advanced remote sensing methods and techniques for environment and ecological studies emphasizing the importance of multi-source data fusion.

Dr. Brook was a recipient of the Dean's Excellence Award, Tel-Aviv University, Israel, in 2003; the Federal Ministry of Education and Research Excellence Scholarship of "Young Scientists Exchange Program", Germany, in 2008; the Rector's Excellence Scholarship, Tel-Aviv, Israel, in 2009. She is a member of the IEEE Women in Engineering (WIE), and the European Geosciences Union (EGU).

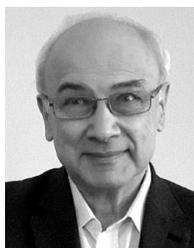

**Alexander N. Gorban** (PhD, ScD, Professor) has held a Personal Chair in applied mathematics at the University of Leicester since 2004. He worked for Russian Academy of Sciences, Siberian Branch (Krasnoyarsk, Russia), and ETH Zurich (Switzerland), was a visiting Professor and Research Scholar at Clay Mathematics Institute (Cambridge, MA), IHES (Bures-sur-Yvette, Île de France), Courant Institute of Mathematical Sciences (New York), and Isaac Newton Institute for Mathematical Sciences (Cambridge, UK). His main research interests are machine learning, data mining and model reduction problems, dynamics of systems of physical, chemical and biological kinetics, and biomathematics. He has been accorded the title of Pioneer of Russian Neuroinformatics (2017) for his extraordinary contribution into theory and applications of artificial neural networks, received Lifetime Achievement Award (MaCKIE-2015) in recognition of outstanding contributions to the research field of (bio)chemical kinetics, and was awarded by Prigogine medal (2003) for achievements in non-equilibrium thermodynamics and physical kinetics.

## Appendixes

**Box 1:** Outliers Analysis Procedure

Outliers analysis was performed separately for each geographic site dataset. Proceeding from each variable distribution, we defined a cut-off separating 1% of points as variable outlier. An observation was considered to be an outlier if either:

(i) It was beyond the cut-off point at the scale of at least one of the 'predictors' while being within the 'usual' interval at the scale of each of dependent variables (We should emphasize that here and hereinafter the notes 'dependent variable', as well as 'independent variable', or 'predictor', when applied to elastic map approach, are used figuratively. Elastic map is set of points, connected via edges and ribs, aimed at approximating points dataset in N-dimensional coordinate system, where N is number of input variables, no matter which of them is implied to be dependent variable.);

(ii) It was beyond the cut-off point at the scale of at least one of the dependent variables while being 'normal' at the scale of each independent variable;

(iii) It was beyond opposite cut-off points (that is, upper/lower or lower/upper) at the scale of predictor and dependent variable under their positive bivariate association;

(iv) It was beyond same-range cut-off points (that is, upper/upper or lower/lower) at the scale of predictor and dependent variable under their negative bivariate association.

Thus, the percentage of excluded outlying observations varied from 2.92% for the Atlanta dataset to 3.90% for the Beijing dataset (see Table A1).

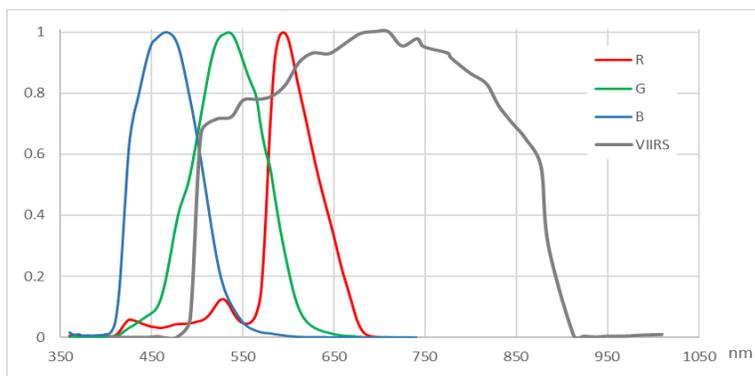

Fig. A1. Relative response of VIIRS/DNB sensor and Nikon D3 DSLR camera from the ISS
(*Source*: Built from data obtained upon request from A. Sánchez de Miguel.)

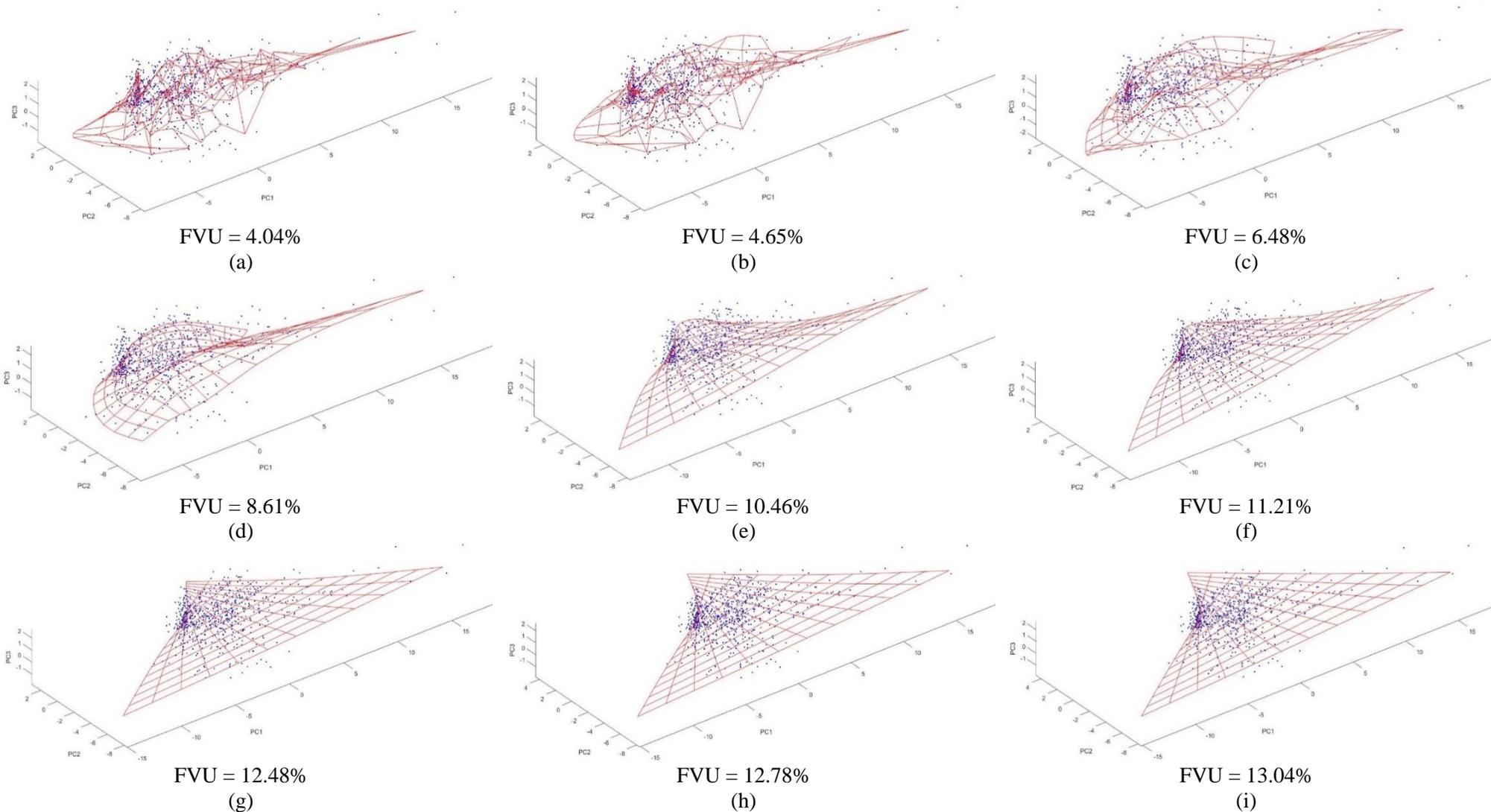

Fig. A2. Examples of elastic maps (depicted by red grid) built for the Haifa's blue light containing dataset (marked by blue dots) using varying bending regimes (see text for explanations): (a) 0.00001; (b) 0.0001; (c) 0.001; (d) 0.01; (e) 0.05; (f) 0.1; (g) 0.3; (h) 0.5, and (i) 1.

*Notes:* The first three principal components (PC1, PC2, and PC3) are used as coordinates for the elastic maps' visualization. Fraction of total (by all six coordinates of the parameter space) variance unexplained (FVU) by elastic maps built under varying bending regimes are reported

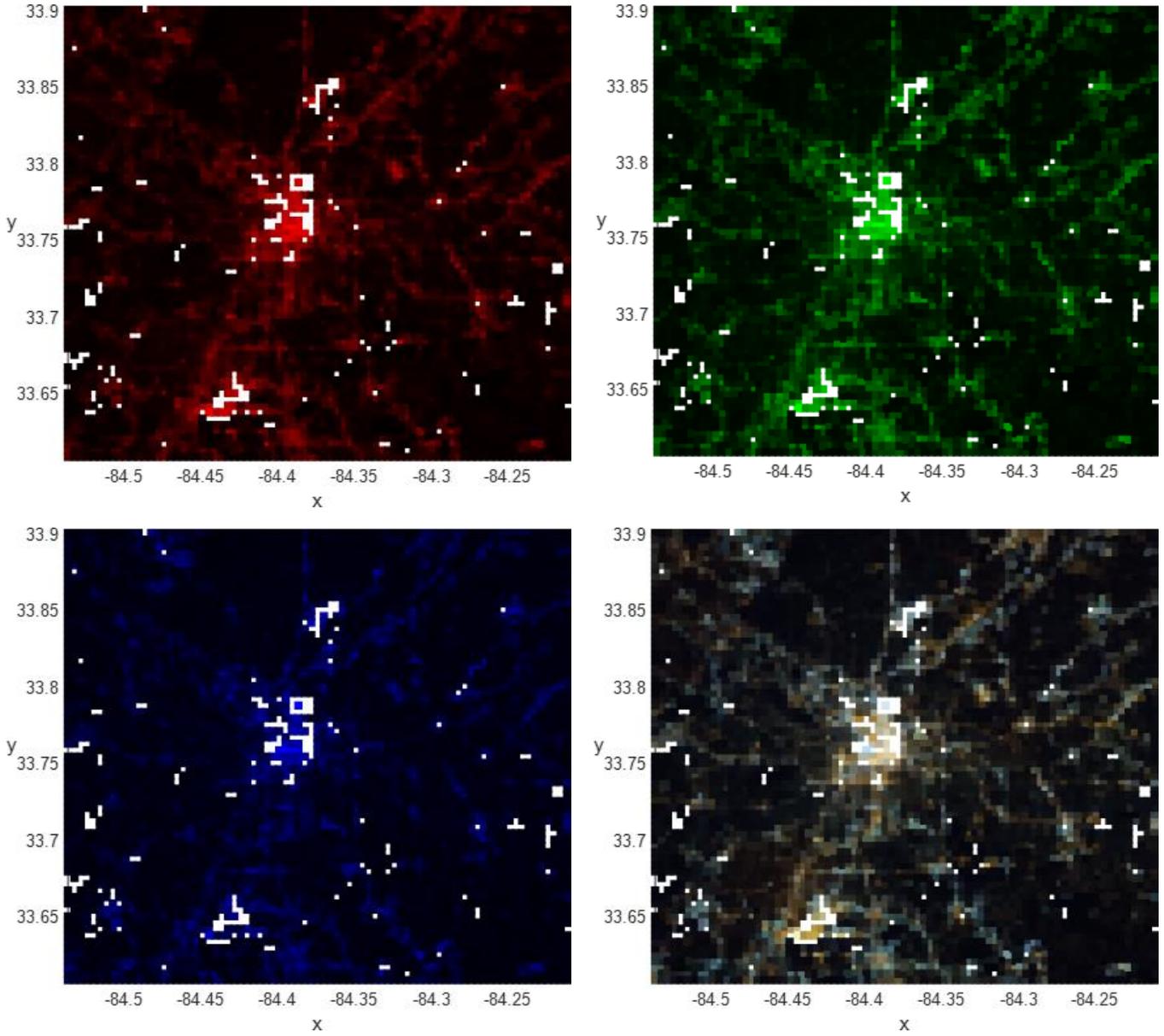

Panel *(a):* ISS-provided resampled to the spatial resolution of VIIRS imagery (see explanation at p.7)

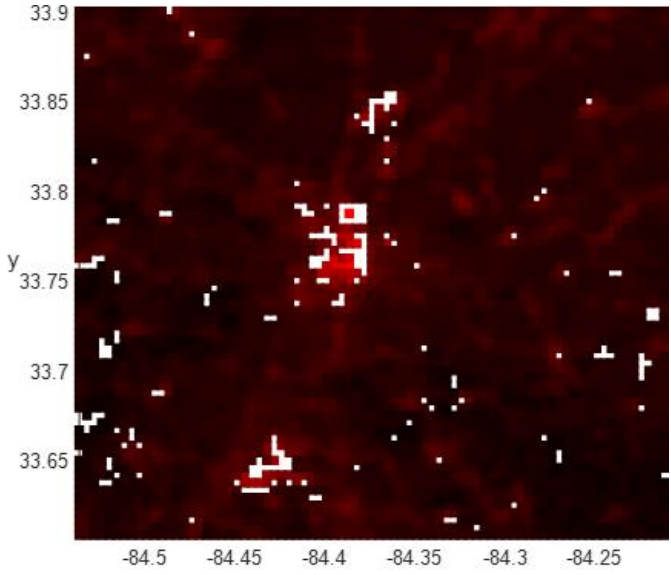

*R* = 0.81; *WMSE* = 15.57

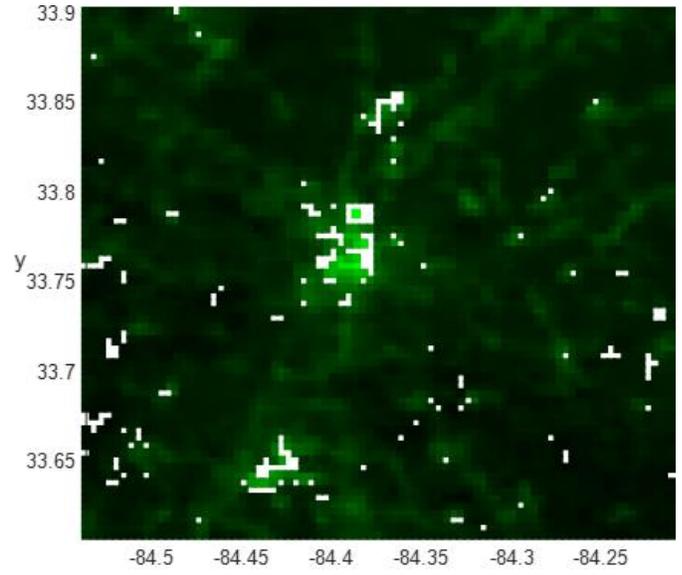

*R* = 0.80; *WMSE* = 9.62

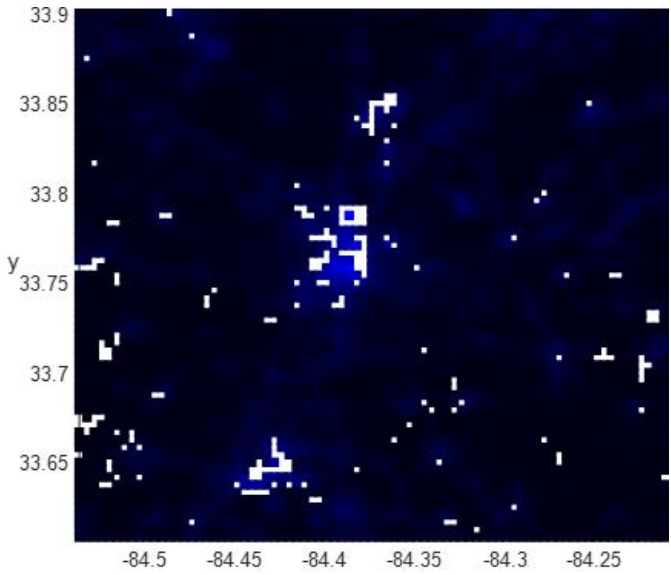

*R* = 0.73; *WMSE* = 1.80

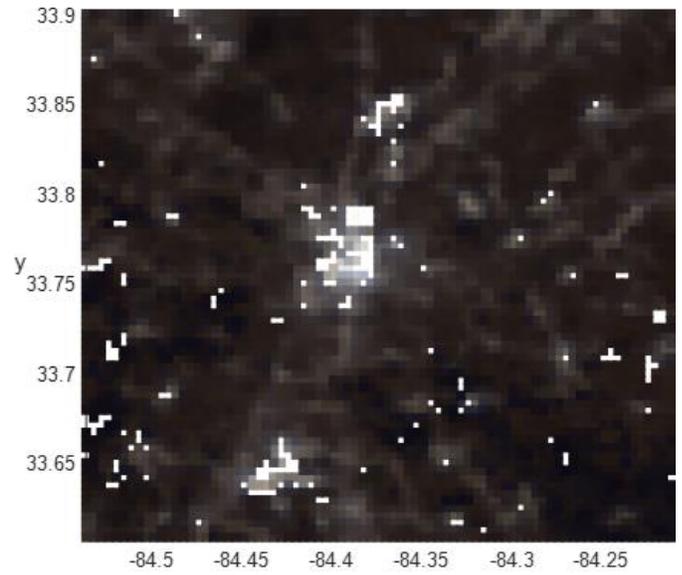

*C_sim* = 0.93

Panel *(b):* Outputs of linear multiple regressions (see explanation at p.7)

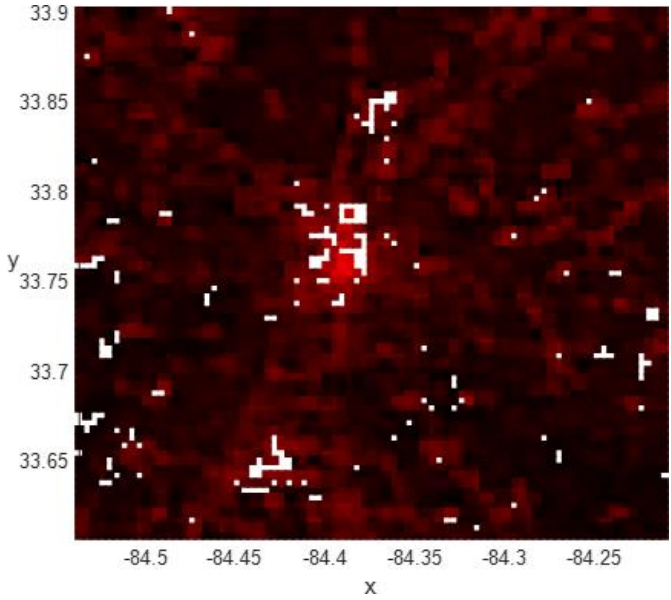

*R* = 0.81; *WMSE* = 10.64

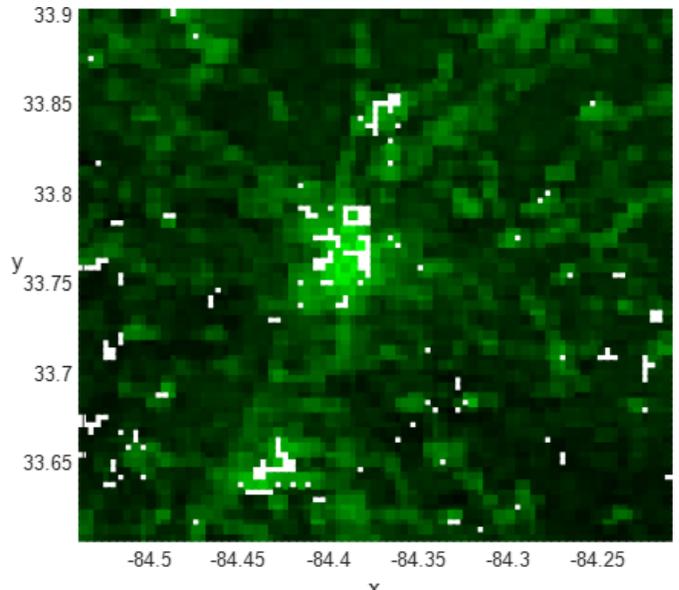

*R* = 0.78; *WMSE* = 9.11

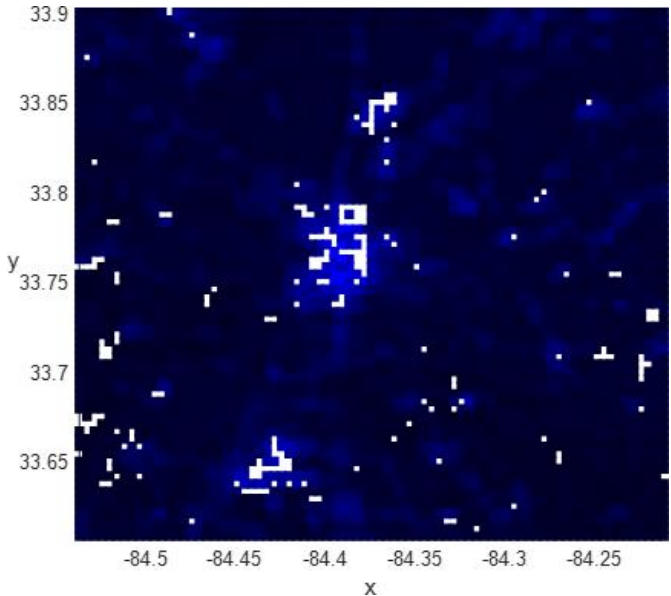

*R* = 0.69; *WMSE* = 2.53

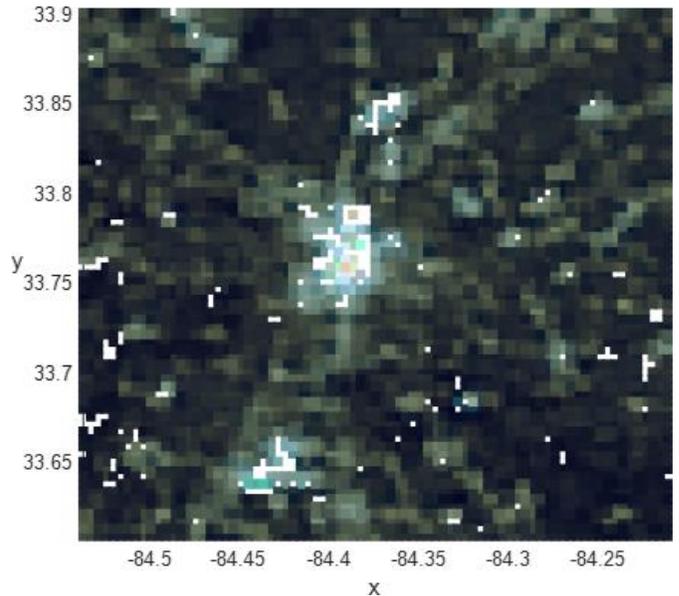

*C_sim* = 0.95

Panel *(c):* Outputs of non-linear kernel regressions (see explanation at p.7)

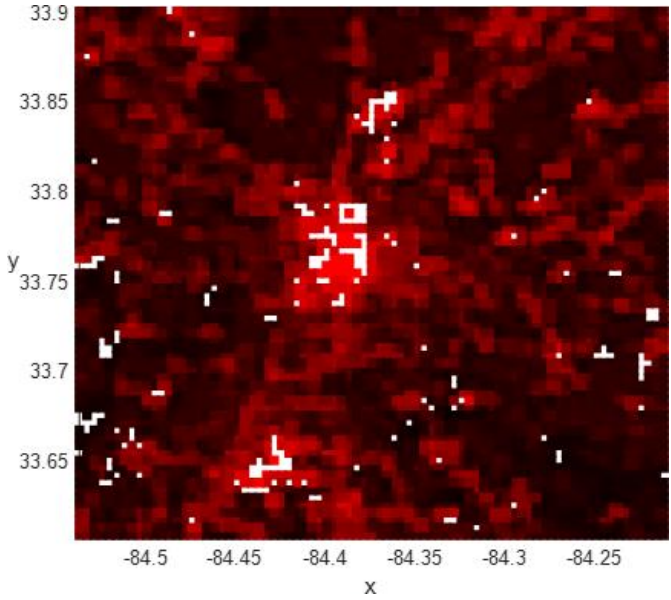

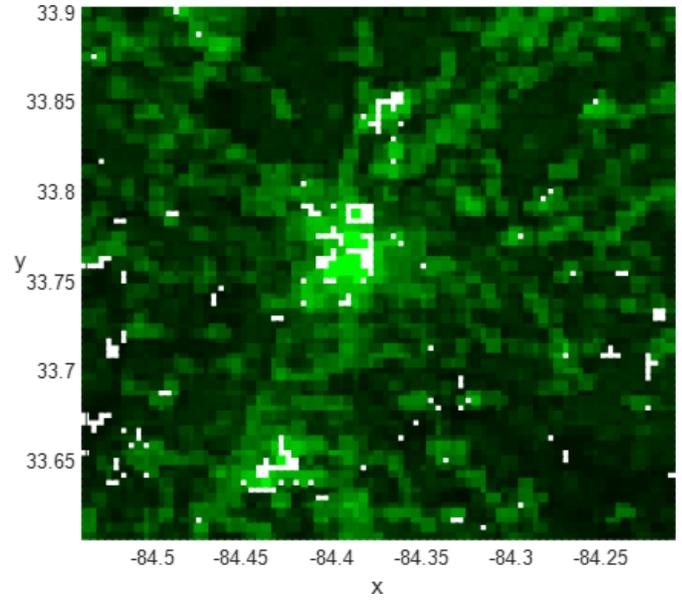

R = 0.78; WMSE =8.55

R = 0.78; WMSE =7.09

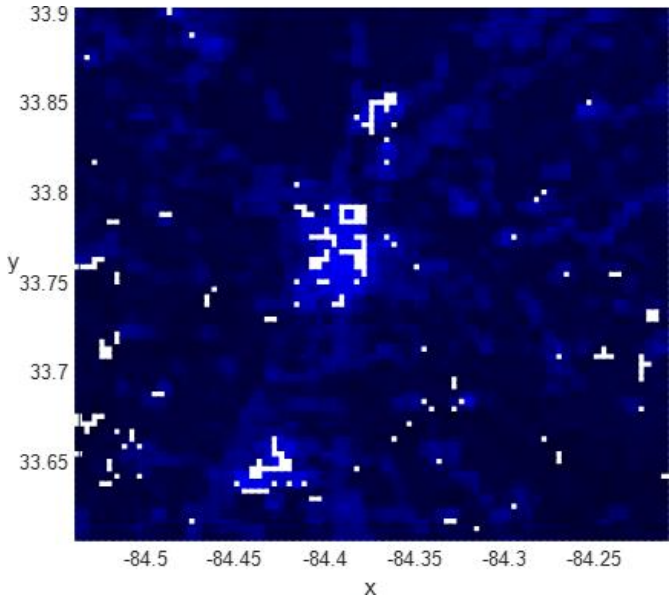

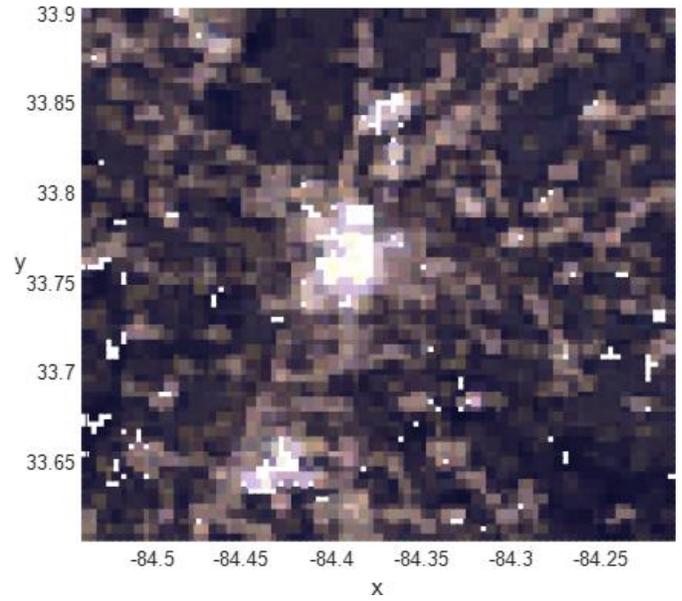

R = 0.73; WMSE =2.59

C_sim = 0.94

Panel *(d):* Outputs of random forest regressions (see explanation at p.7)

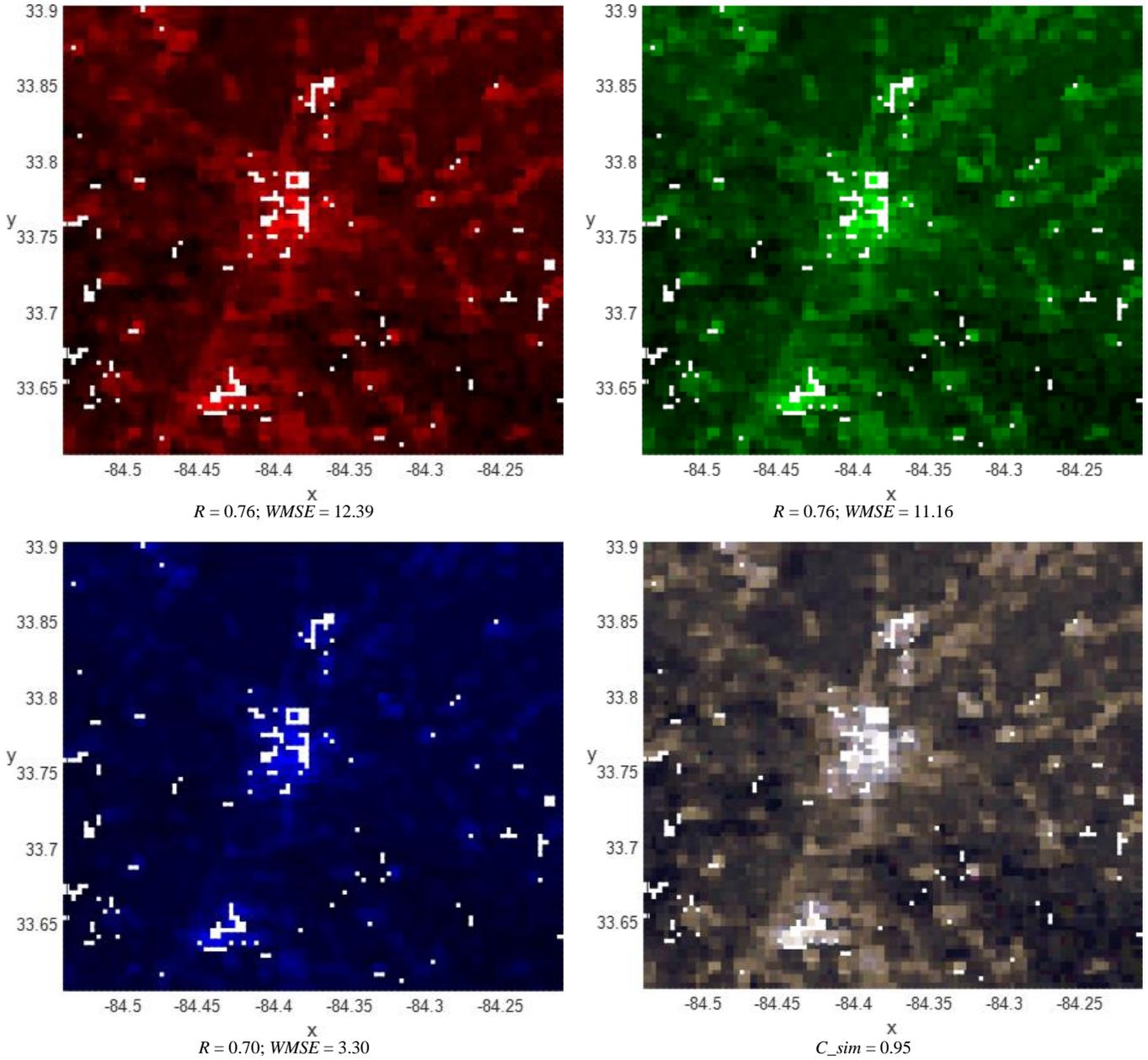

Panel *(e):* Outputs of elastic map models

Fig. A3. *Atlanta metropolitan area (the US):* Red, Green, Blue bands, and RGB images, provided by ISS and resampled to the spatial resolution of VIIRS imagery (panel *(a)*), and outputs of four models trained on *Haifa datasets*: linear multiple regressions (panel *(b)*), non-linear kernel regressions (panel *(c)*), random forest regressions (panel *(d)*), and elastic map models (panel *(e)*).

*Notes:* Output generated by elastic maps, built under 0.05 bending penalty, is reported. *R* and *WMSE* denote correspondingly for Pearson's correlation and weighted mean squared error of the red, green, and blue lights' estimates, *C_sim* – for contrast similarity between restored and original RGB images. White points in the city area correspond to outliers.

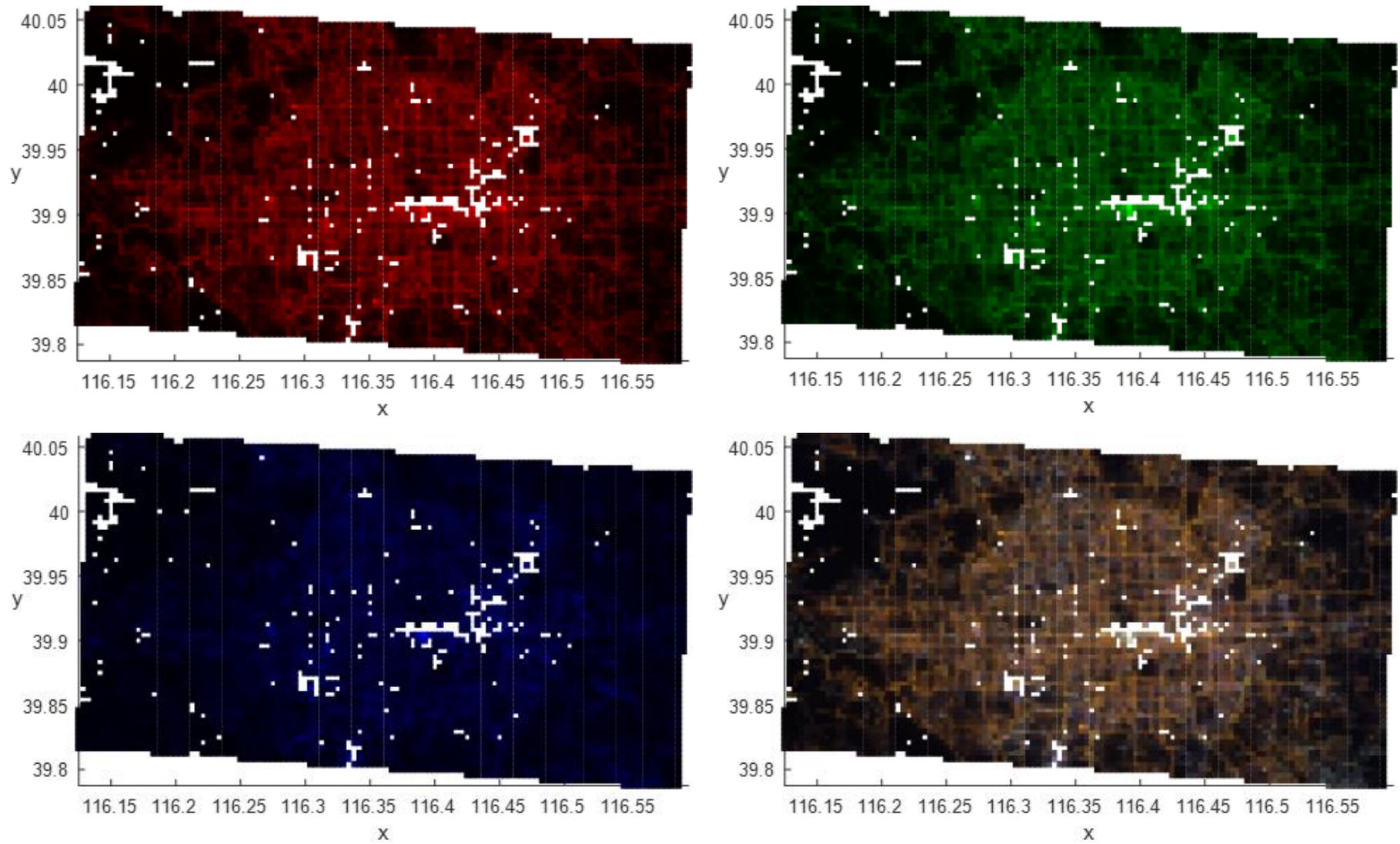

Panel *(a):* ISS-provided and resampled to the spatial resolution of VIIRS imagery (see explanation at p.12)

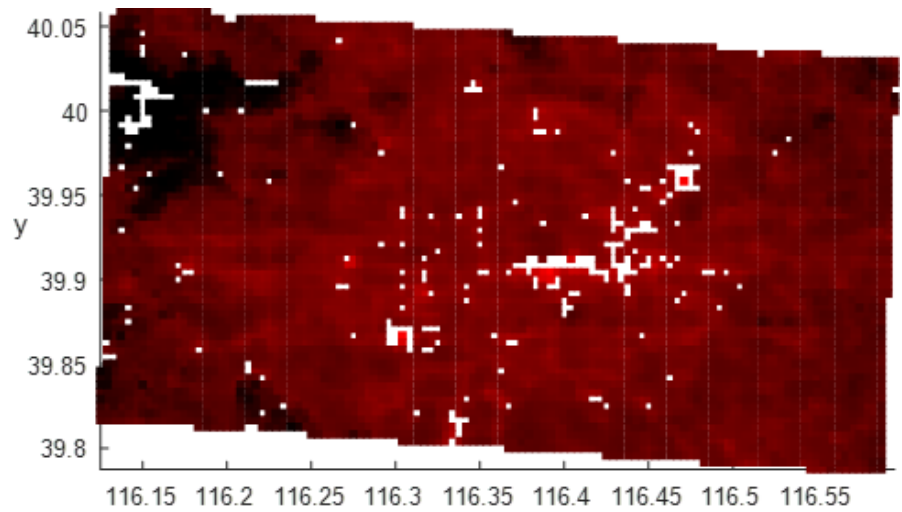

$R = 0.79$; $WMSE = 3.31$

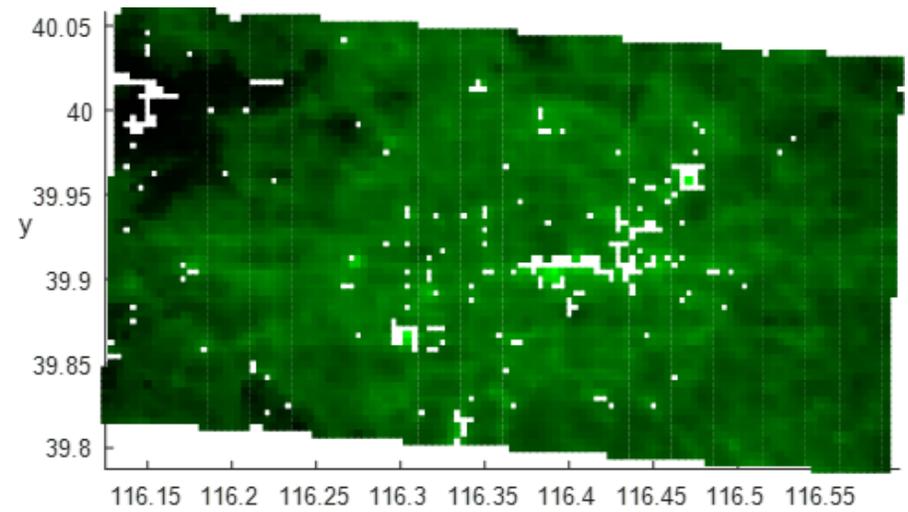

$R = 0.85$; $WMSE = 3.60$

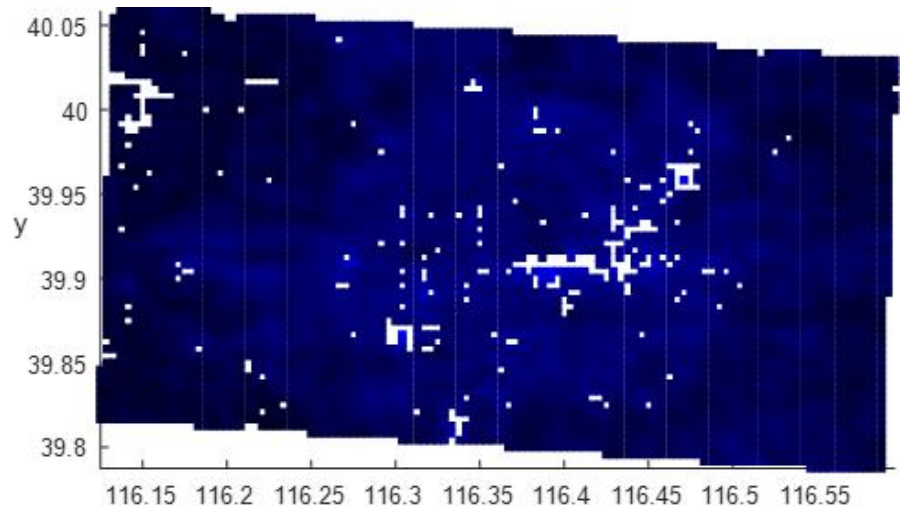

$R = 0.78$; $WMSE = 2.58$

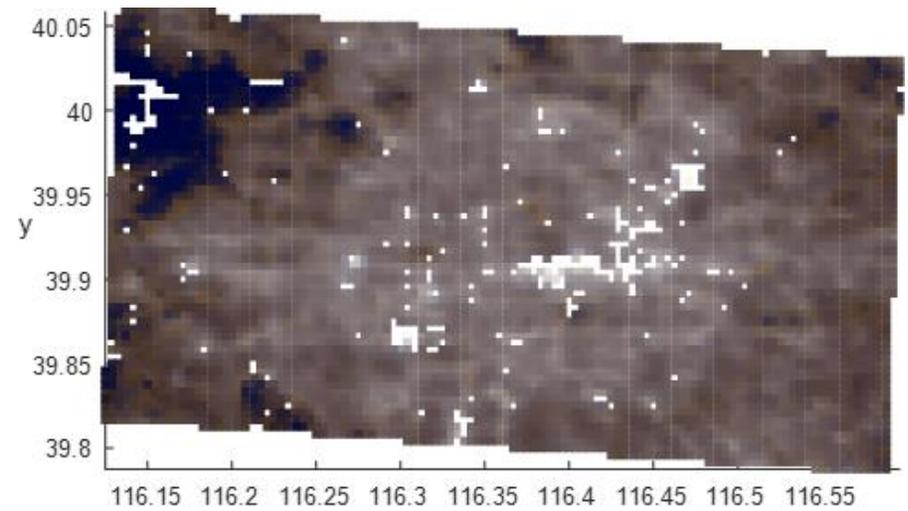

$C\_sim = 0.94$

Panel *(b):* Outputs of linear multiple regressions (see explanation at p.12)

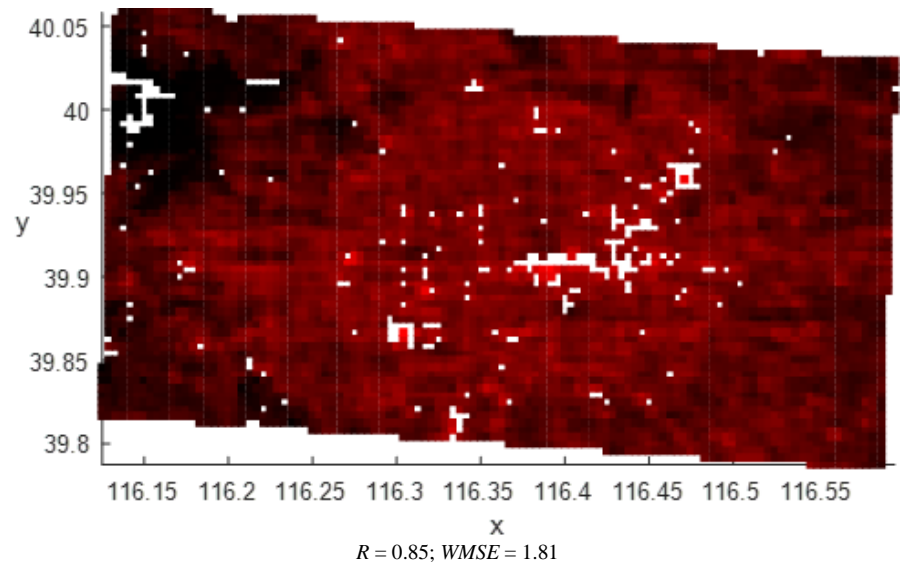

*R* = 0.85; *WMSE* = 1.81

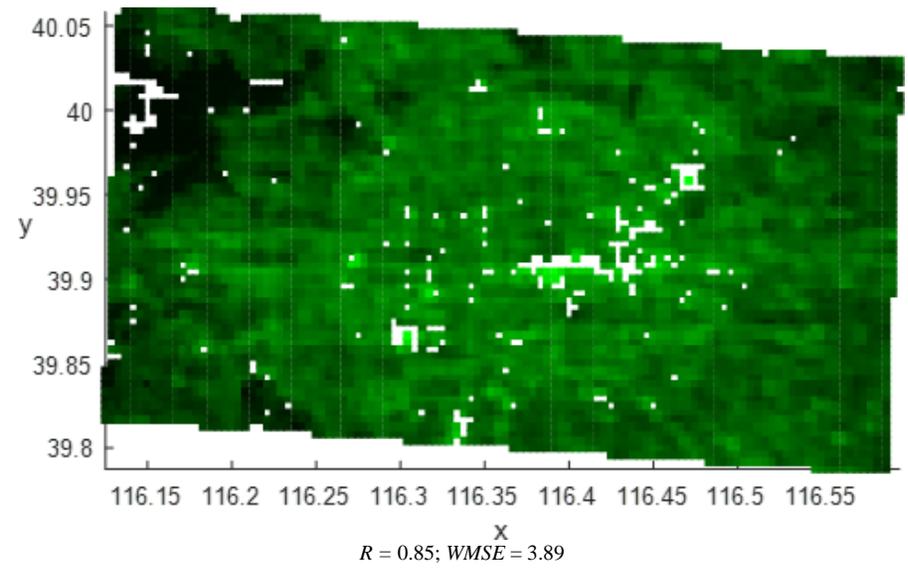

*R* = 0.85; *WMSE* = 3.89

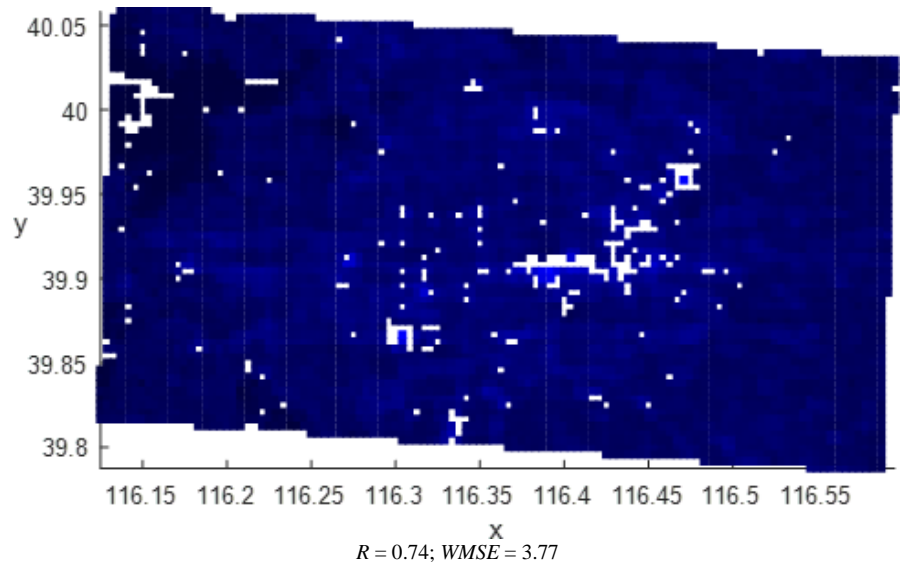

*R* = 0.74; *WMSE* = 3.77

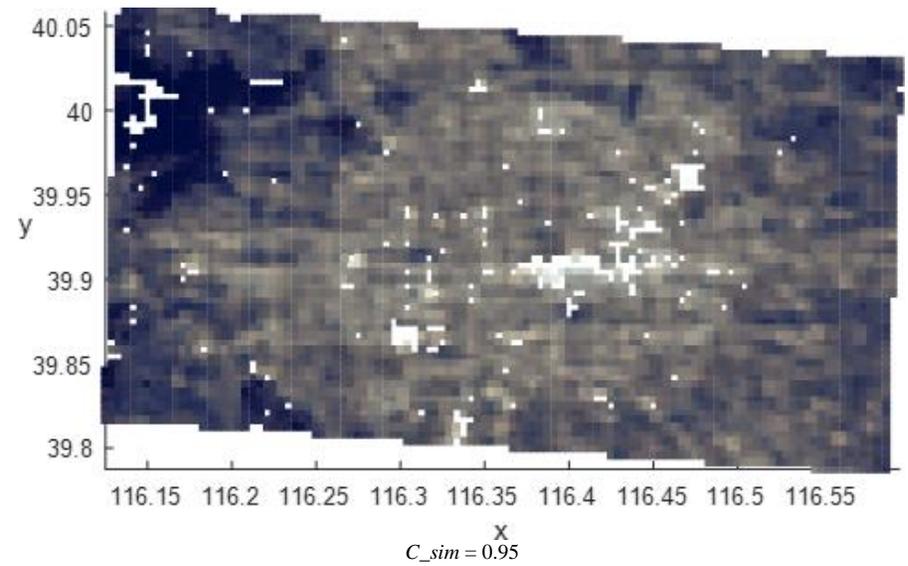

*C_sim* = 0.95

Panel *(c):* Outputs of non-linear kernel regressions (see explanation at p.12)

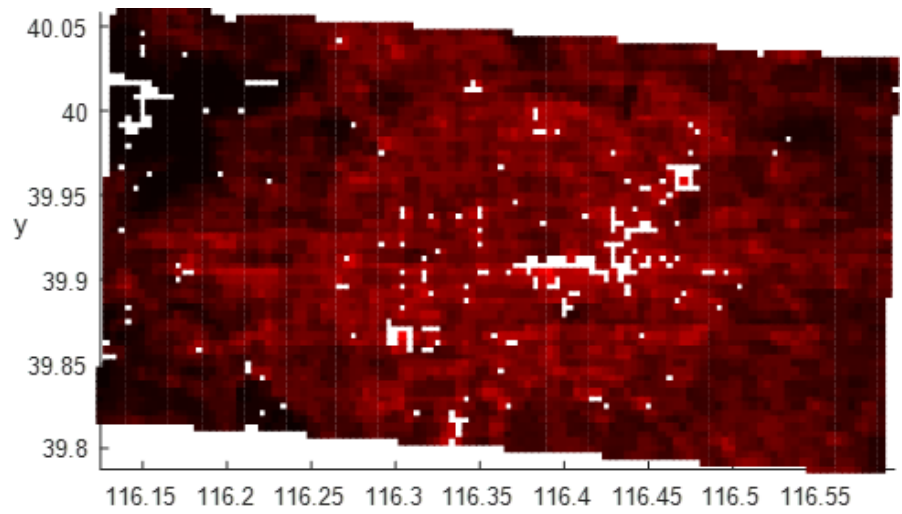

*R* = 0.86; *WMSE* = 1.02

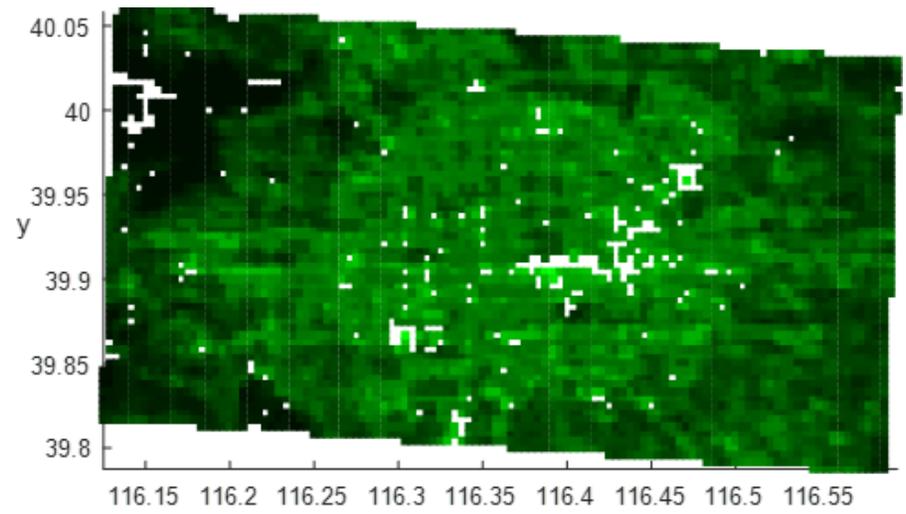

*R* = 0.86; *WMSE* = 2.34

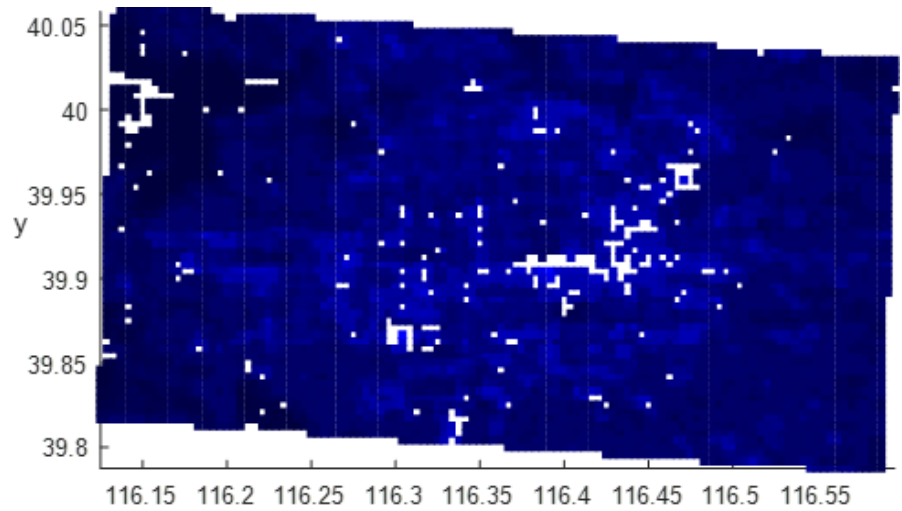

*R* = 0.72; *WMSE* =4.25

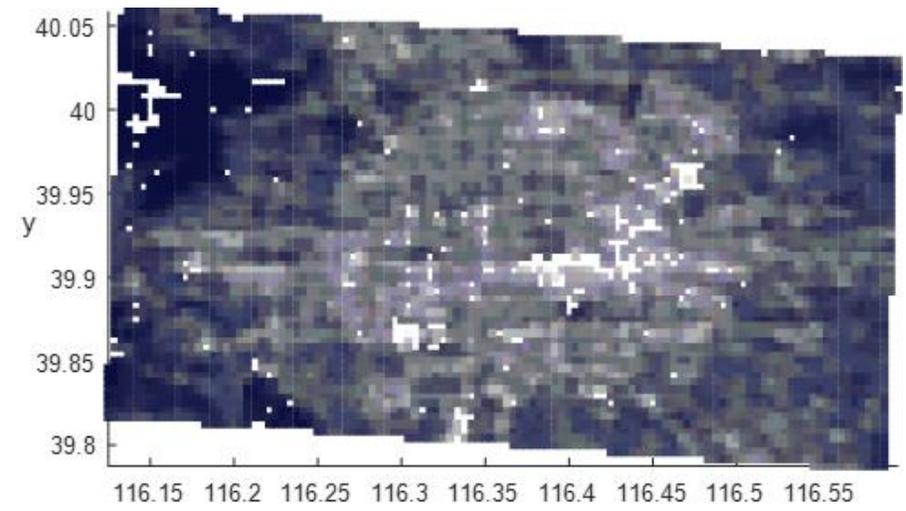

*C_sim* = 0.97

Panel *(d):* Outputs of random forest regressions (see explanation at p.12)

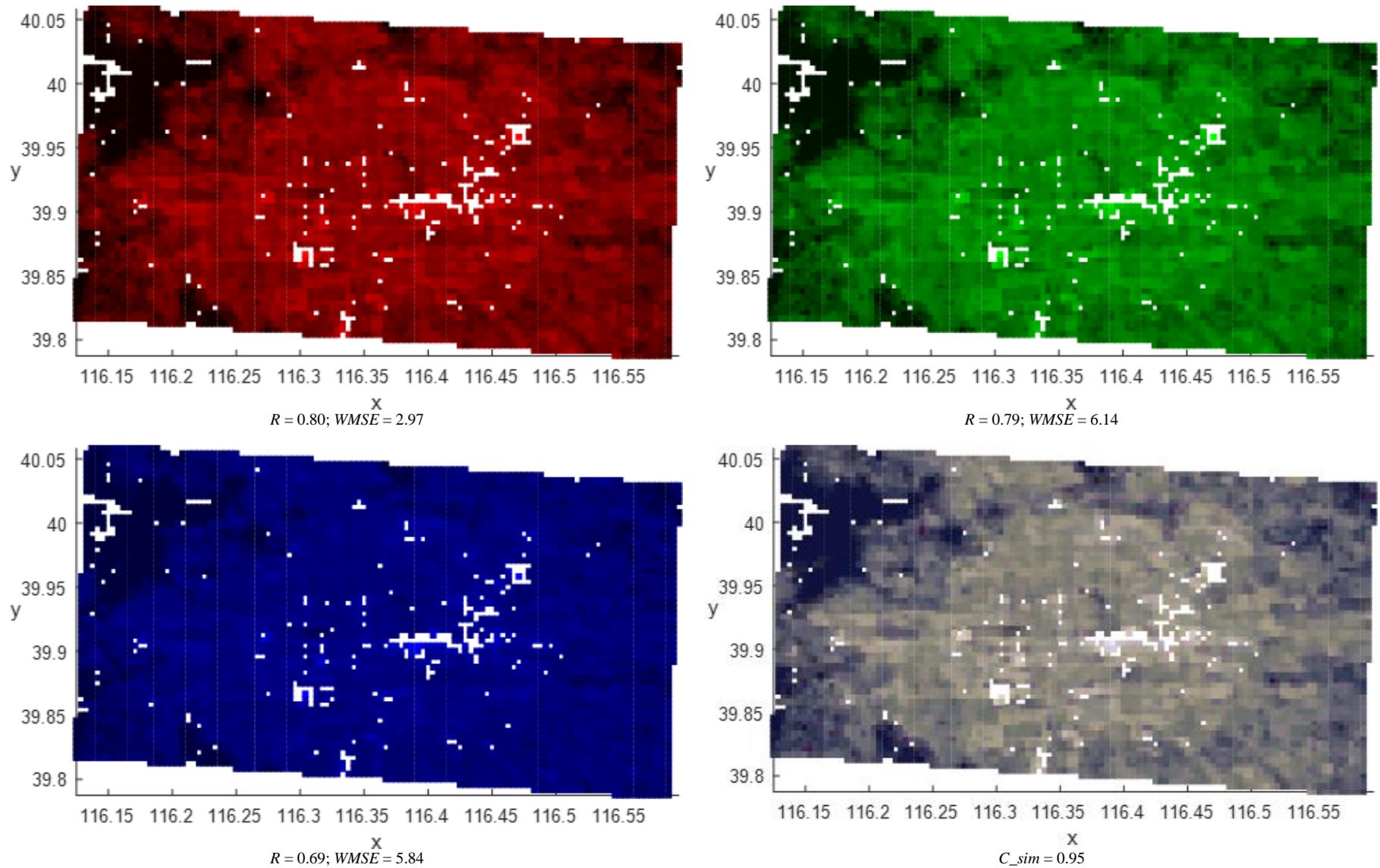

Panel *(e):* Outputs of elastic map models

Fig. A4. *Beijing metropolitan area (China):* Red (the first column), Green (the second column), Blue (the third column)) bands, and RGB images (the fourth column); ISS-provided, resampled to the spatial resolution of VIIRS imagery (the first row), and outputs of four models trained on *Haifa datasets*: linear multiple regressions (the second row), non-linear kernel regressions (the third row), random forest regressions (the fourth row), and elastic map models (the fifth row).

*Notes:* Output generated by elastic maps, built under 0.05 bending penalty, is reported. *R* and *WMSE* denote correspondingly for Pearson's correlation and weighted mean squared error of the red, green, and blue lights' estimates, *C_sim* – for contrast similarity between restored and original RGB images. White points in the city area correspond to outliers.

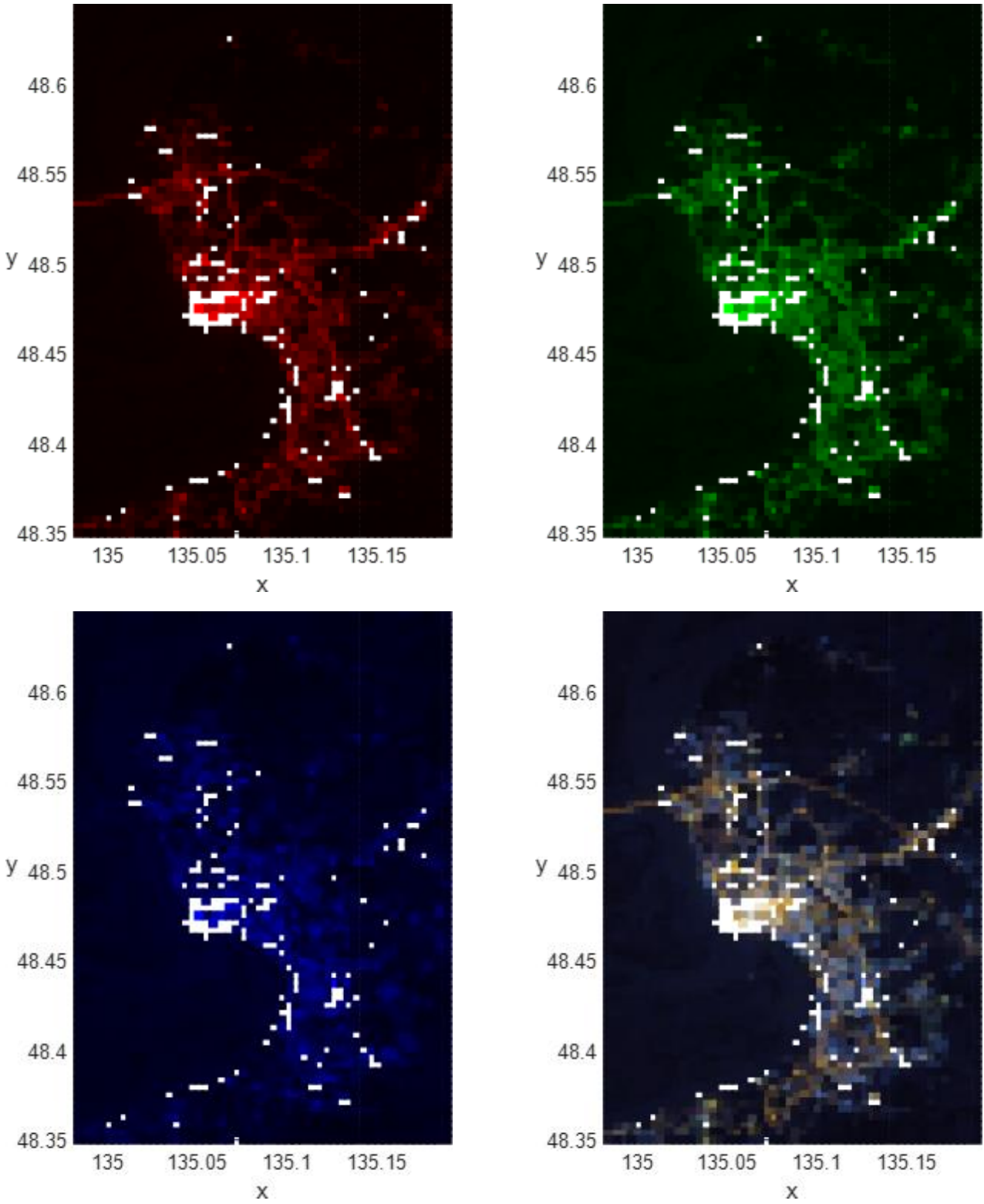

Panel *(a):* ISS-provided and resampled to the spatial resolution of VIIRS imagery (see explanation at p.17)

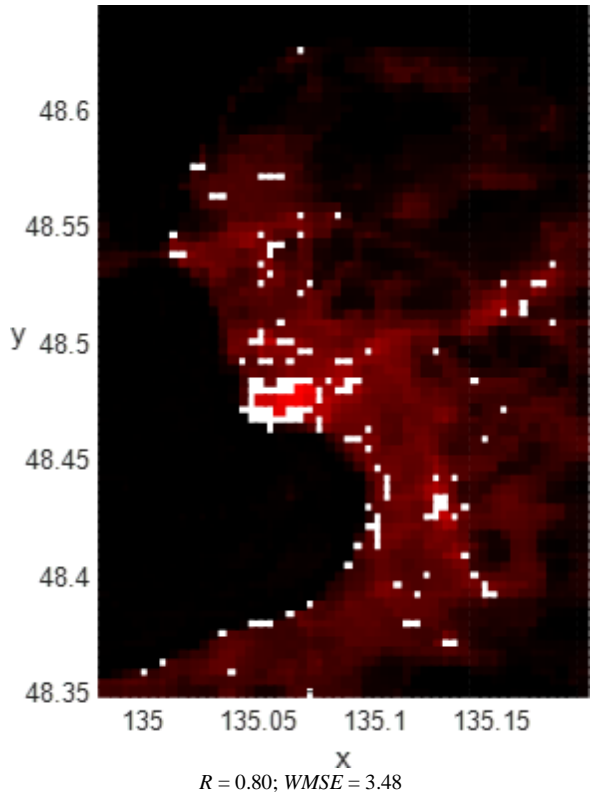

$R = 0.80$; $WMSE = 3.48$

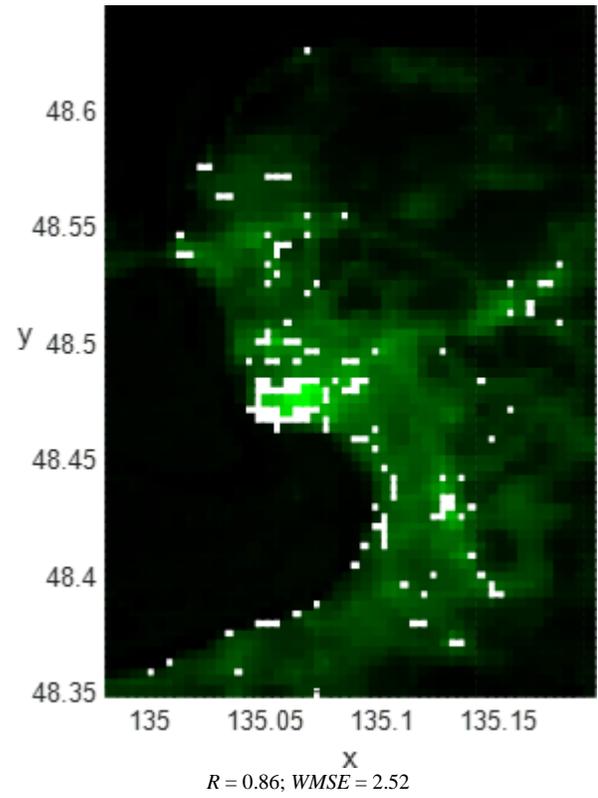

$R = 0.86$; $WMSE = 2.52$

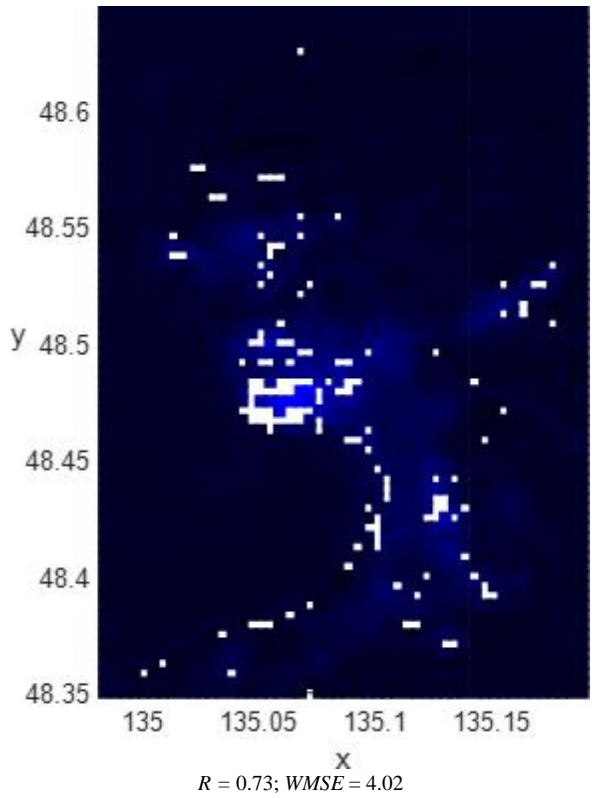

$R = 0.73$; $WMSE = 4.02$

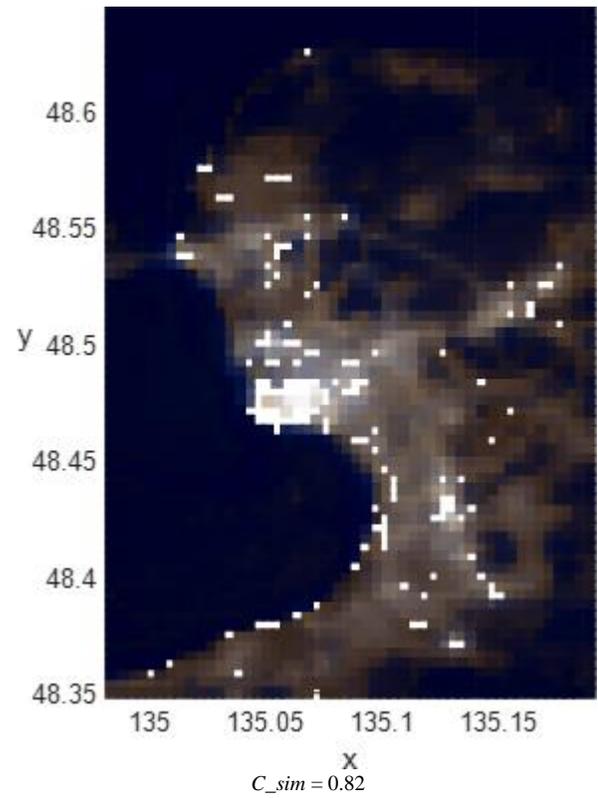

$C\_sim = 0.82$

Panel *(b):* Outputs of linear multiple regressions (see explanation at p.17)

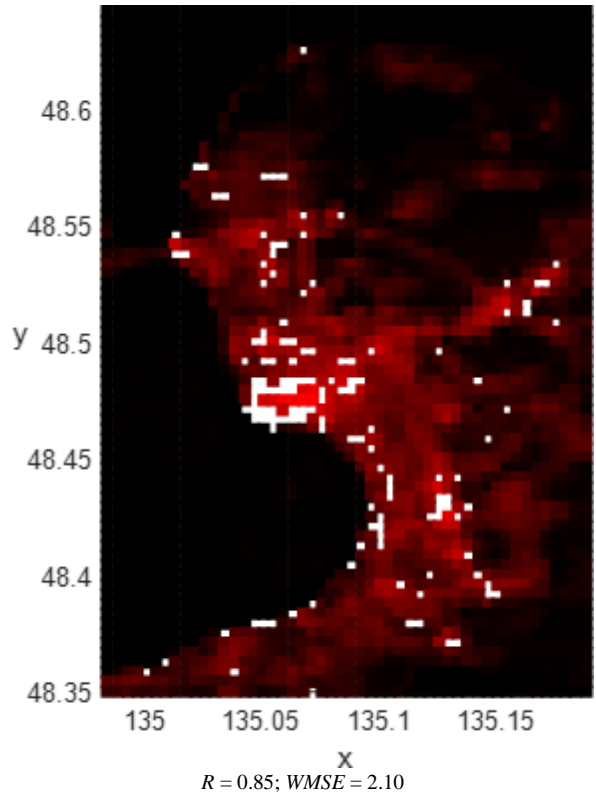
$R = 0.85; \ WMSE = 2.10$

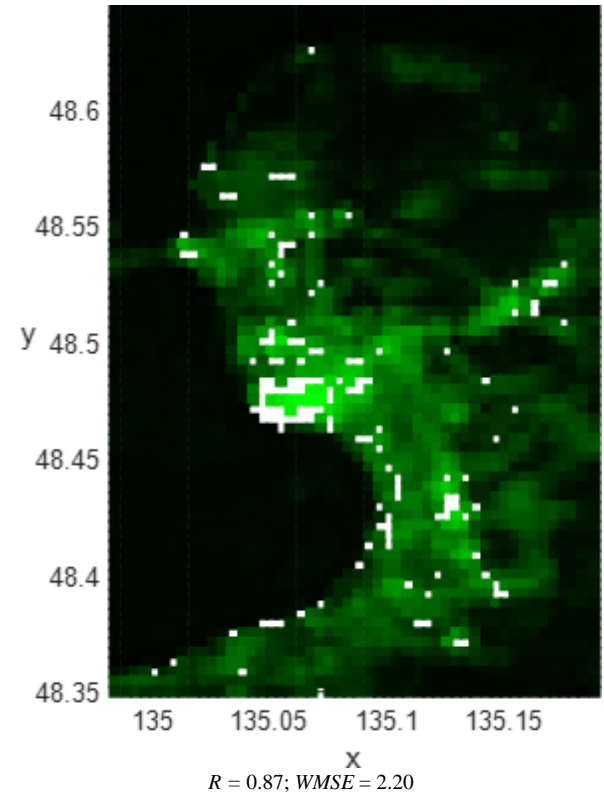
$R = 0.87; \ WMSE = 2.20$

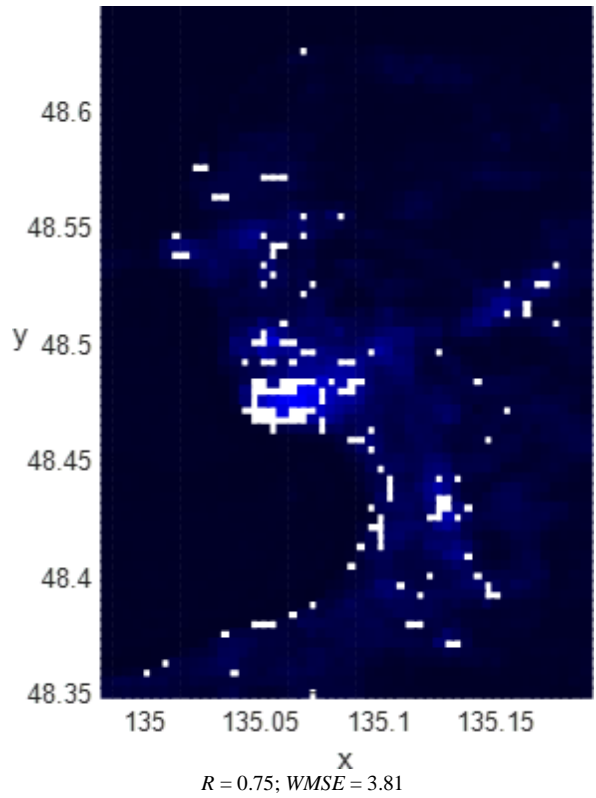
$R = 0.75; \ WMSE = 3.81$

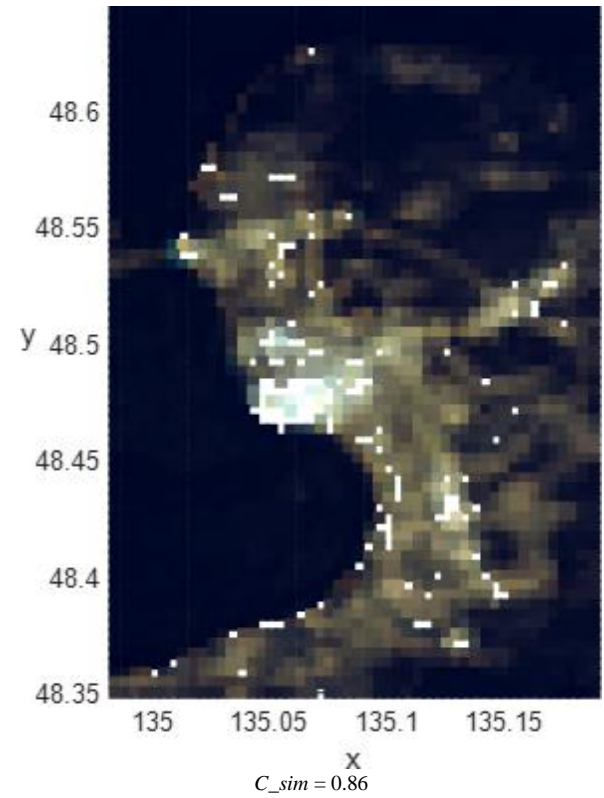
$C\_sim = 0.86$

Panel *(c):* Outputs of non-linear kernel regressions (see explanation at p.17)

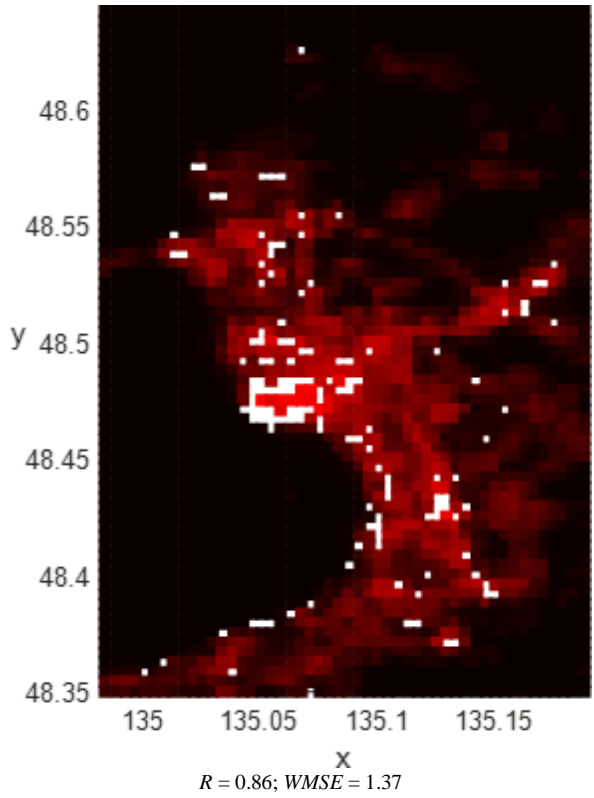

*R* = 0.86; *WMSE* = 1.37

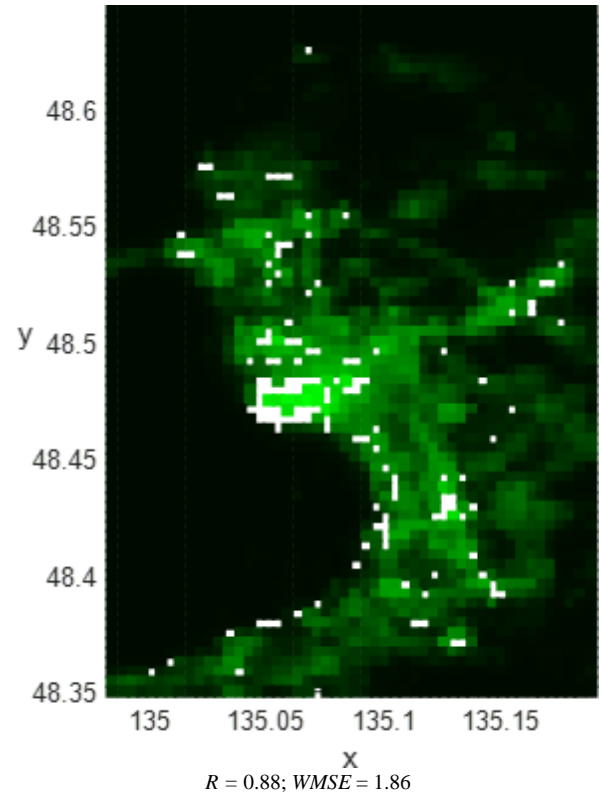

*R* = 0.88; *WMSE* = 1.86

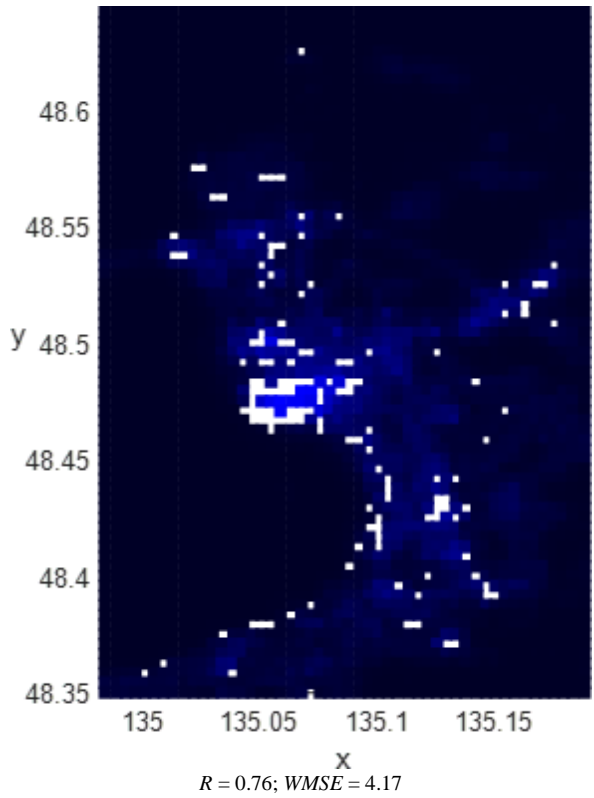

*R* = 0.76; *WMSE* = 4.17

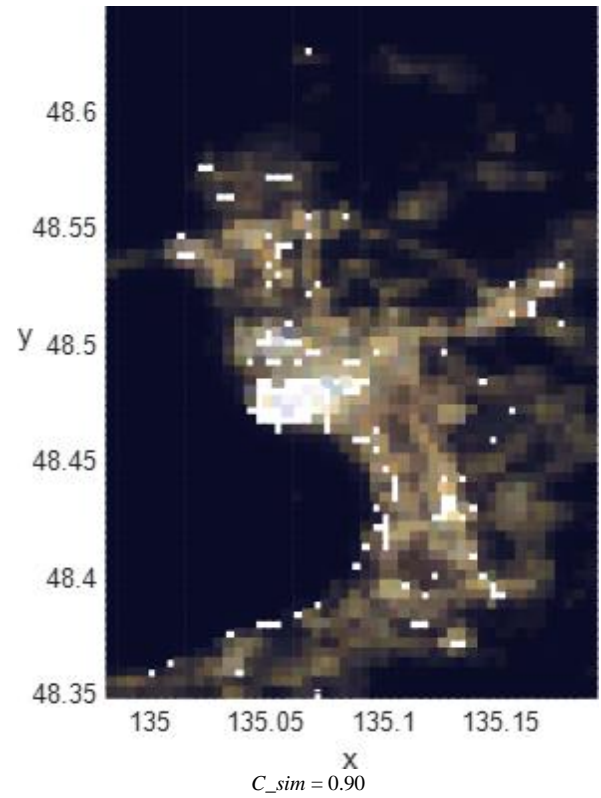

*C_sim* = 0.90

Panel *(d):* Outputs of random forest regressions (see explanation at p.17)

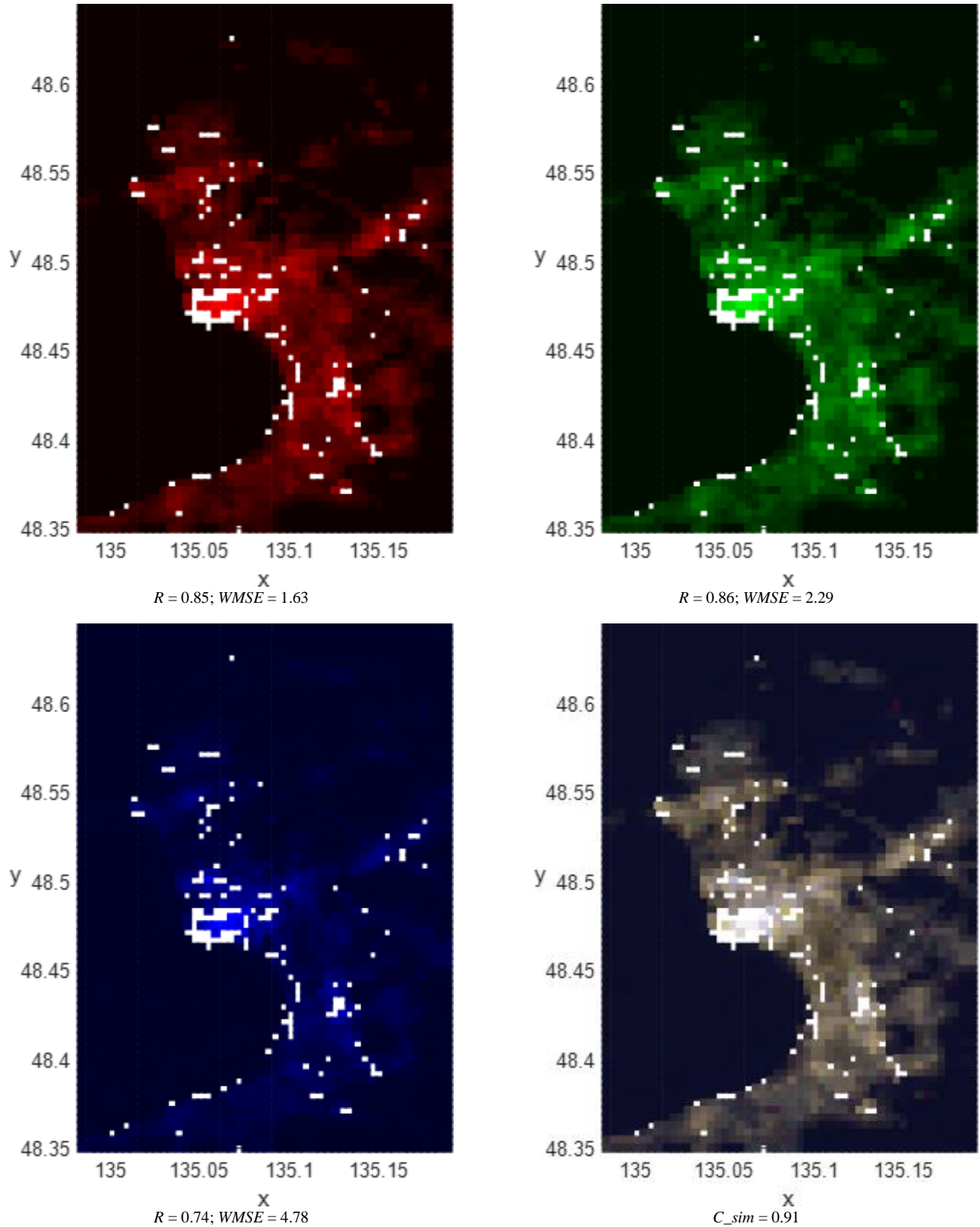

Panel *(e):* Outputs of elastic map models

Fig. A5. *Khabarovsk (Russia):* Red (the first column), Green (the second column), Blue (the third column)) bands, and RGB images (the fourth column); ISS-provided, resampled to the spatial resolution of VIIRS imagery (the first row), and outputs of four models trained on *Haifa datasets*: linear multiple regressions (the second row), non-linear kernel regressions (the third row), random forest regressions (the fourth row), and elastic map models (the fifth row).
*Notes:* Output generated by elastic maps, built under 0.05 bending penalty, is reported. *R* and *WMSE* denote correspondingly for Pearson's correlation and weighted mean squared error of the red, green, and blue lights' estimates, *C_sim* – for contrast similarity between restored and original RGB images. White points in the city area correspond to outliers.

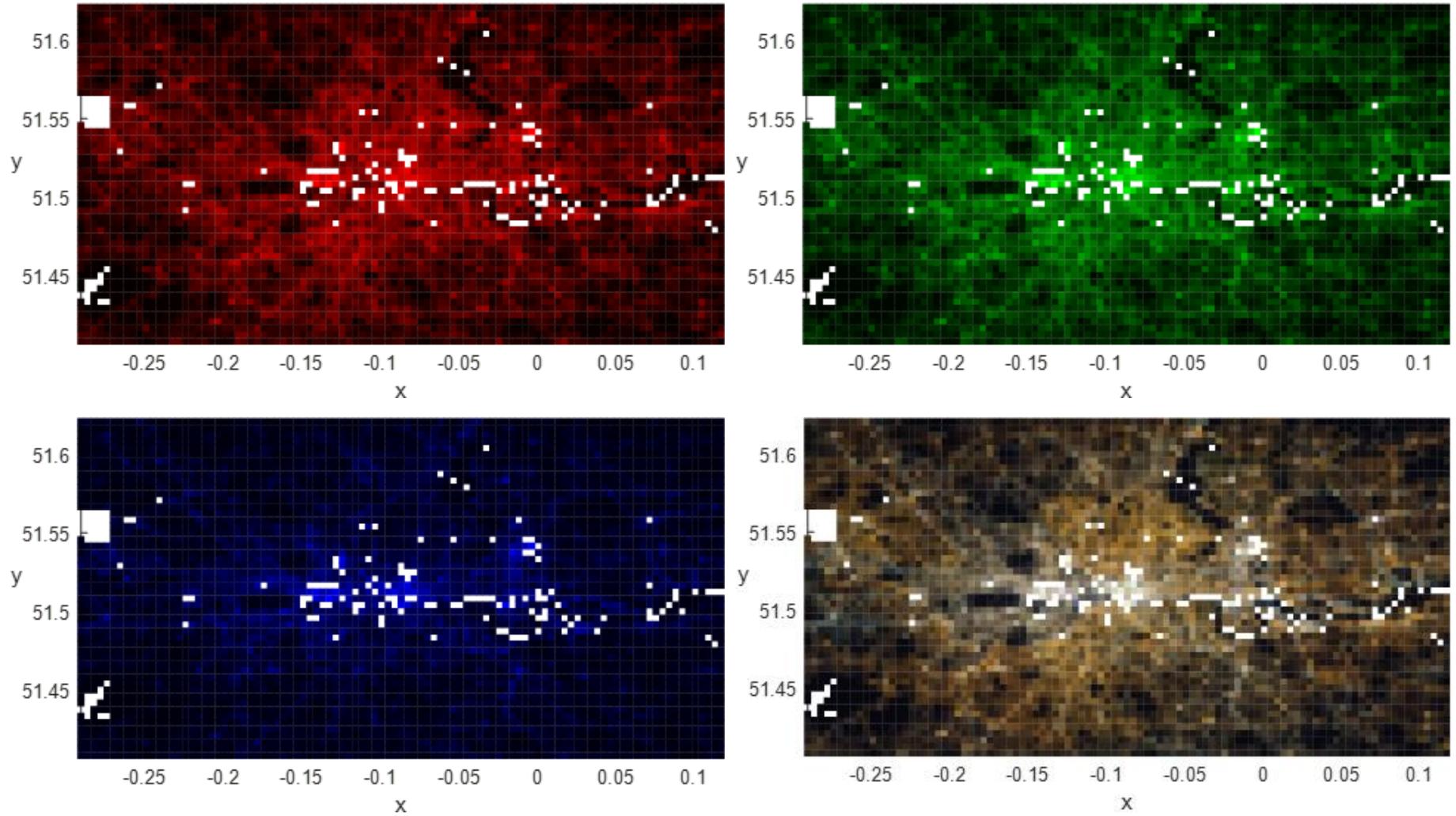

Panel *(a):* ISS-provided and resampled to the spatial resolution of VIIRS imagery (see explanation at p.22)

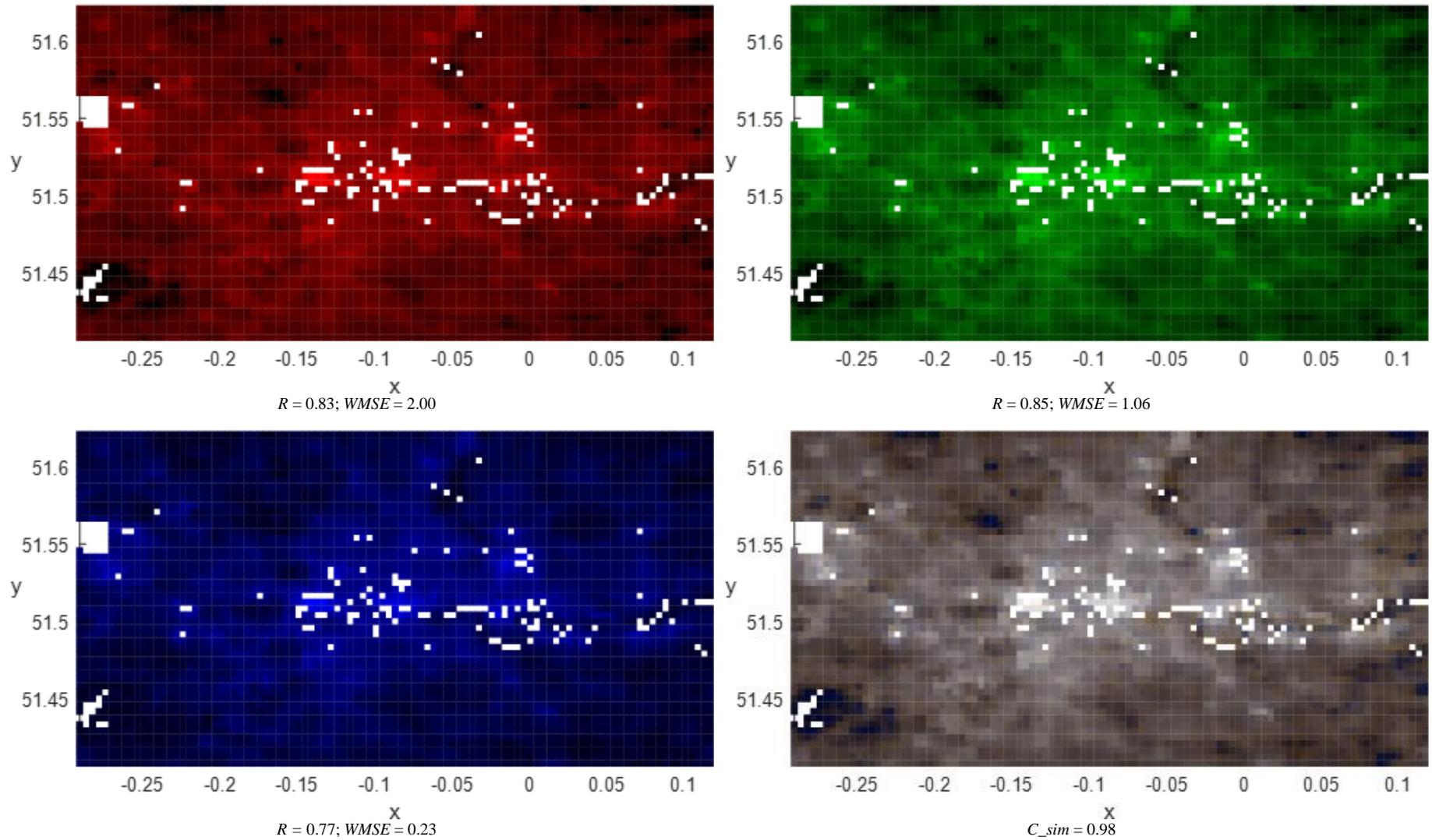

$R = 0.83; WMSE = 2.00$

$R = 0.85; WMSE = 1.06$

$R = 0.77; WMSE = 0.23$

$C\_sim = 0.98$

Panel *(b):* Outputs of linear multiple regressions (see explanation at p.22)

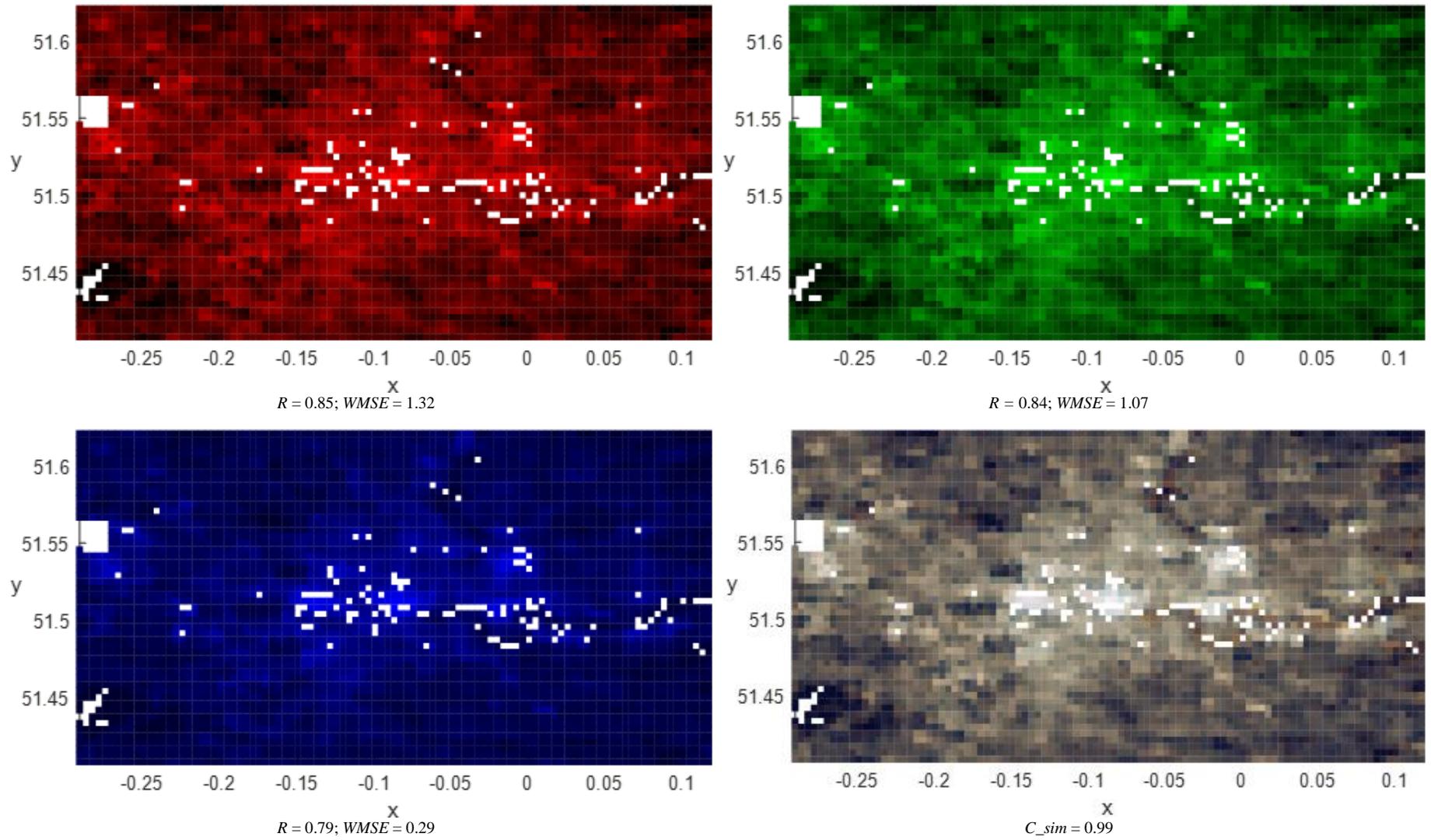

*R* = 0.85; *WMSE* = 1.32

*R* = 0.84; *WMSE* = 1.07

*R* = 0.79; *WMSE* = 0.29

*C_sim* = 0.99

Panel *(c):* Outputs of non-linear kernel regressions (see explanation at p.22)

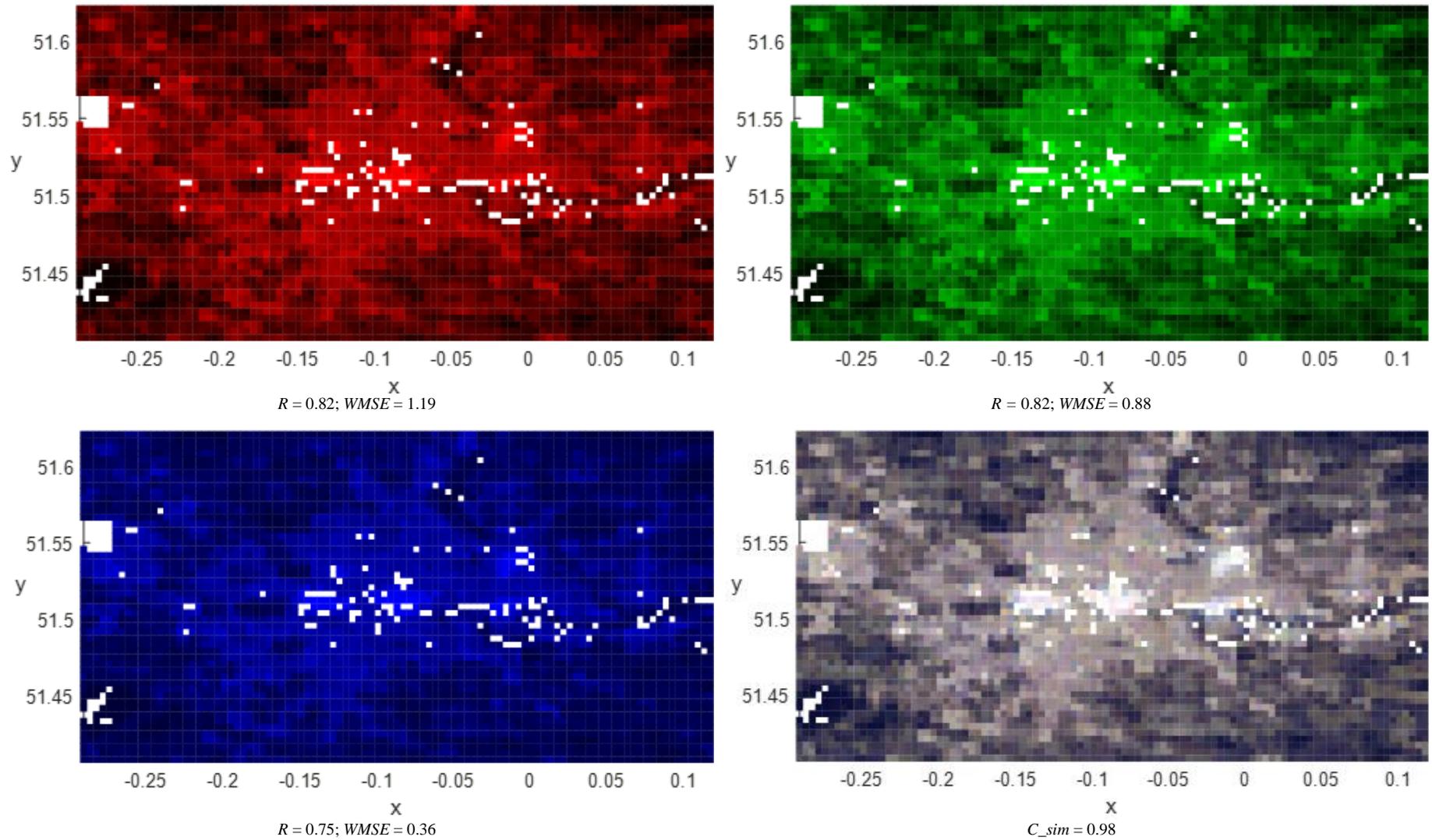

Panel *(d):* Outputs of random forest regressions (see explanation at p.22)

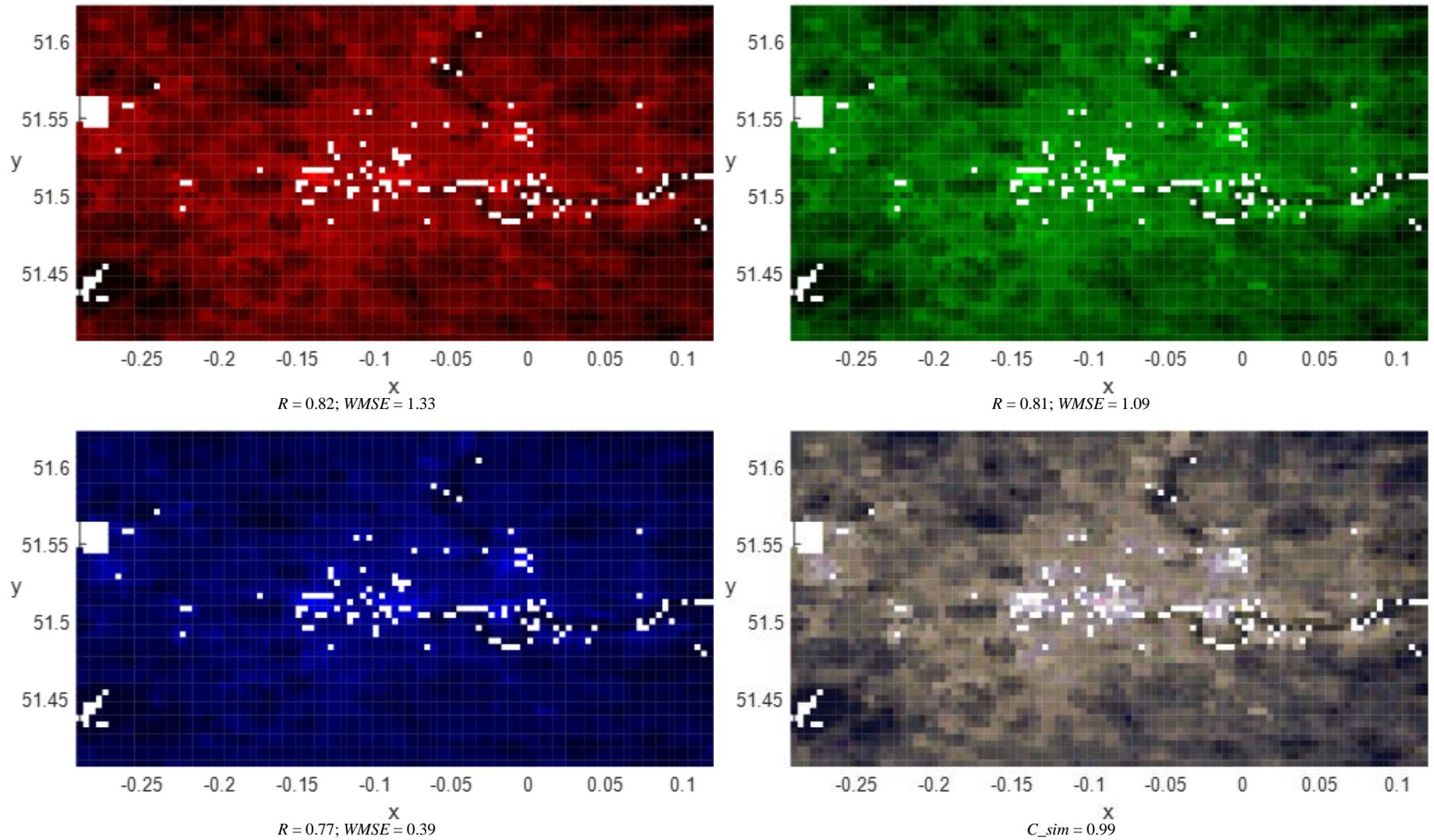

Panel *(e):* Outputs of elastic map models

Fig. A6. *London metropolitan area (the UK):* Red (the first column), Green (the second column), Blue (the third column)) bands, and RGB images (the fourth column); ISS-provided, resampled to the spatial resolution of VIIRS imagery (the first row), and outputs of four models trained on *Haifa datasets*: linear multiple regressions (the second row), non-linear kernel regressions (the third row), random forest regressions (the fourth row), and elastic map models (the fifth row).

*Notes:* Output generated by elastic maps, built under 0.05 bending penalty, is reported. *R* and *WMSE* denote correspondingly for Pearson's correlation and weighted mean squared error of the red, green, and blue lights' estimates, *C_sim* – for contrast similarity between restored and original RGB images. White points in the city area correspond to outliers.

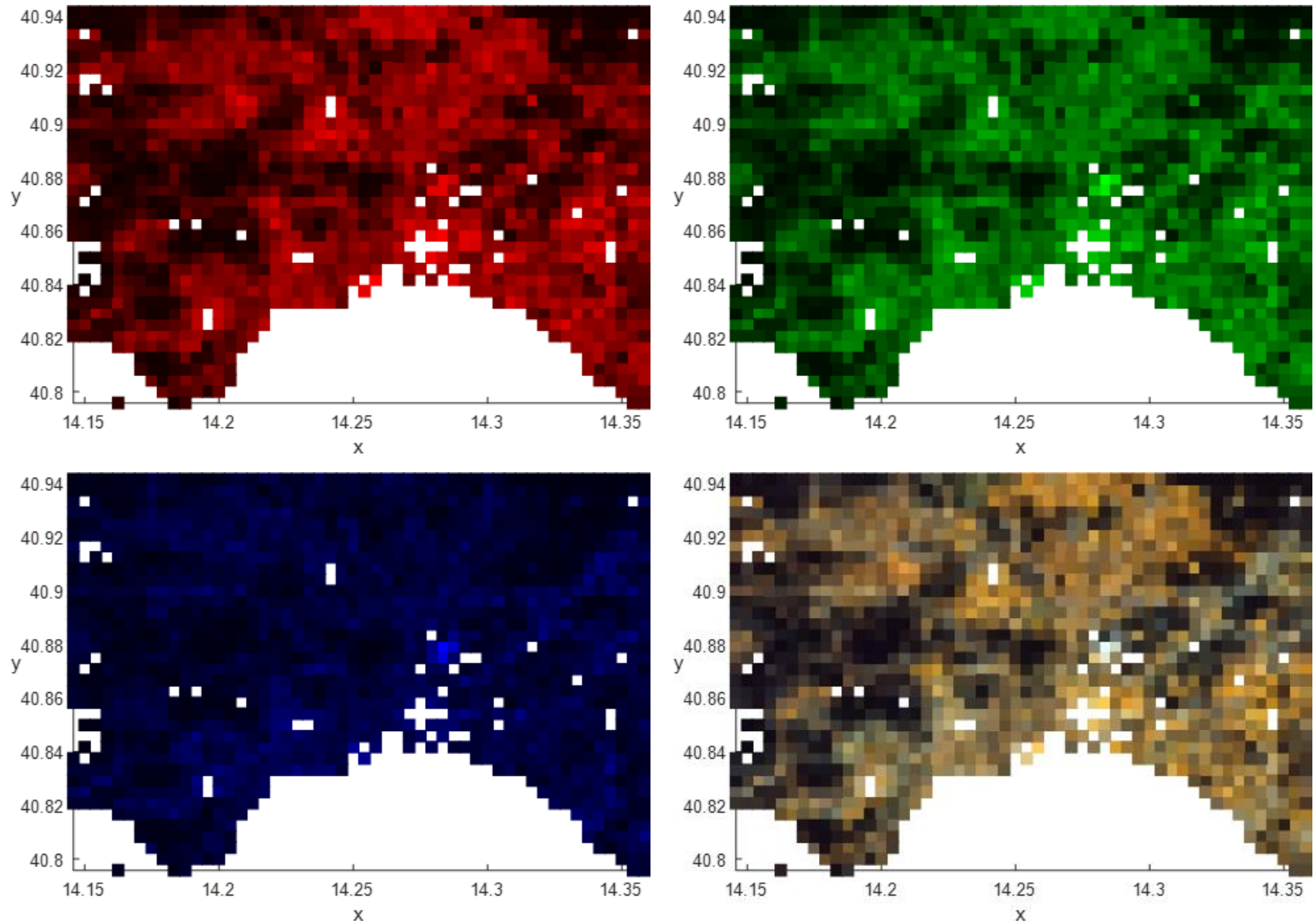

Panel *(a):* ISS-provided and resampled to the spatial resolution of VIIRS imagery (see explanation at p.27)

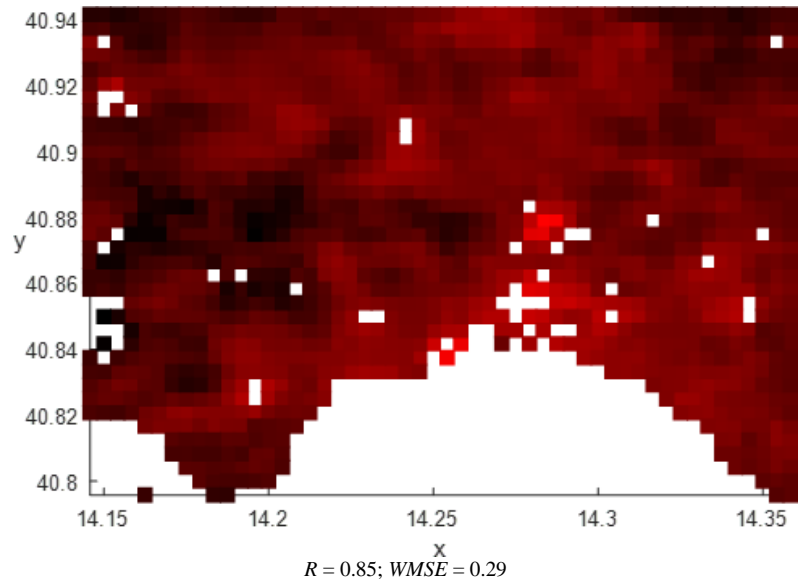

$R = 0.85; WMSE = 0.29$

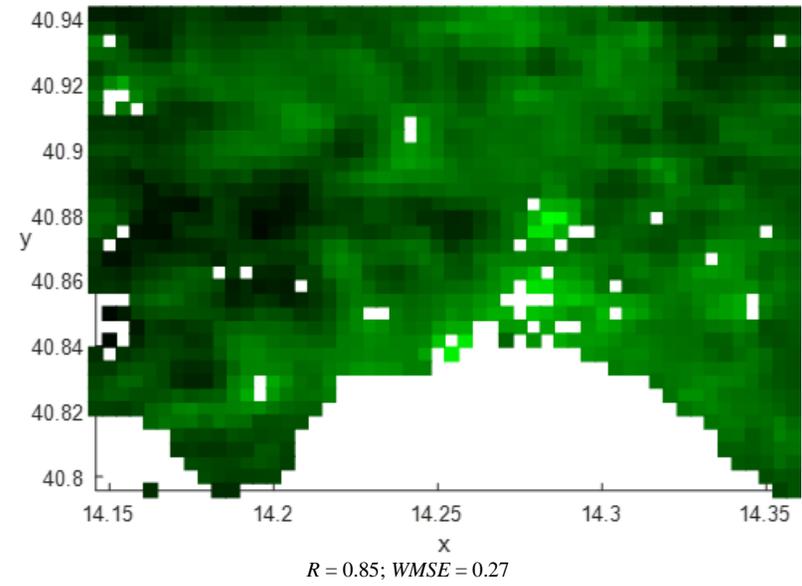

$R = 0.85; WMSE = 0.27$

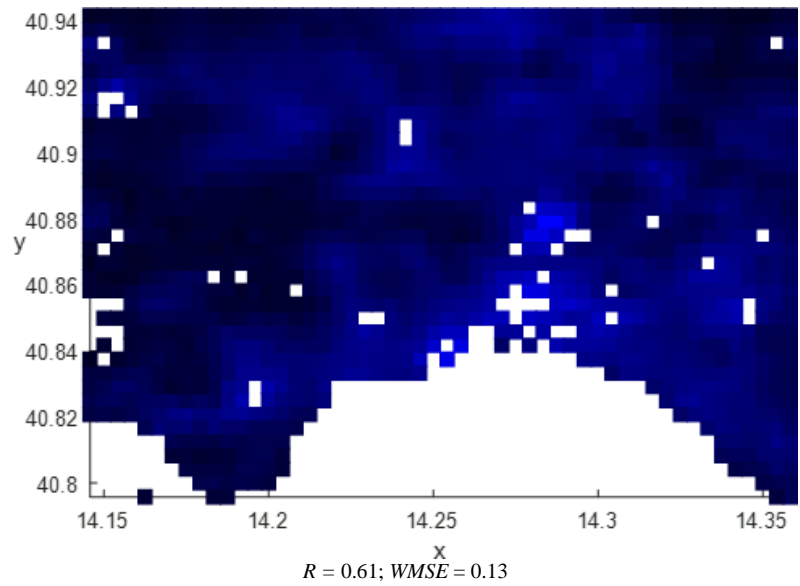

$R = 0.61; WMSE = 0.13$

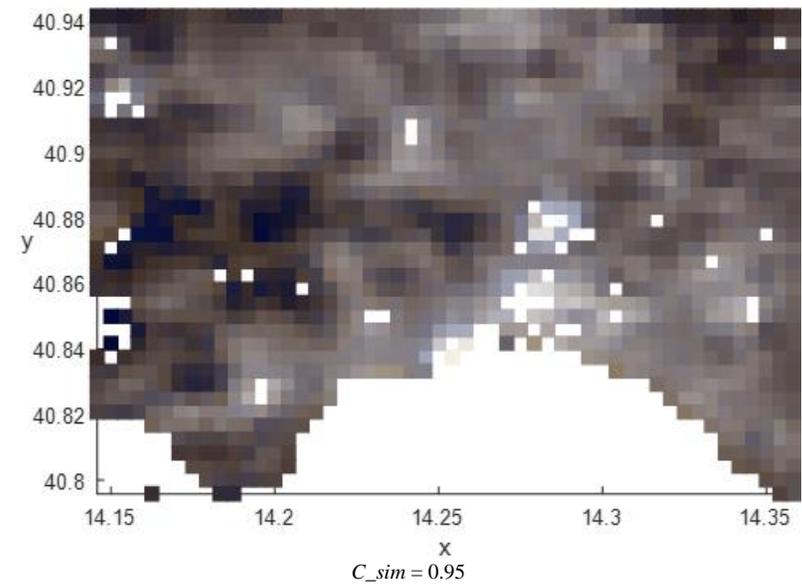

$C\_sim = 0.95$

Panel *(b):* Outputs of linear multiple regressions (see explanation at p.27)

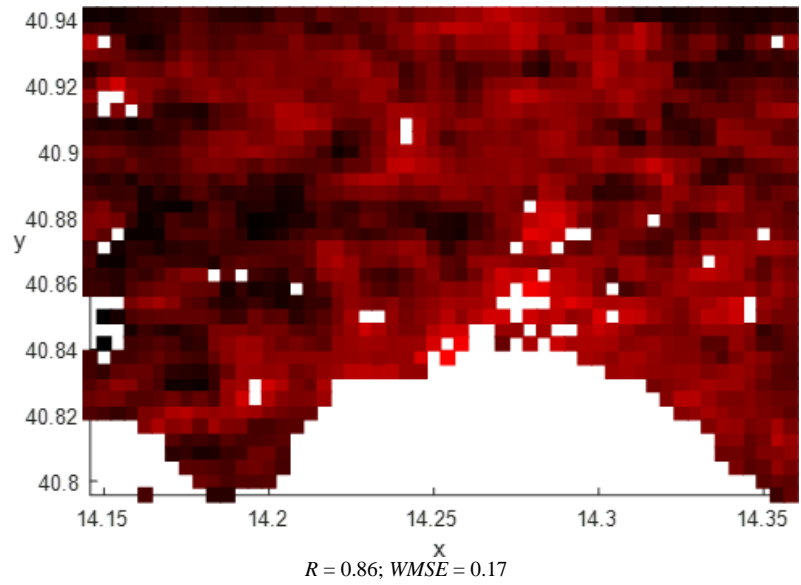

$R = 0.86; WMSE = 0.17$

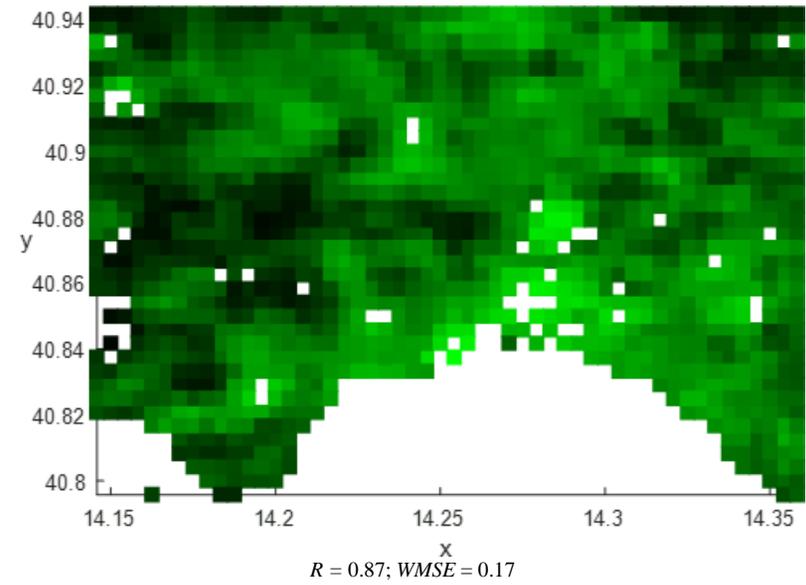

$R = 0.87; WMSE = 0.17$

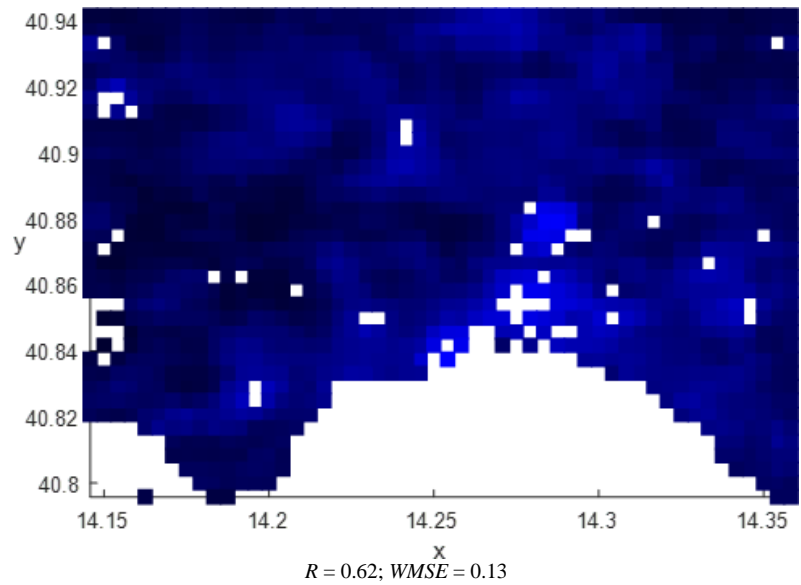

$R = 0.62; WMSE = 0.13$

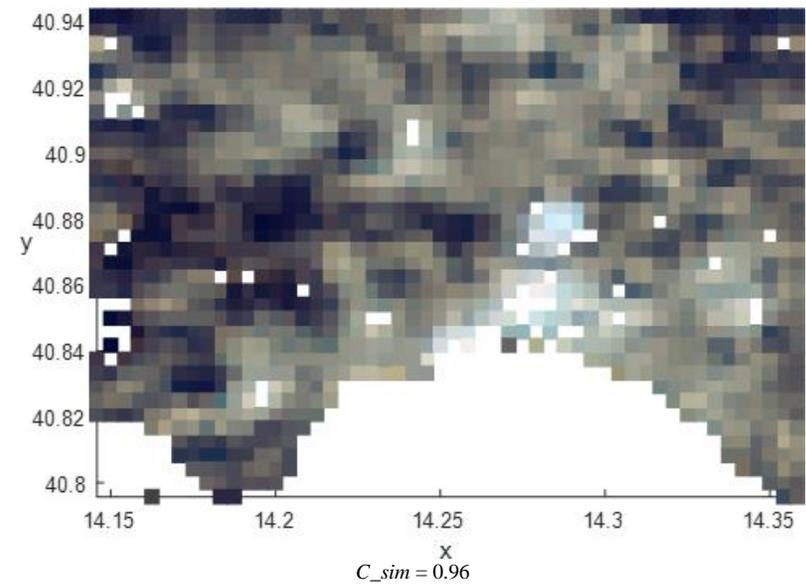

$C\_sim = 0.96$

Panel *(c):* Outputs of non-linear kernel regressions (see explanation at p.27)

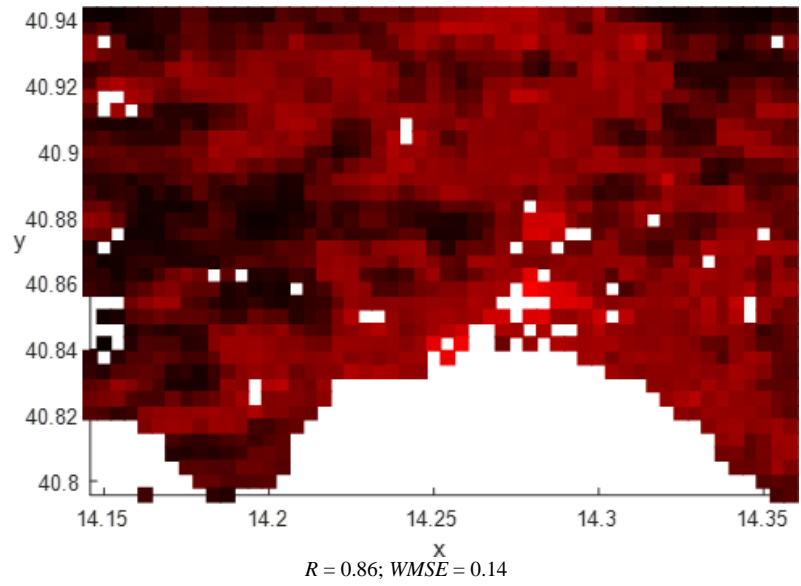

$R = 0.86;\ WMSE = 0.14$

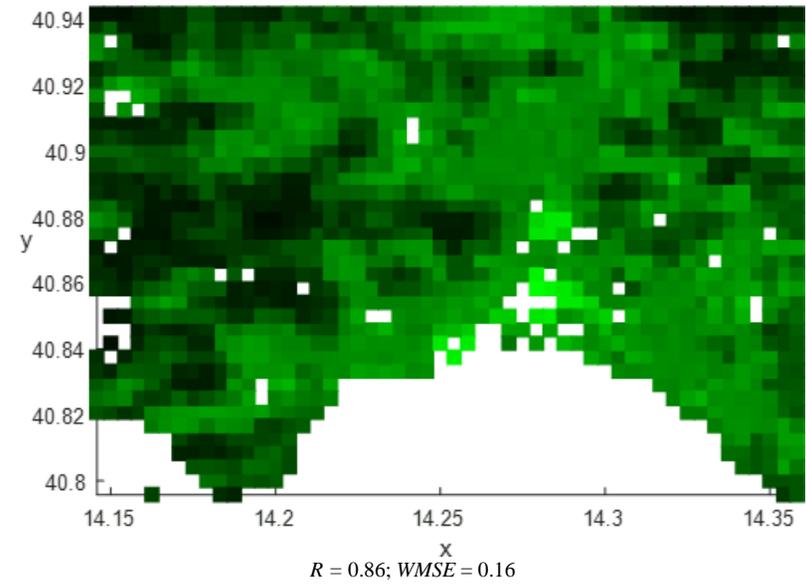

$R = 0.86;\ WMSE = 0.16$

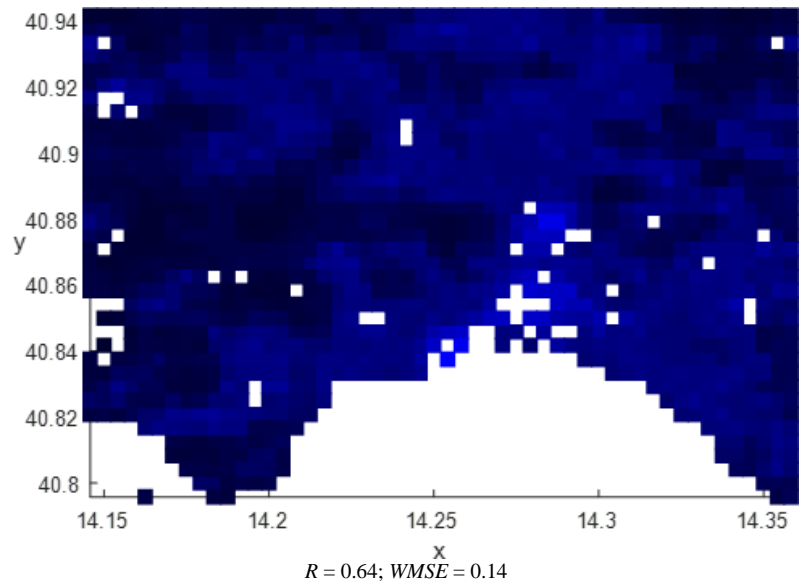

$R = 0.64;\ WMSE = 0.14$

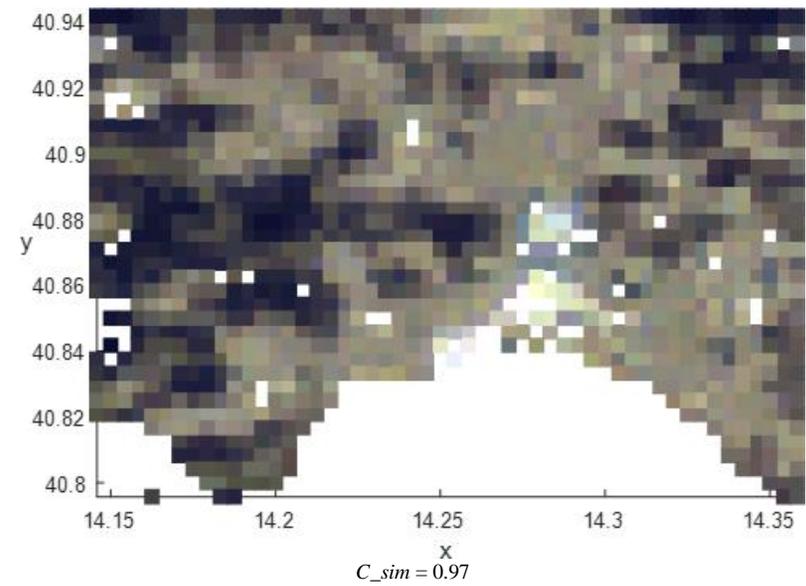

$C\_sim = 0.97$

Panel *(d):* Outputs of random forest regressions (see explanation at p.27)

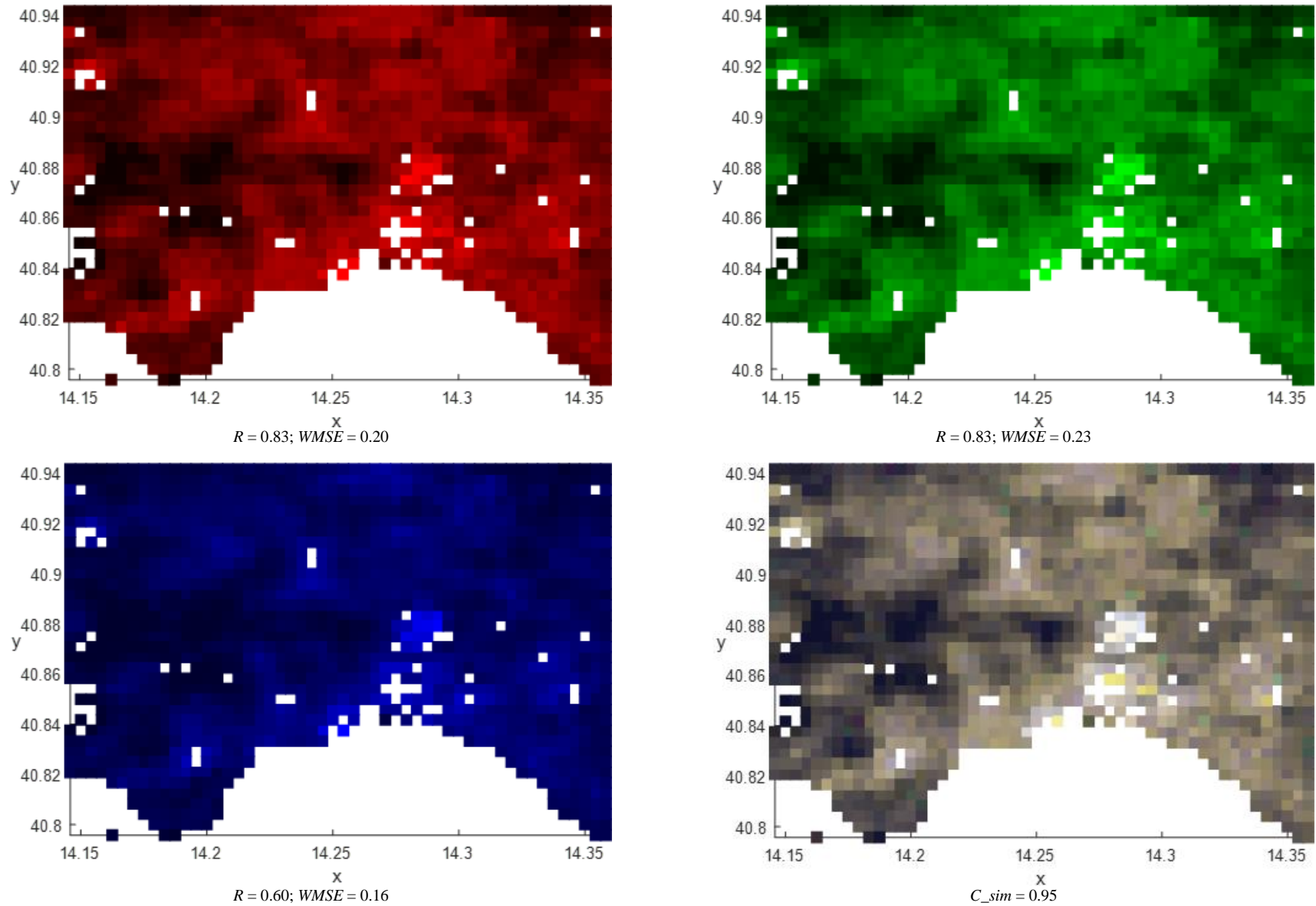

Panel *(e)*: Outputs of elastic map models

Fig. A7. *Naples metropolitan area (Italy)*: Red (the first column), Green (the second column), Blue (the third column)) bands, and RGB images (the fourth column); ISS-provided, resampled to the spatial resolution of VIIRS imagery (the first row), and outputs of four models trained on *Haifa datasets*: linear multiple regressions (the second row), non-linear kernel regressions (the third row), random forest regressions (the fourth row), and elastic map models (the fifth row).

*Notes:* Output generated by elastic maps, built under 0.05 bending penalty, is reported. *R* and *WMSE* denote correspondingly for Pearson's correlation and weighted mean squared error of the red, green, and blue lights' estimates, *C_sim* – for contrast similarity between restored and original RGB images. White points in the city area correspond to outliers.

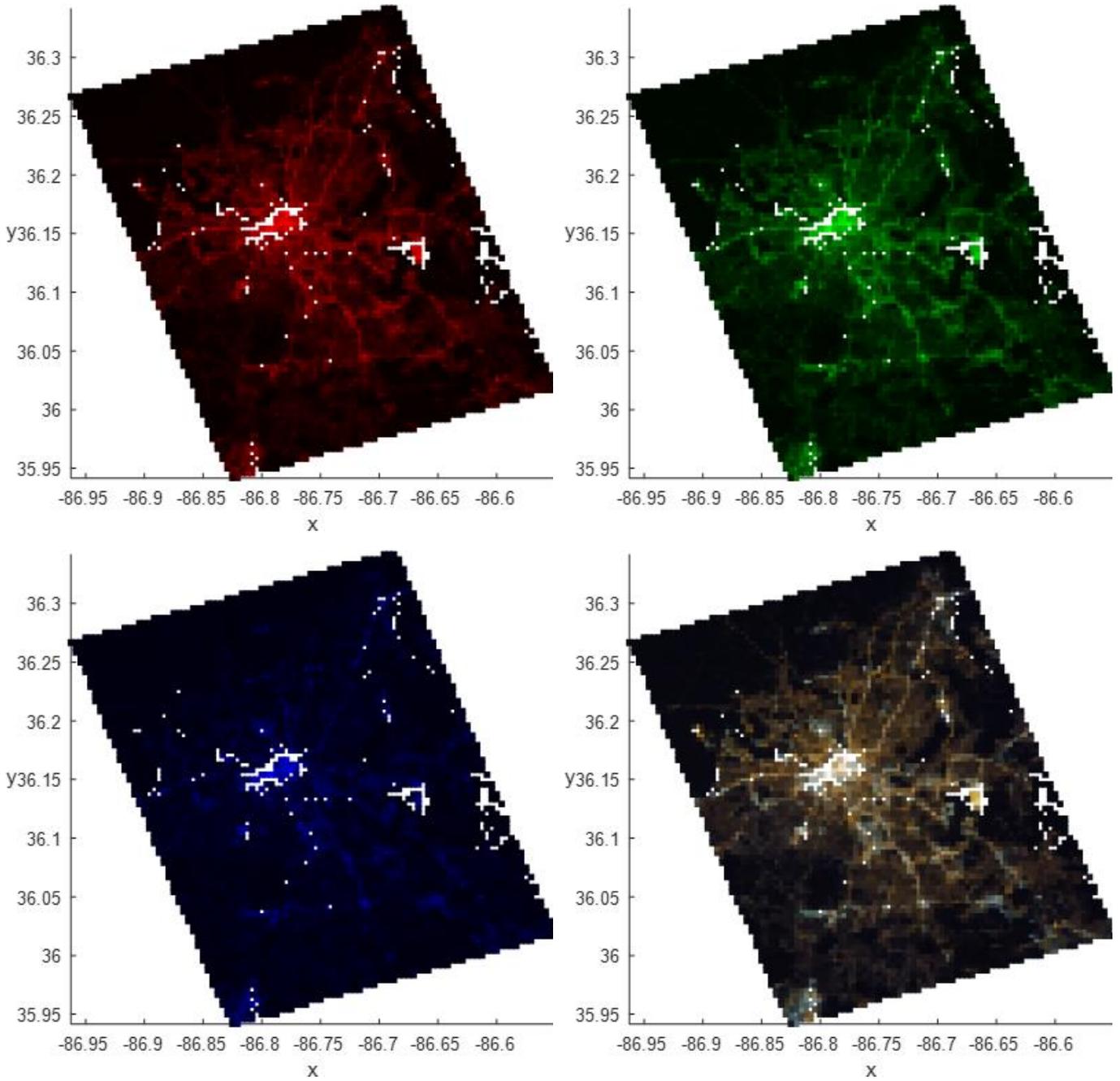

Panel *(a):* ISS-provided and resampled to the spatial resolution of VIIRS imagery (see explanation at p.32)

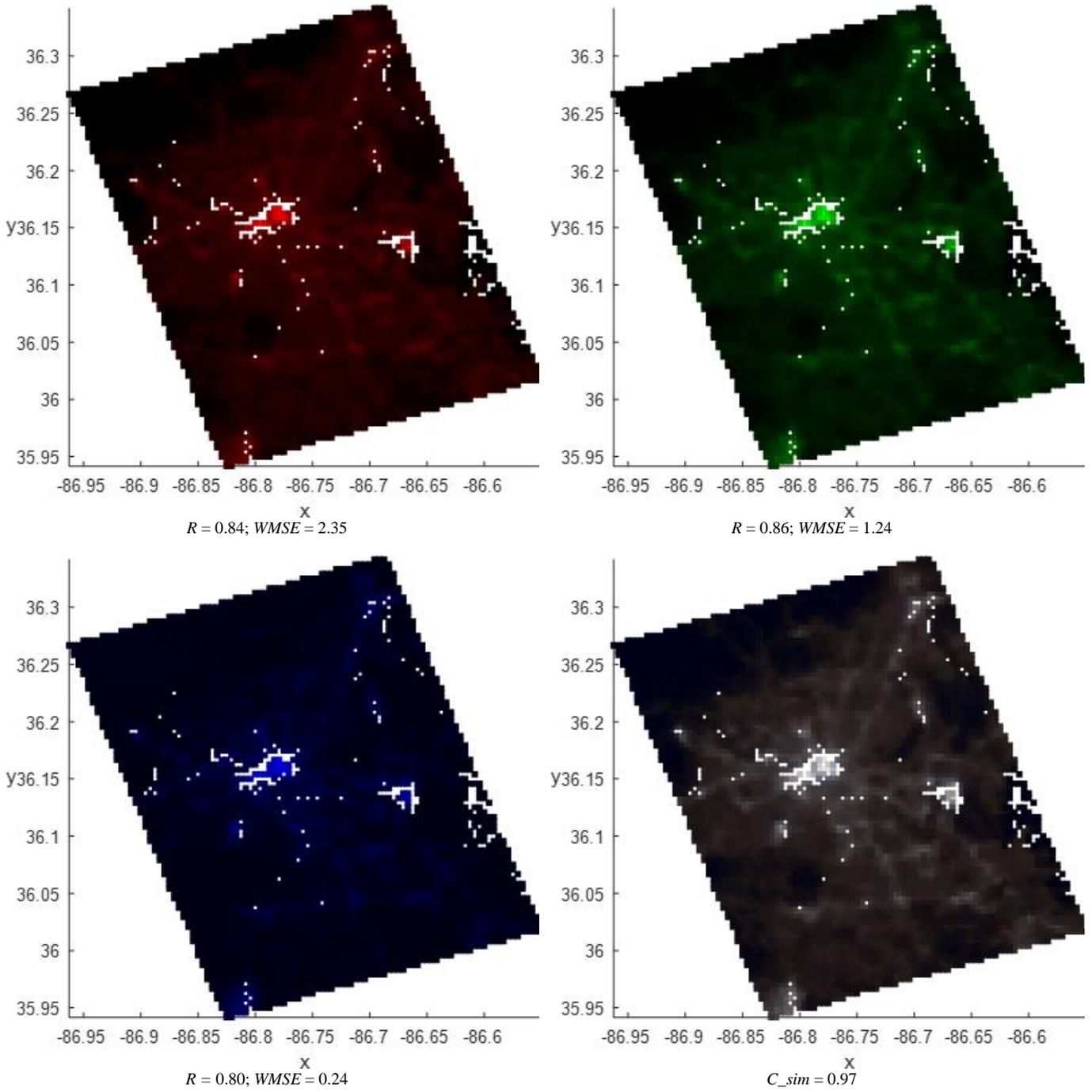

$R = 0.84$; $WMSE = 2.35$

$R = 0.86$; $WMSE = 1.24$

$R = 0.80$; $WMSE = 0.24$

$C\_sim = 0.97$

Panel *(b):* Outputs of linear multiple regressions (see explanation at p.32)

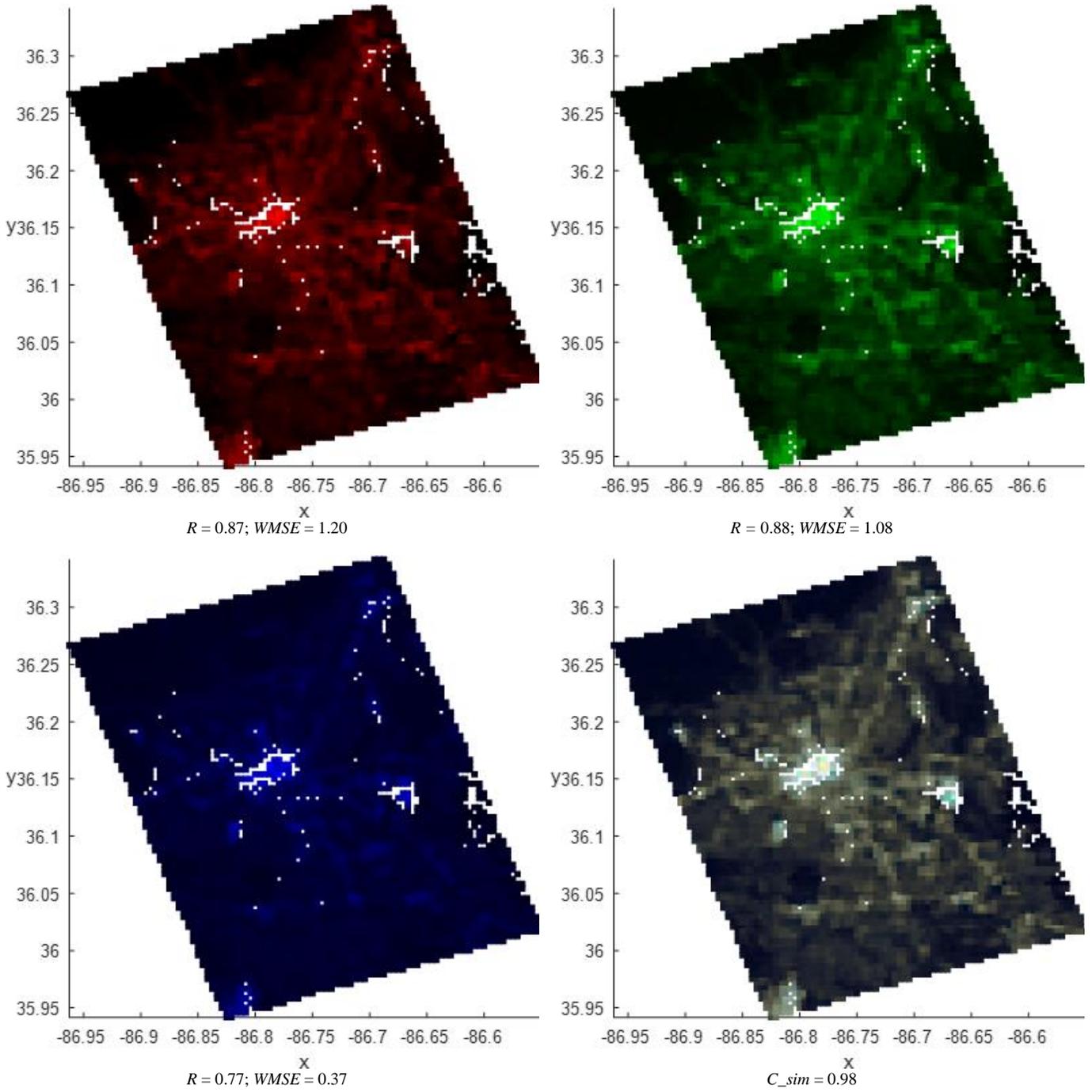

Panel *(c):* Outputs of non-linear kernel regressions (see explanation at p.32)

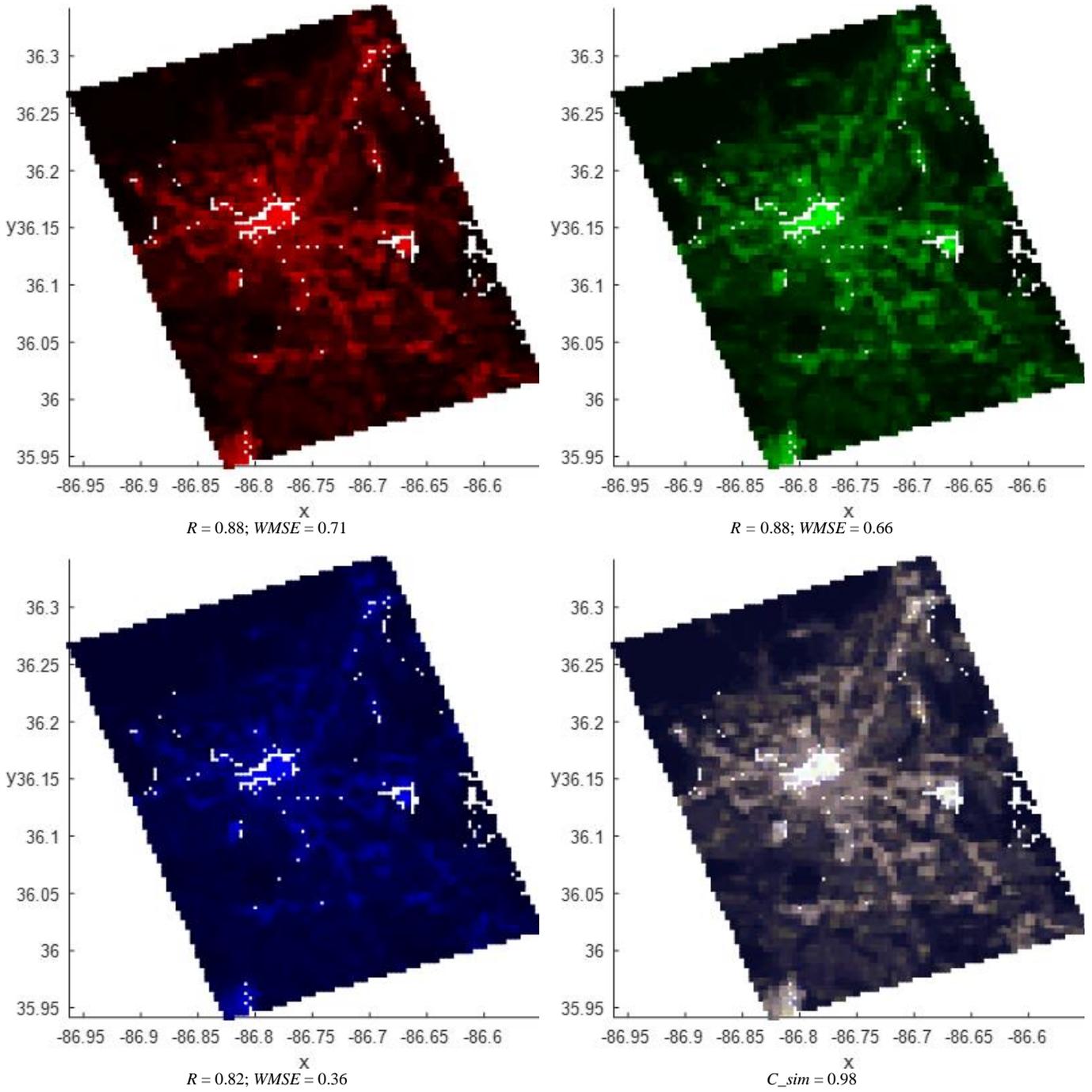

Panel *(d):* Outputs of random forest regressions (see explanation at p.32)

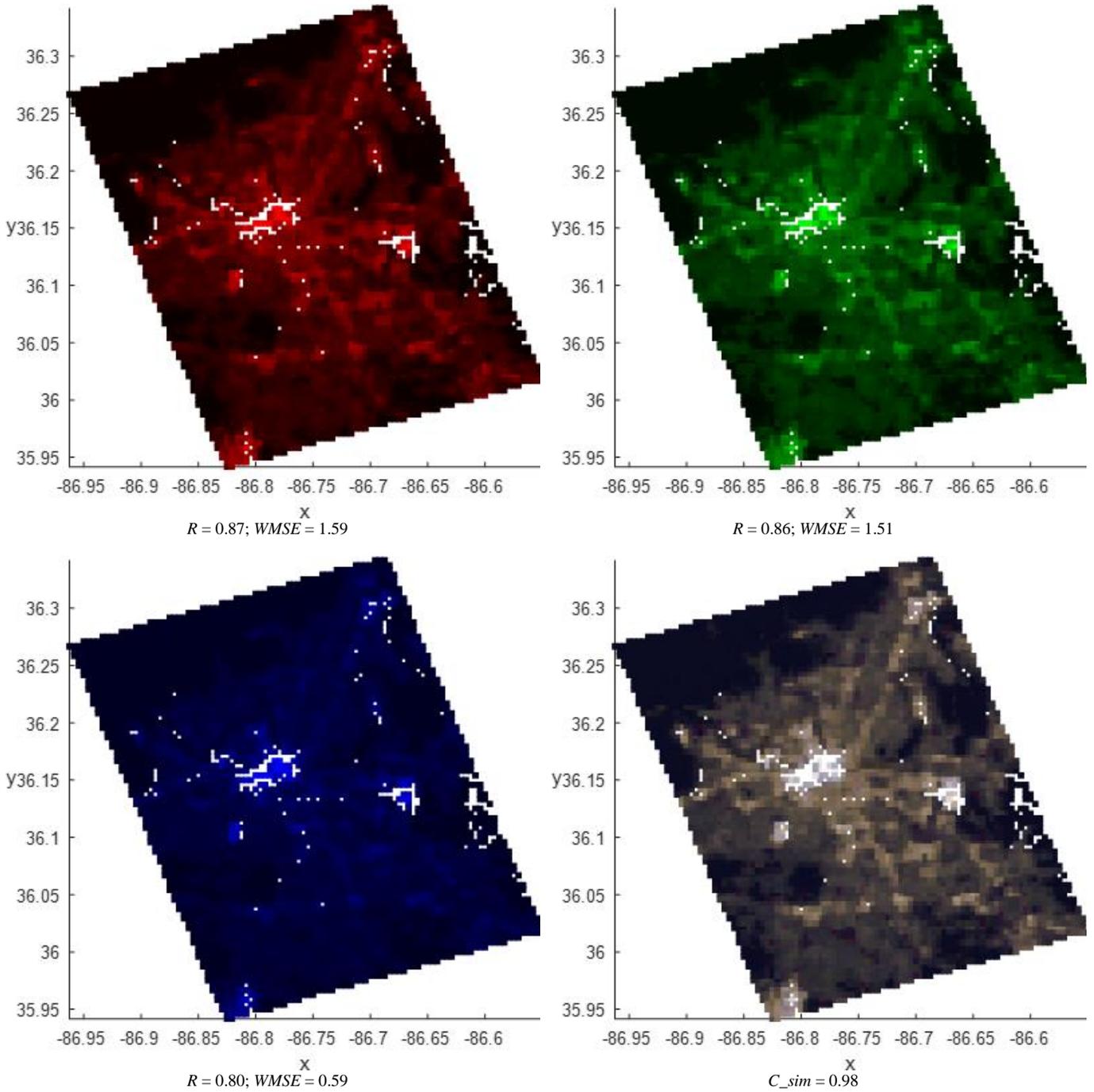

Panel *(e):* Outputs of elastic map models

Fig. A8. *Nashville metropolitan area (the US):* Red (the first column), Green (the second column), Blue (the third column)) bands, and RGB images (the fourth column); ISS-provided, resampled to the spatial resolution of VIIRS imagery (the first row), and outputs of four models trained on *Haifa datasets*: linear multiple regressions (the second row), non-linear kernel regressions (the third row), random forest regressions (the fourth row), and elastic map models (the fifth row). *Notes:* Output generated by elastic maps, built under 0.05 bending penalty, is reported. *R* and *WMSE* denote correspondingly for Pearson's correlation and weighted mean squared error of the red, green, and blue lights' estimates, *C_sim* – for contrast similarity between restored and original RGB images. White points in the city area correspond to outliers.

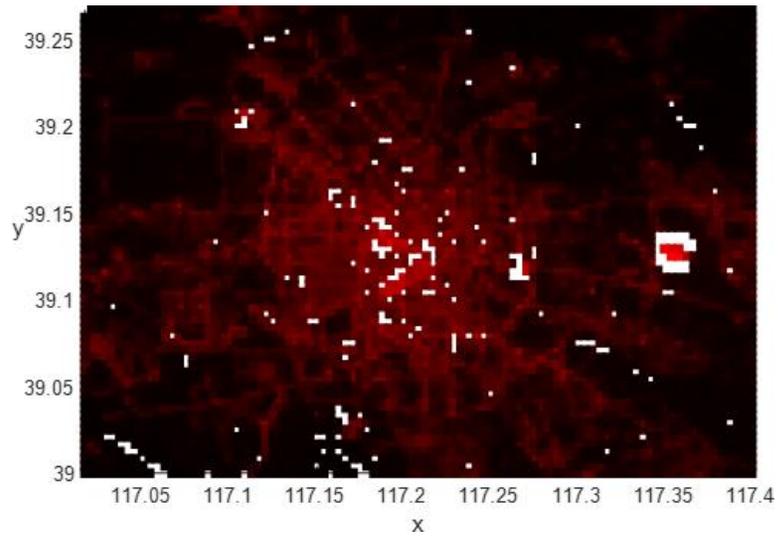
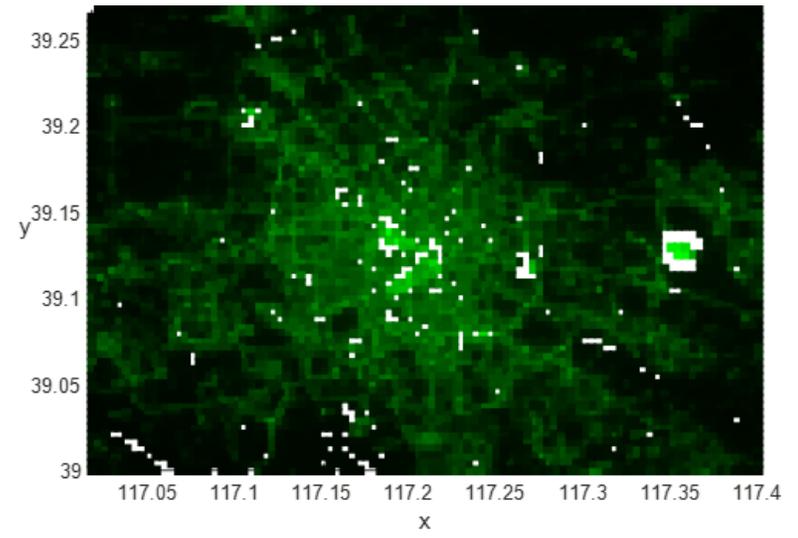
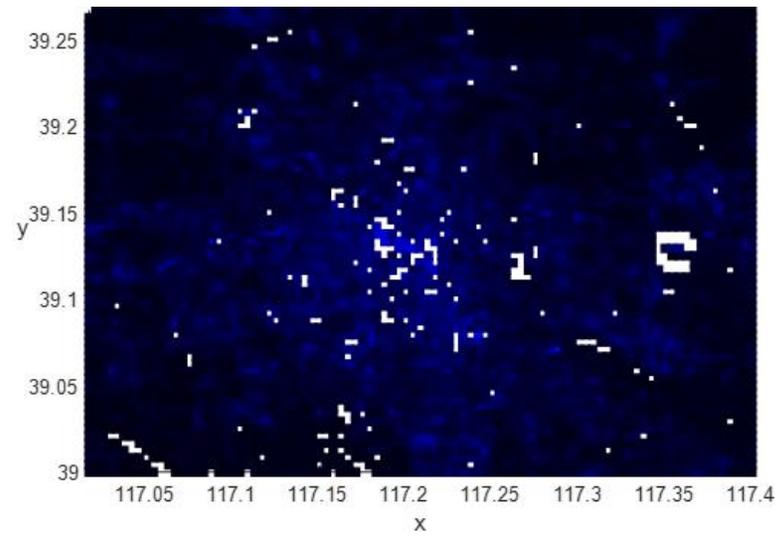
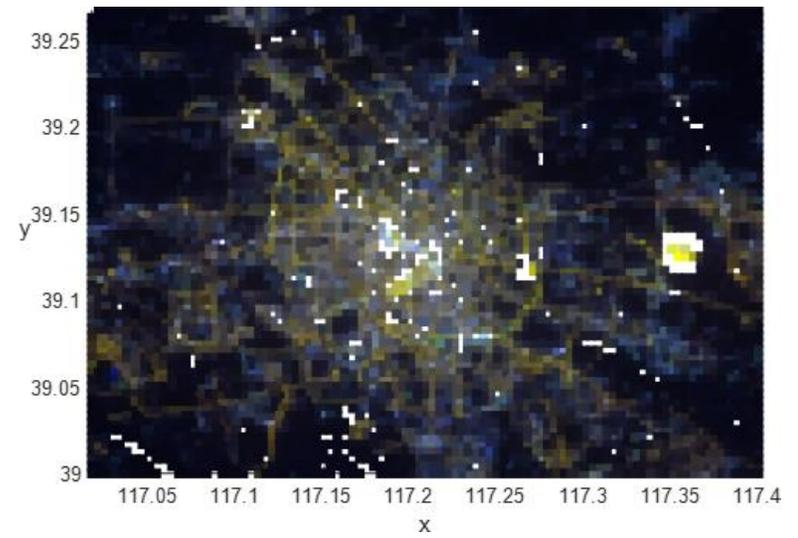

Panel *(a):* ISS-provided and resampled to the spatial resolution of VIIRS imagery (see explanation at p.37)

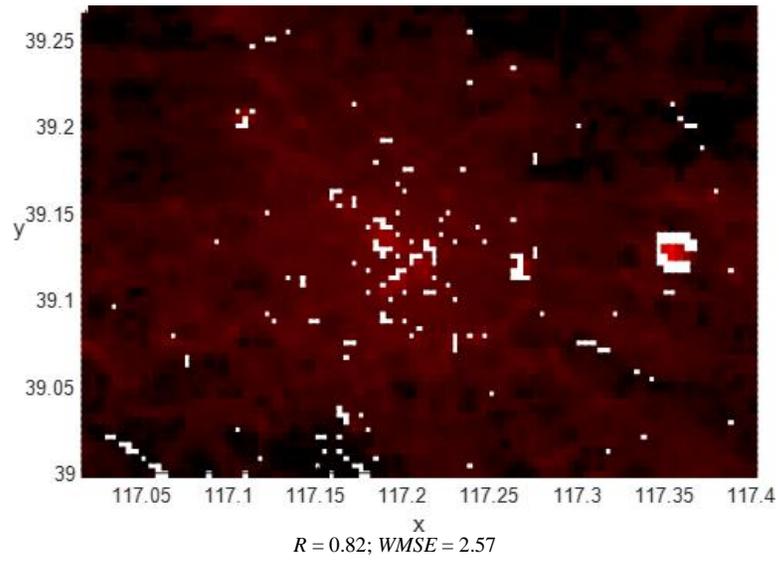

*R* = 0.82; *WMSE* = 2.57

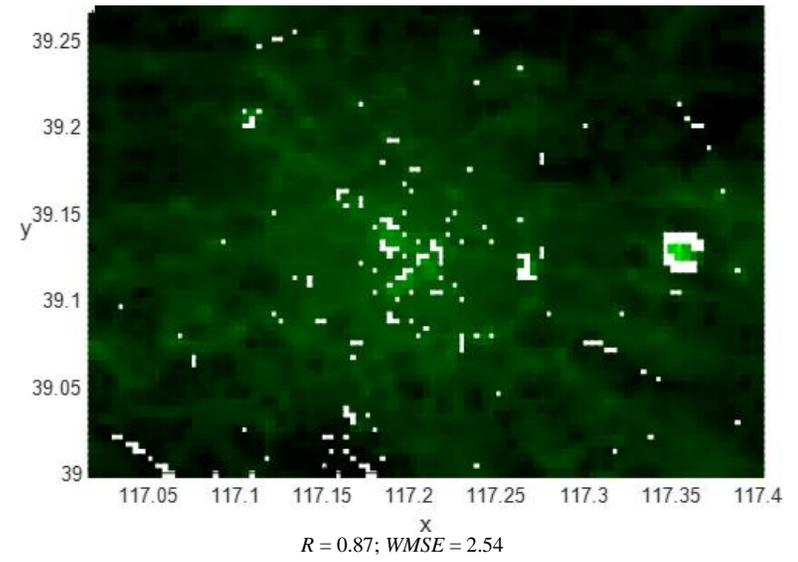

*R* = 0.87; *WMSE* = 2.54

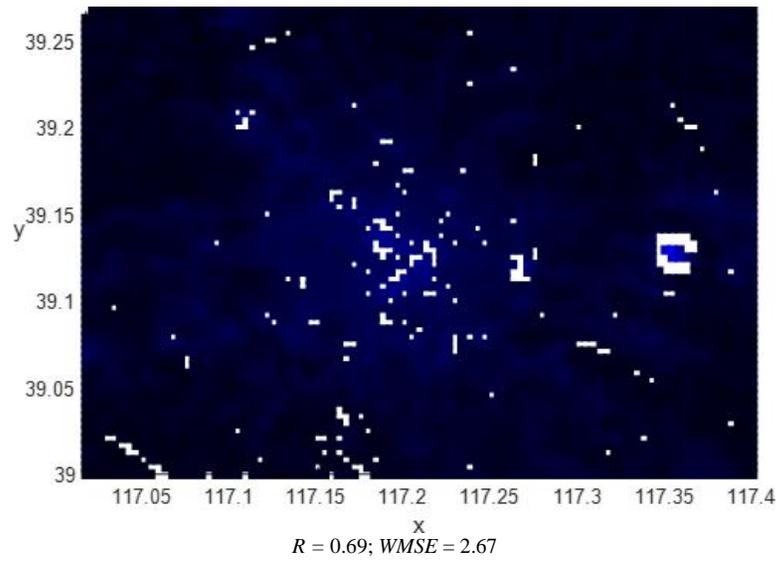

*R* = 0.69; *WMSE* = 2.67

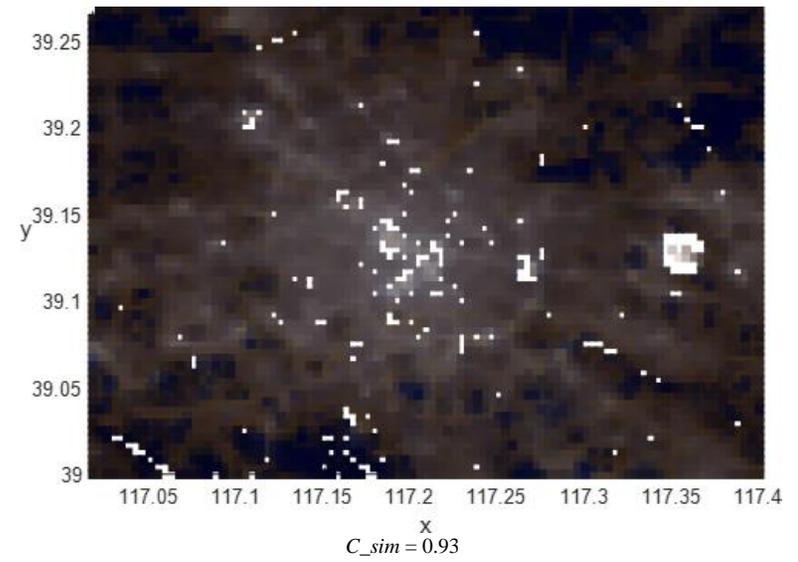

*C_sim* = 0.93

Panel *(b):* Outputs of linear  multiple regressions (see explanation at p.37)

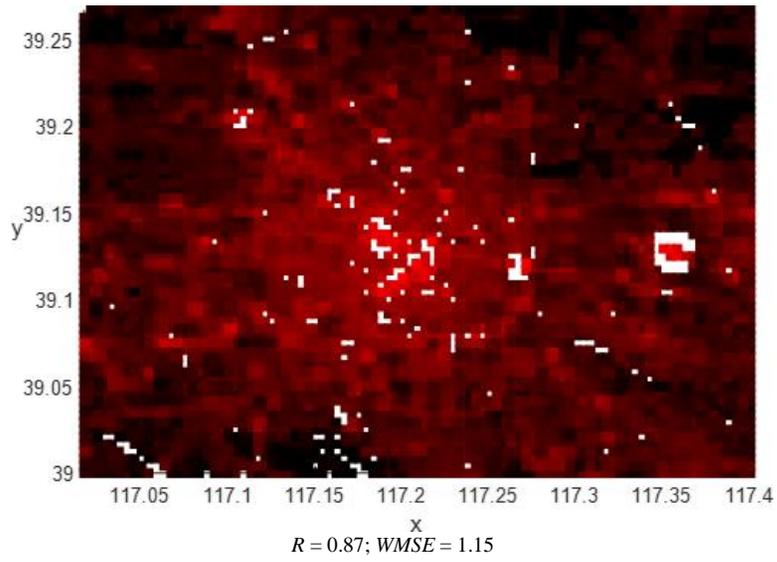

R = 0.87; *WMSE* = 1.15

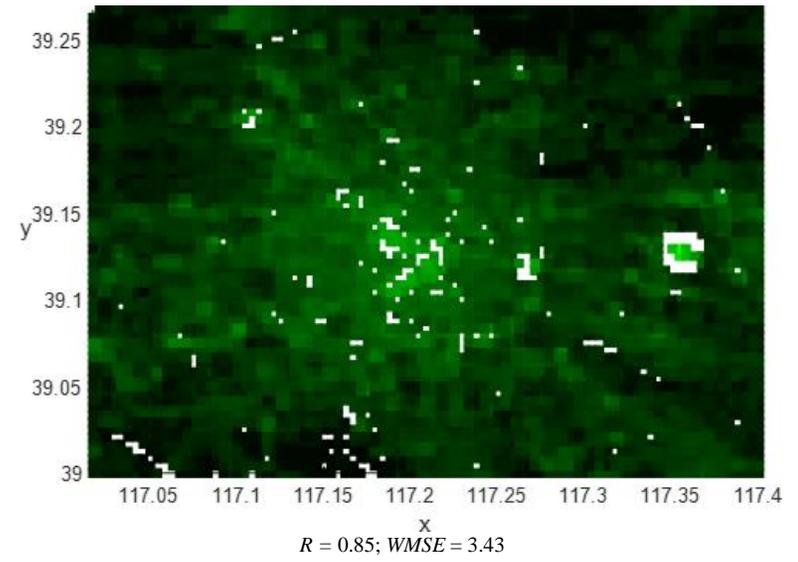

R = 0.85; *WMSE* = 3.43

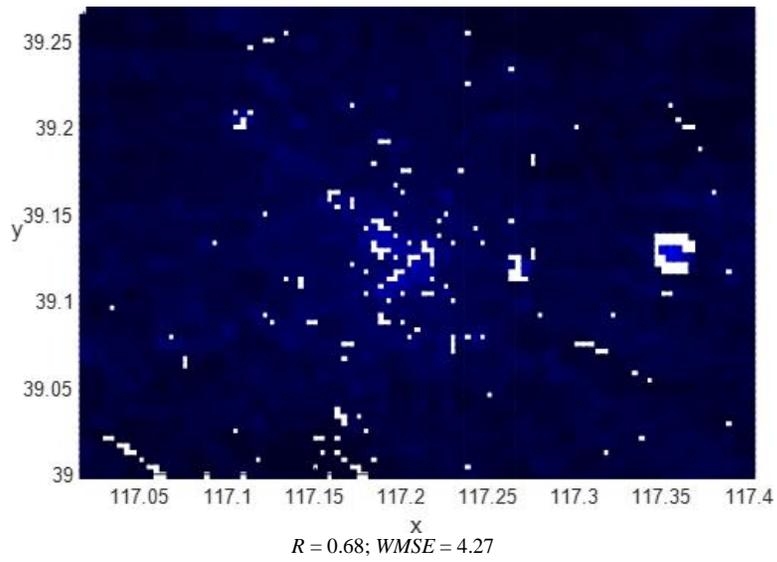

R = 0.68; *WMSE* = 4.27

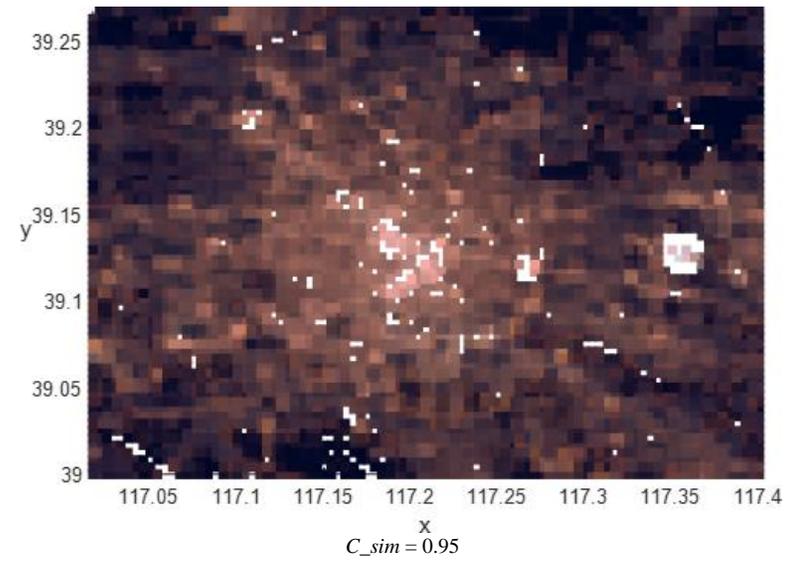

C_sim = 0.95

Panel *(c):* Outputs of non-linear kernel regressions (see explanation at p.37)

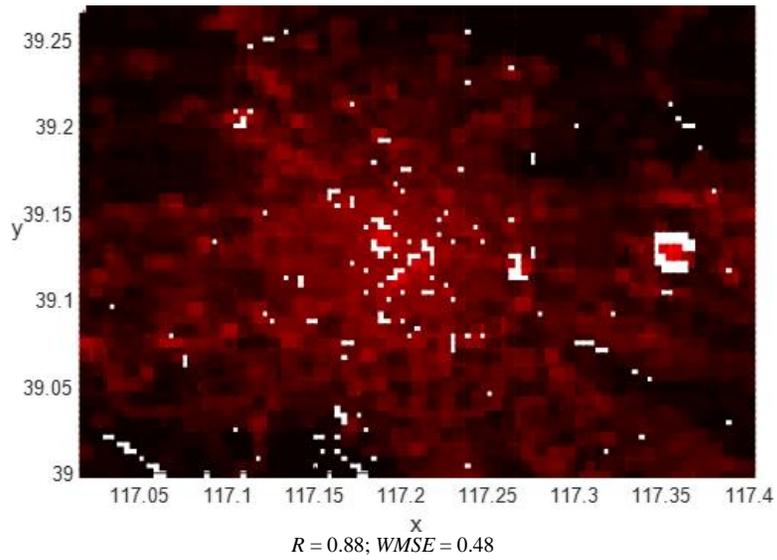

$R = 0.88; WMSE = 0.48$

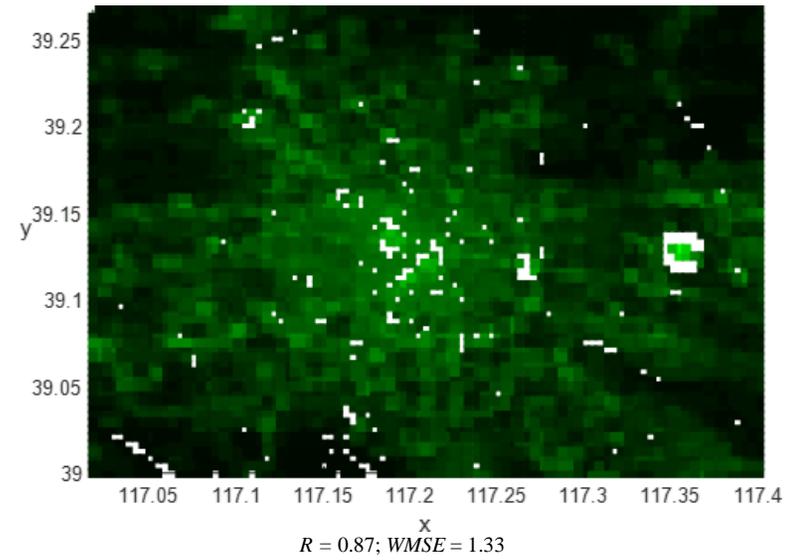

$R = 0.87; WMSE = 1.33$

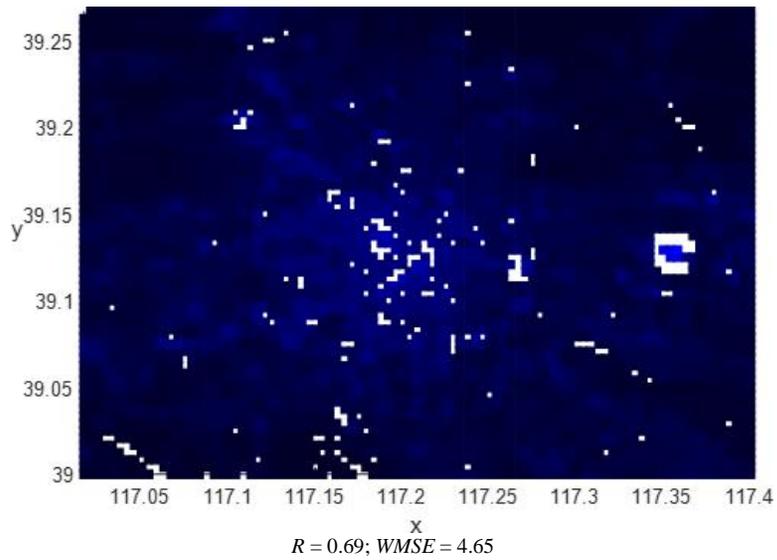

$R = 0.69; WMSE = 4.65$

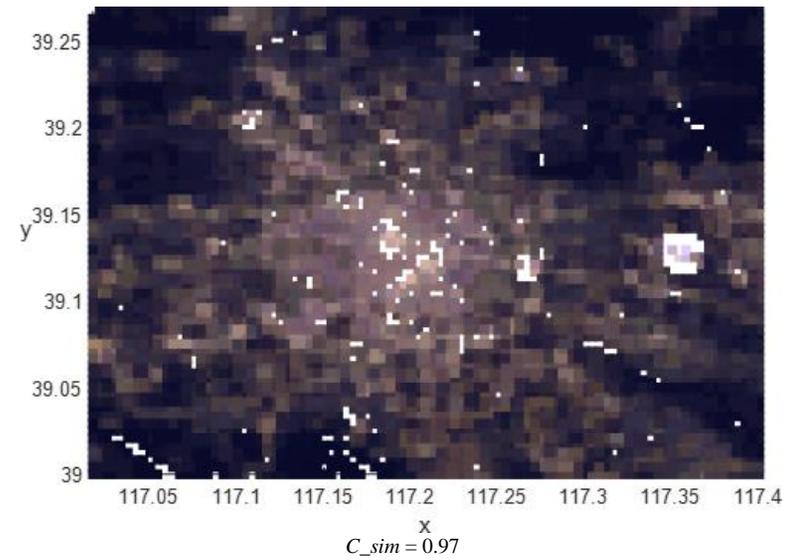

$C\_sim = 0.97$

Panel *(d)*: Outputs of random forest regressions (see explanation at p.37)

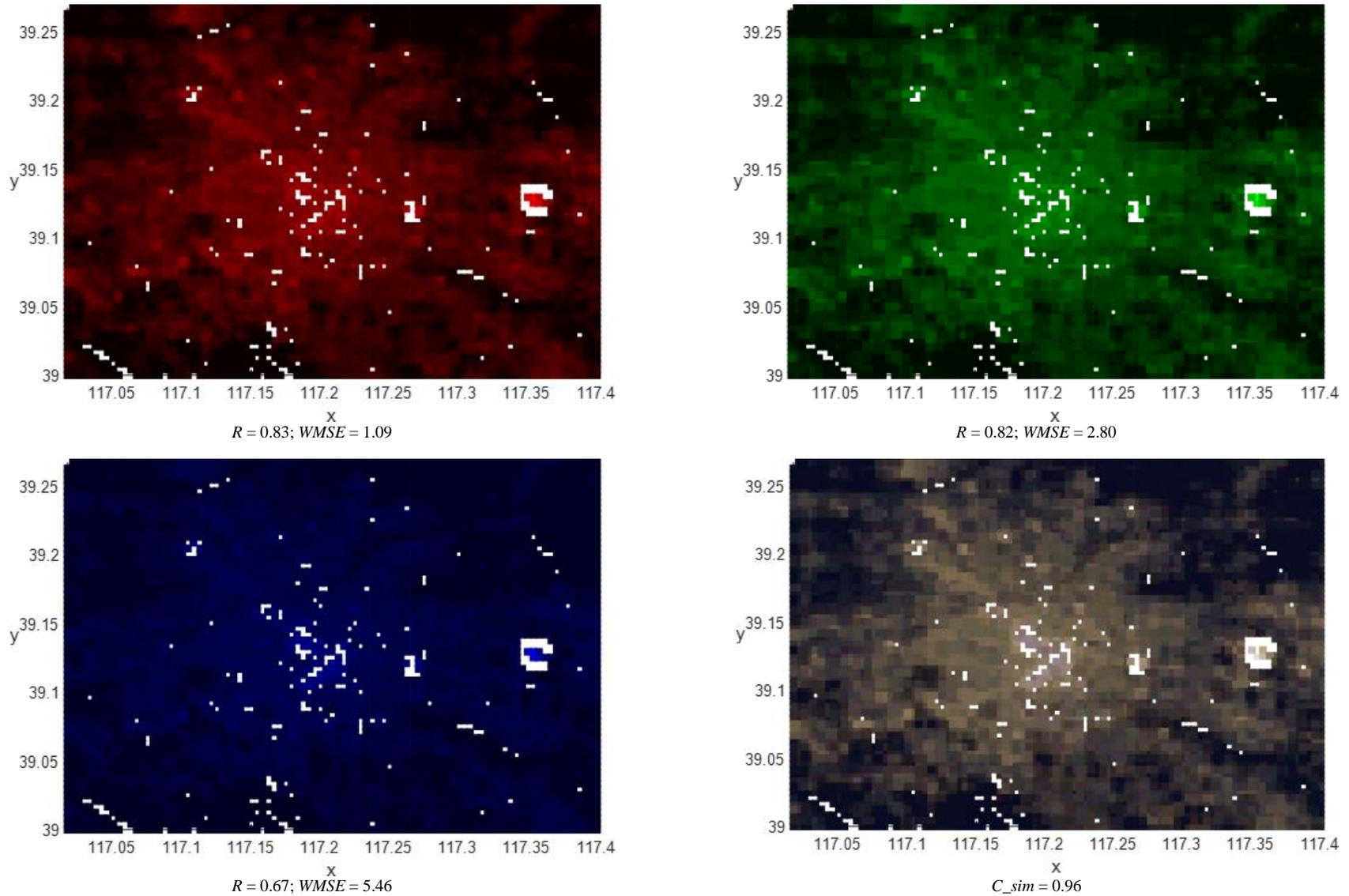

Panel *(e):* Outputs of elastic map models

Fig. A9. *Tianjing metropolitan area (China):* Red (the first column), Green (the second column), Blue (the third column)) bands, and RGB images (the fourth column); ISS-provided, resampled to the spatial resolution of VIIRS imagery (the first row), and outputs of four models trained on *Haifa datasets*: linear multiple regressions (the second row), non-linear kernel regressions (the third row), random forest regressions (the fourth row), and elastic map models (the fifth row).
*Notes:* Output generated by elastic maps, built under 0.05 bending penalty, is reported. *R* and *WMSE* denote correspondingly for Pearson's correlation and weighted mean squared error of the red, green, and blue lights' estimates, *C_sim* – for contrast similarity between restored and original RGB images. White points in the city area correspond to outliers.

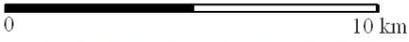

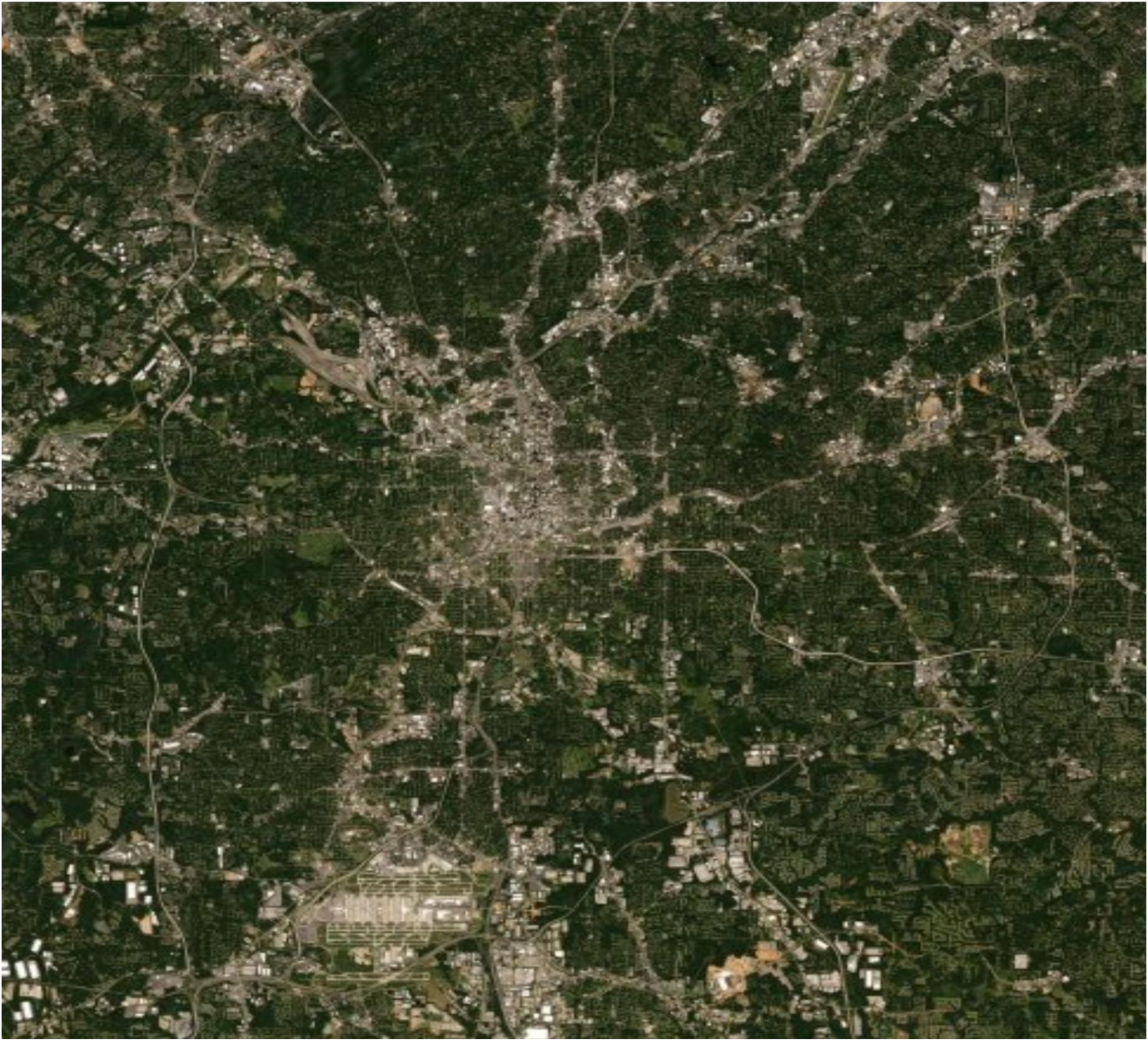

*(a)   Atlanta metropolitan area (the US)*

N

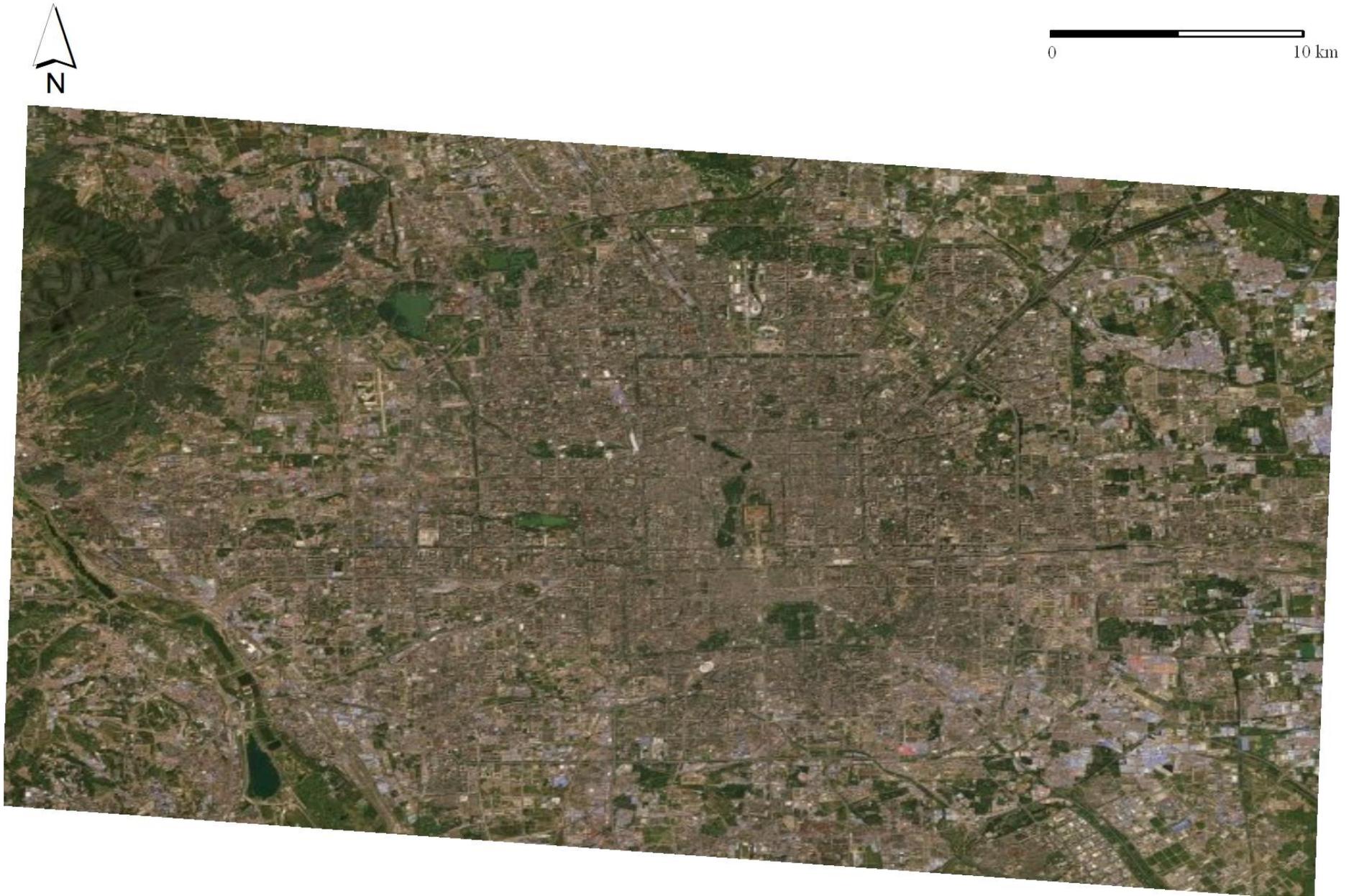

*(b)   Beijing metropolitan area (China)*

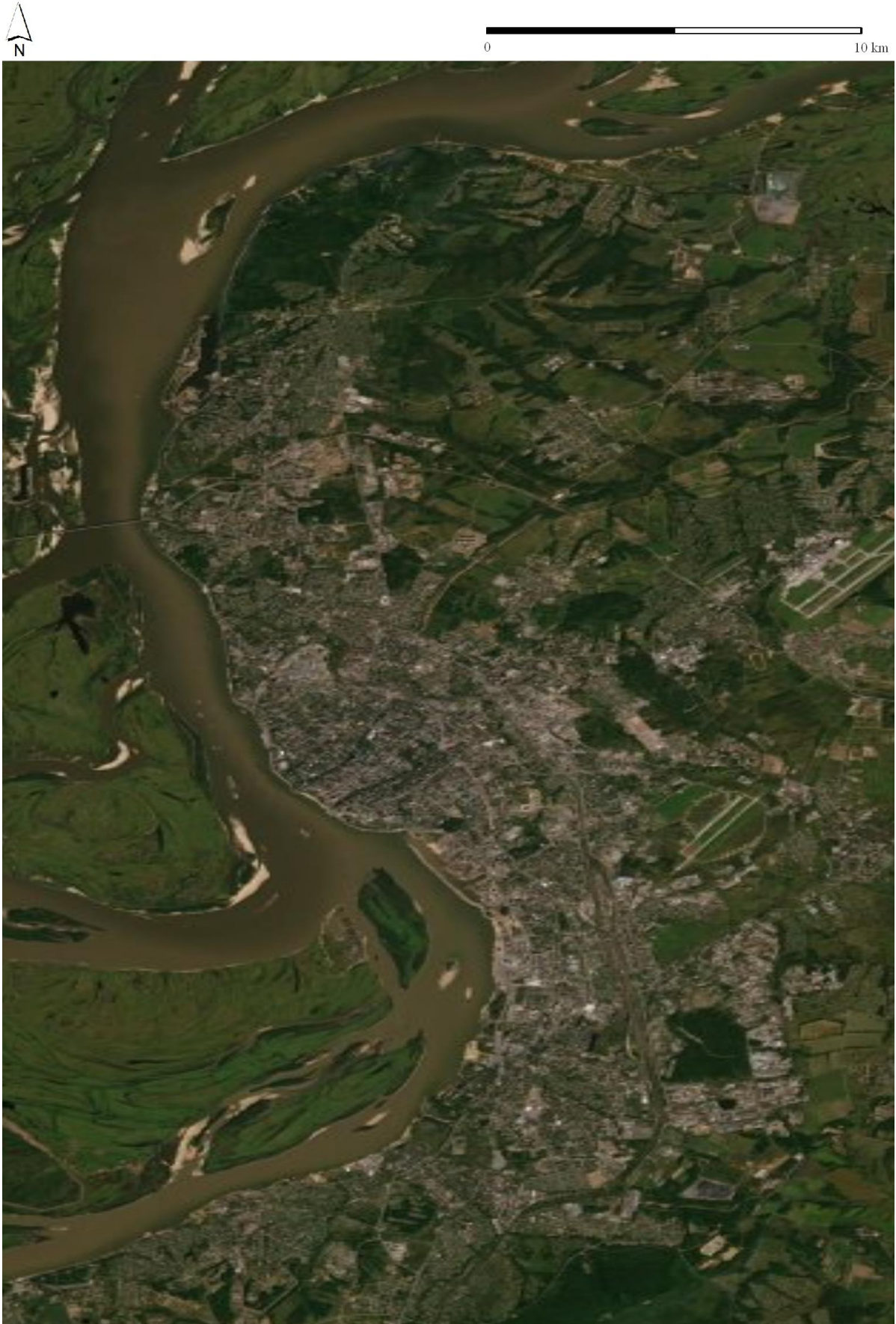

N

0                    10 km

*(c)   Khabarovsk (Russia)*

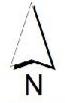

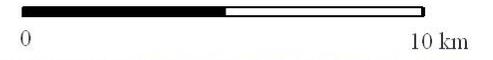

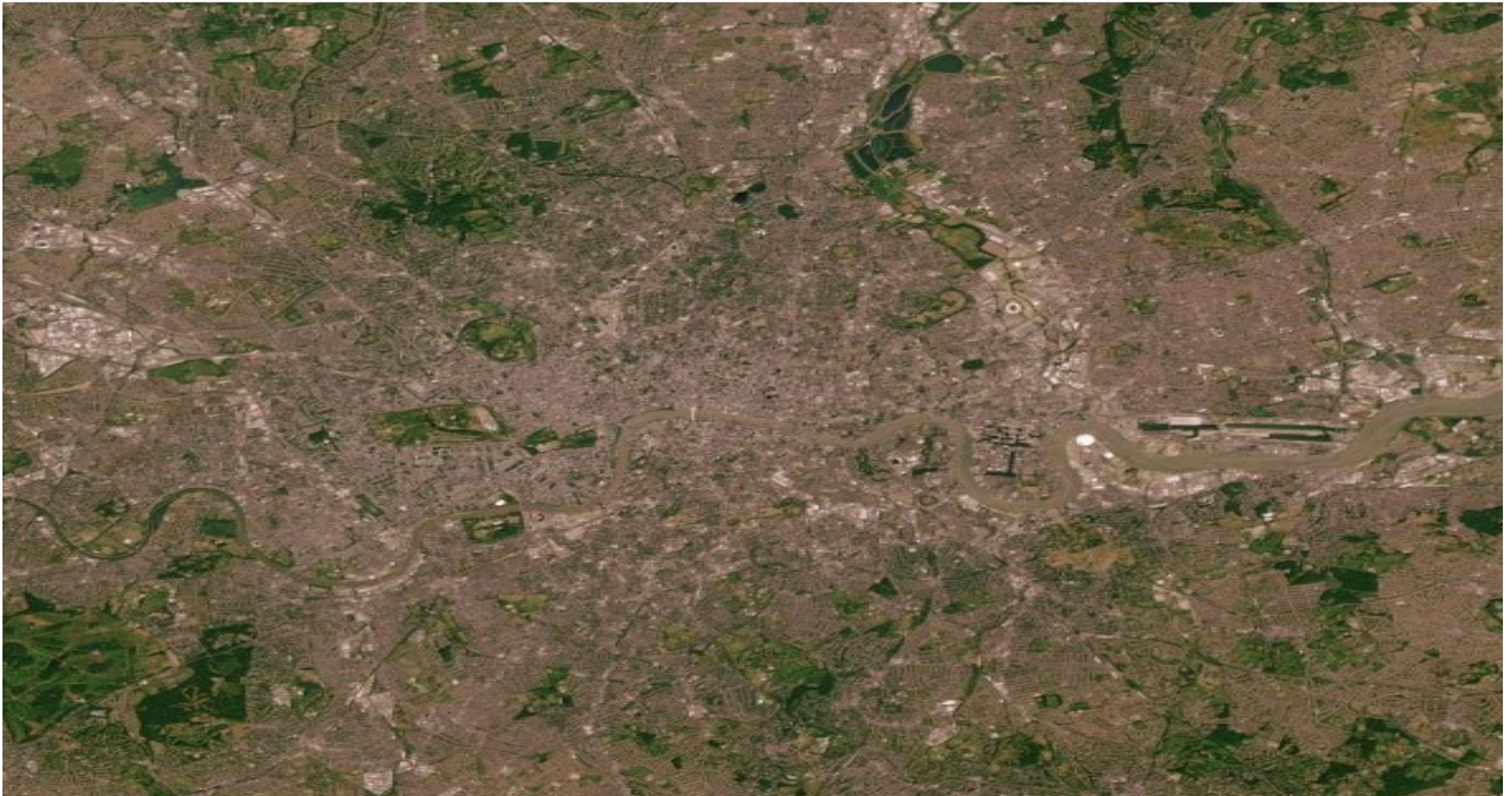

*(d)   London metropolitan area (the UK)*

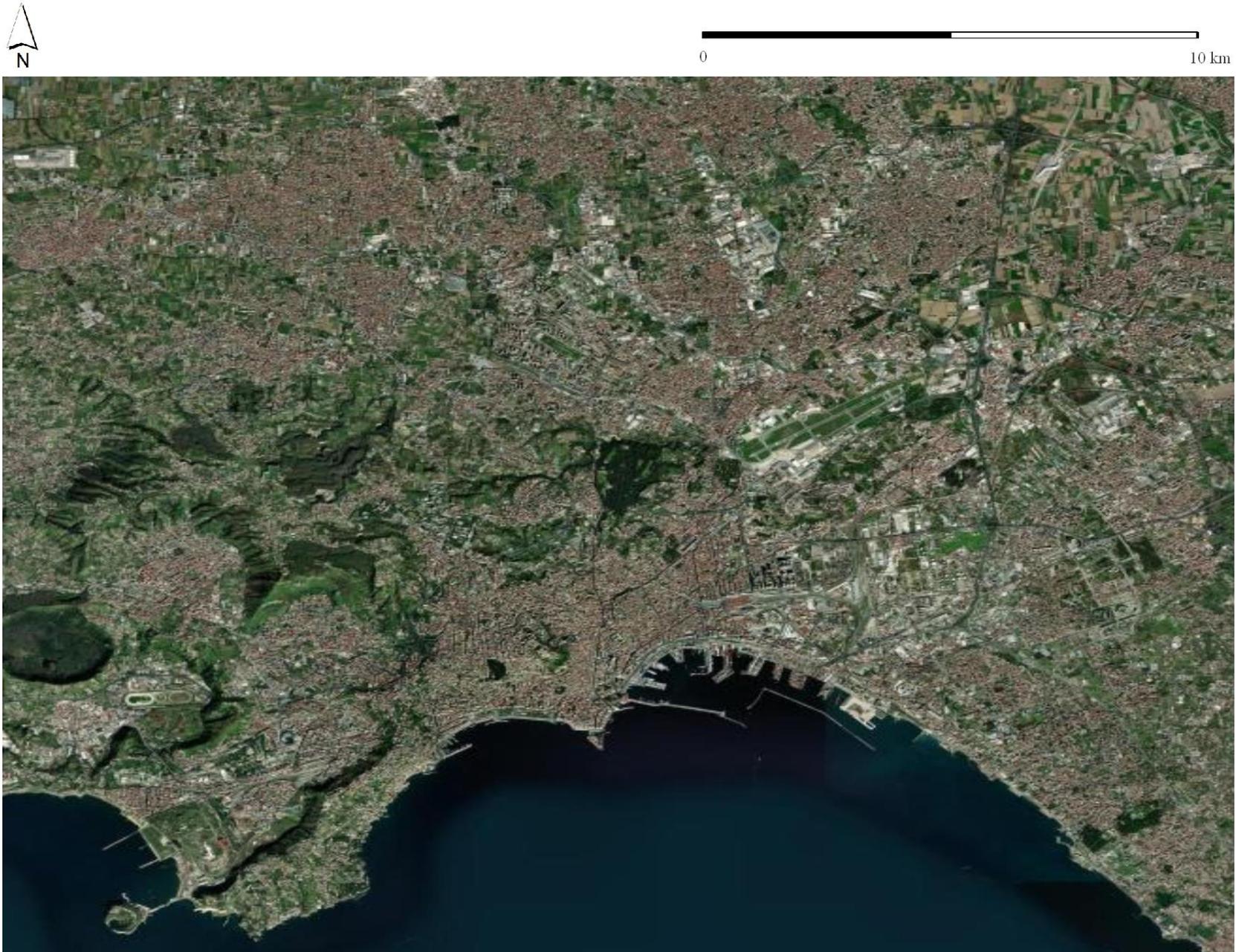

N

0                                                                    10 km

*(e)    Naples metropolitan area (Italy)*

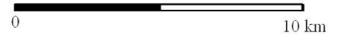

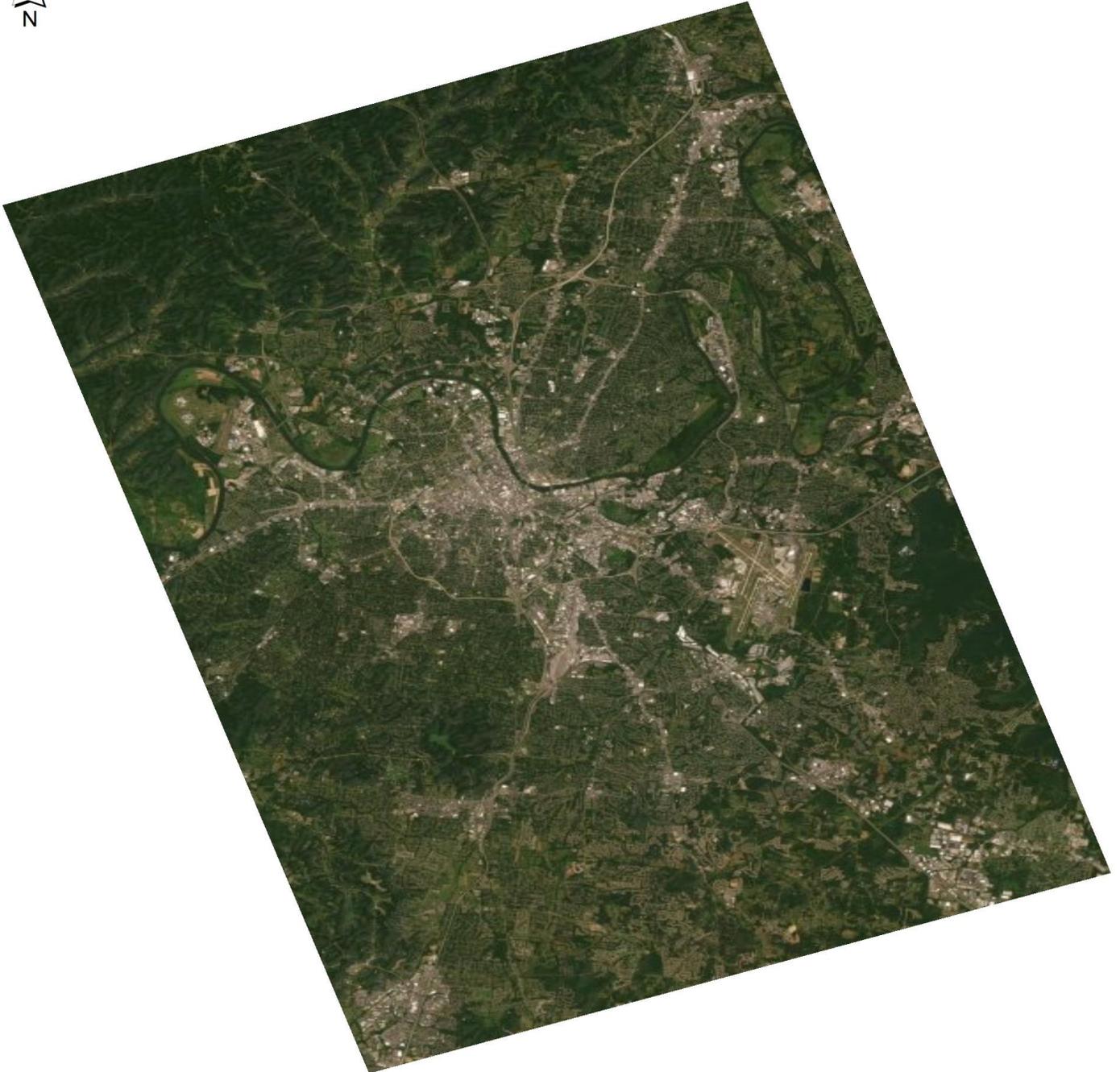

*(f)    Nashville metropolitan area (the US)*

N

0                    10 km

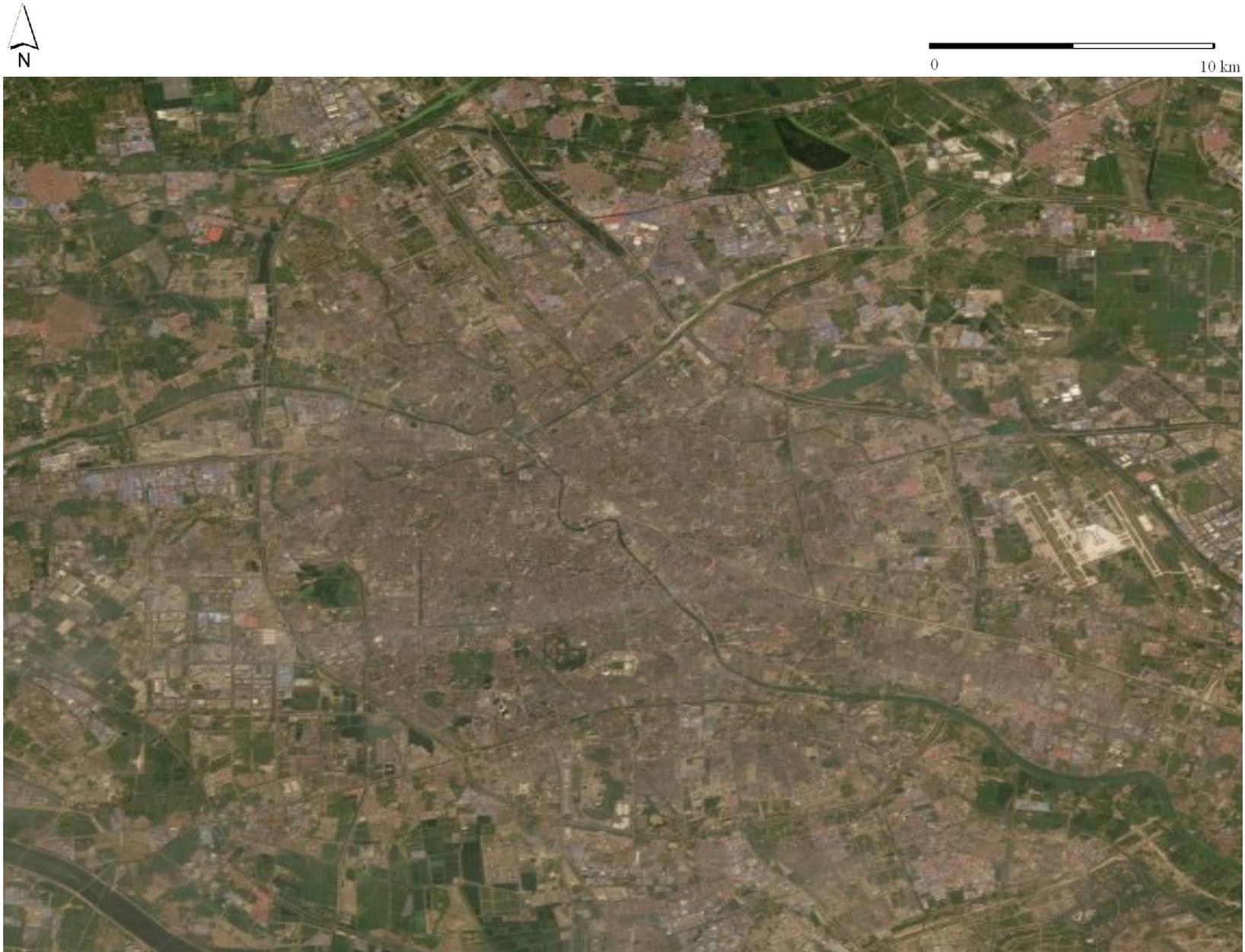

*(g)    Tianjing metropolitan area (China)*

Fig. A10. Day-time satellite images for the cities under analysis (*Source*: Imagery basemaps provided by ArcGIS v.10.x software)

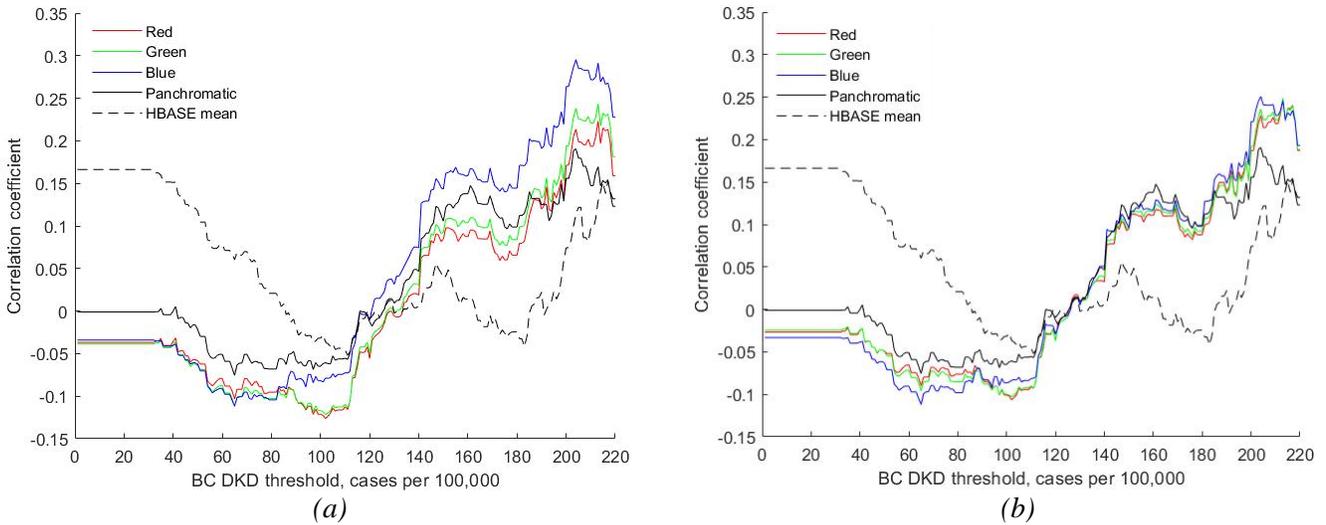

*(a)*        *(b)*

Fig. A11. The association between BC DKD rates and ALAN, estimated for different ALAN bands for the Haifa Bay metropolitan area (see text for explanations): *(a)* the augmented random forest model with all explanatory variables included; *(b)* random forest model without HBASE-based predictors.

*Notes:* The BC DKD levels are estimated as the number of new cases per km$^2$ divided by the population density in 100,000 persons per km$^2$. The data are drawn from [62]; the vertical axes feature values of the Pearson correlation coefficient between the BC DKD rate and ALAN emissions in different ALAN bands, estimated by the proposed modeling approach. The correlations are calculated for observations with BC DKD rates *above* a certain threshold. Overall correlations correspond to zero BC DKD threshold and $r$=-3.40E-02 (B), $r$=-3.88E-02 (G), $r$=-3.69E-02 (R) and $r$=-1.1E-03 (Panchromatic) for the left panel diagram and $r$=3.32E-02 (B), $r$=-2.47E-02 (G), $r$=-2.68E-02 (R) and $r$=-1.1E-03 (Panchromatic) for the right panel diagram.

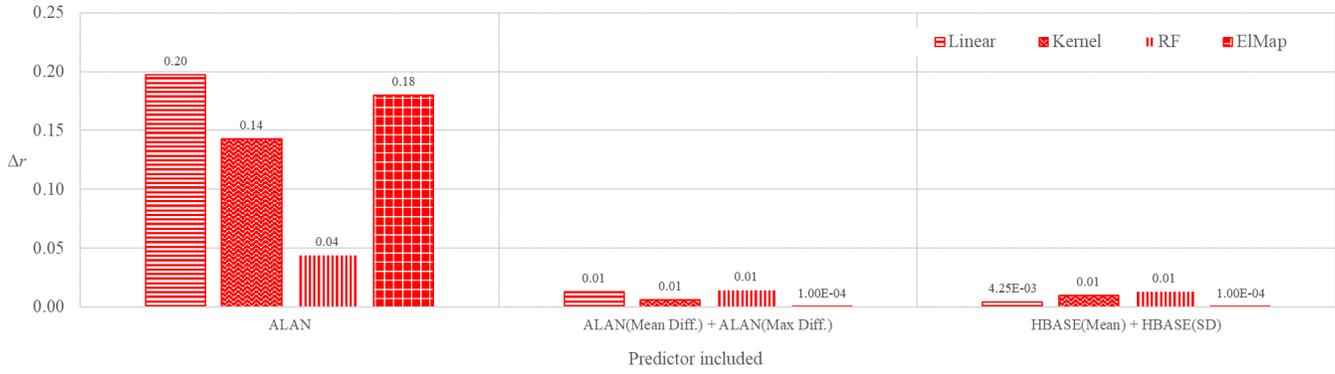

*(a)*

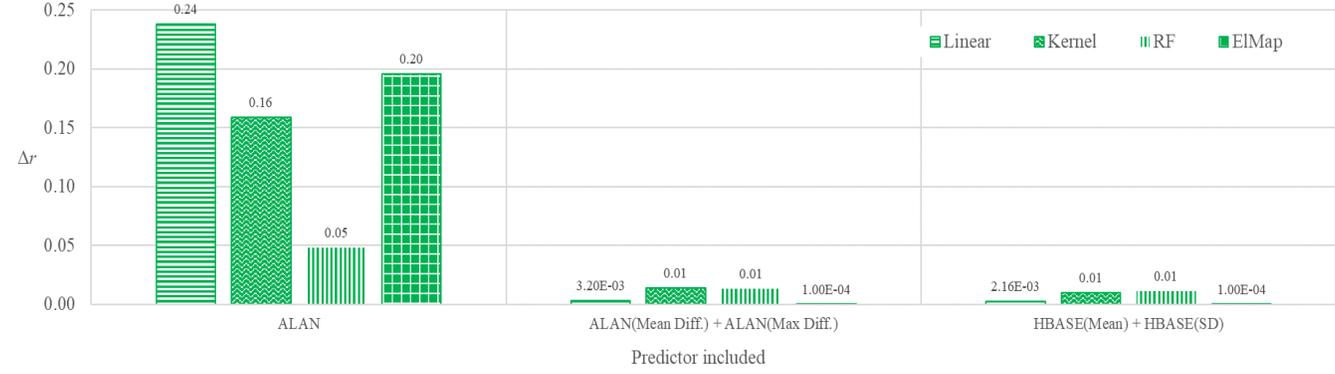

*(b)*

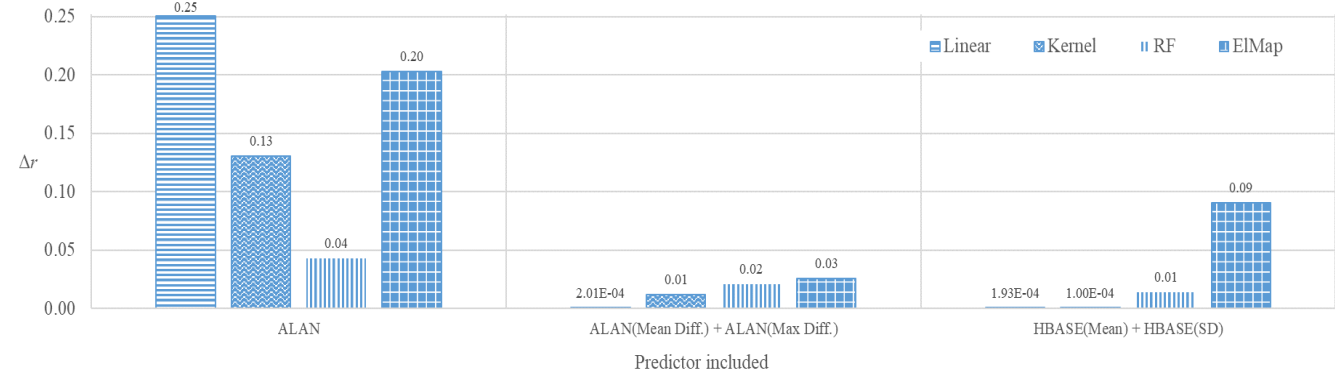

*(c)*

Fig. A12. Changes in the models' performance (Δ*r*), attributed to the exclusion of different groups of variables from the set of predictors, estimated separately for different model types (Study dataset: all metropolitan areas under analysis; N. of pixels/obs. = 33,846); the models are estimated separately for the Red *(a)*, Green *(b)*, and Blue *(c)* spectra

TABLE AI

Datasets representation: Number of observations and numbers of contributing points from middle-resolution HBASE layer and high-resolution RGB layer to each observation

| Dataset | N. of observations (incl. % of outliers in brackets) | Average N. of contributing points from HBASE image | Average N. of contributing points from RGB image |
|---|---|---|---|
| Atlanta | 5589 (2.9%) | 239.2 | 1663 |
| Beijing | 6588 (3.9%) | 239.1 | 2080 |
| Haifa | 685 (3.5%) | 239.0 | 3677 |
| Khabarovsk | 3550 (3.7%) | 239.1 | 1433 |
| London | 5148 (2.9%) | 239.1 | 1810 |
| Naples | 1539 (3.6%) | 238.8 | 1472 |
| Nashville | 5852 (2.9%) | 239.1 | 1576 |
| Tianjing | 6044 (3.2%) | 239.1 | 1158 |
| *Total* | *34995 (3.3%)* | *239.1* | *1669* |

TABLE AII

Descriptive statistics of variables

| Variable | Min | Max | Mean |
|---|---|---|---|
| **All datasets (n=33,846)** | | | |
| Red ALAN level (dn) | 2.493 | 237.283 | 37.484 |
| Green ALAN level (dn) | 1.779 | 198.602 | 26.843 |
| Blue ALAN level (dn) | 2.522 | 180.178 | 17.860 |
| Panchromatic ALAN level (nW/cm2/sr) | 0.360 | 361.520 | 27.568 |
| Panchromatic ALAN (mean difference) (nW/cm2/sr) | -37.499 | 174.290 | -0.025 |
| Panchromatic ALAN (max difference) (nW/cm2/sr) | -152.750 | 268.780 | -3.737 |
| HBASE (mean) (%) | 0.000 | 100.000 | 76.272 |
| HBASE (standard deviation) | 0.000 | 43.991 | 9.788 |
| **Atlanta dataset (n=5,426)** | | | |
| Red ALAN level (dn) | 2.949 | 201.203 | 25.235 |
| Green ALAN level (dn) | 2.388 | 167.117 | 21.345 |
| Blue ALAN level (dn) | 4.166 | 136.365 | 17.243 |
| Panchromatic ALAN level (nW/cm2/sr) | 2.640 | 357.470 | 32.261 |
| Panchromatic ALAN (mean difference) (nW/cm2/sr) | -33.824 | 174.290 | -0.031 |
| Panchromatic ALAN (max difference) (nW/cm2/sr) | -129.640 | 268.780 | -7.327 |
| HBASE (mean) (%) | 38.642 | 100.000 | 83.334 |
| HBASE (standard deviation) | 0.000 | 25.222 | 11.063 |
| **Beijing dataset (n=6,331)** | | | |
| Red ALAN level (dn) | 2.493 | 184.672 | 41.140 |
| Green ALAN level (dn) | 1.779 | 151.764 | 24.322 |
| Blue ALAN level (dn) | 2.522 | 100.429 | 12.913 |
| Panchromatic ALAN level (nW/cm2/sr) | 1.160 | 115.870 | 25.876 |
| Panchromatic ALAN (mean difference) (nW/cm2/sr) | -8.977 | 45.115 | -0.079 |
| Panchromatic ALAN (max difference) (nW/cm2/sr) | -42.250 | 78.790 | -1.691 |
| HBASE (mean) (%) | 8.892 | 100.000 | 89.844 |
| HBASE (standard deviation) | 0.000 | 29.064 | 4.946 |
| **Haifa dataset (n=661)** | | | |
| Red ALAN level (dn) | 5.686 | 237.283 | 73.389 |
| Green ALAN level (dn) | 5.686 | 198.602 | 63.088 |
| Blue ALAN level (dn) | 15.249 | 157.615 | 40.651 |
| Panchromatic ALAN level (nW/cm2/sr) | 2.380 | 288.840 | 55.450 |
| Panchromatic ALAN (mean difference) (nW/cm2/sr) | -37.499 | 101.203 | 0.216 |
| Panchromatic ALAN (max difference) (nW/cm2/sr) | -152.750 | 179.620 | -10.234 |
| HBASE (mean) (%) | 14.633 | 99.904 | 64.053 |
| HBASE (standard deviation) | 0.000 | 34.345 | 12.504 |
| **Khabarovsk dataset (n=3,417)** | | | |
| Red ALAN level (dn) | 3.802 | 149.199 | 15.584 |
| Green ALAN level (dn) | 2.217 | 102.652 | 10.850 |
| Blue ALAN level (dn) | 2.935 | 49.388 | 8.166 |
| Panchromatic ALAN level (nW/cm2/sr) | 0.360 | 189.300 | 13.938 |
| Panchromatic ALAN (mean difference) (nW/cm2/sr) | -14.268 | 41.678 | -0.075 |
| Panchromatic ALAN (max difference) (nW/cm2/sr) | -60.490 | 116.010 | -2.876 |
| HBASE (mean) (%) | 0.000 | 96.229 | 37.044 |
| HBASE (standard deviation) | 0.000 | 35.988 | 9.874 |
| **London dataset (n=4,996)** | | | |
| Red ALAN level (dn) | 4.836 | 181.283 | 51.068 |
| Green ALAN level (dn) | 4.087 | 176.981 | 40.740 |
| Blue ALAN level (dn) | 7.907 | 180.178 | 29.369 |
| Panchromatic ALAN level (nW/cm2/sr) | 3.220 | 170.590 | 39.760 |
| Panchromatic ALAN (mean difference) (nW/cm2/sr) | -21.001 | 45.999 | -0.101 |
| Panchromatic ALAN (max difference) (nW/cm2/sr) | -78.500 | 82.990 | -3.142 |
| HBASE (mean) (%) | 9.600 | 100.000 | 87.243 |
| HBASE (standard deviation) | 0.000 | 43.991 | 9.660 |
| **Naples dataset (n=1,484)** | | | |
| Red ALAN level (dn) | 8.900 | 197.201 | 80.248 |
| Green ALAN level (dn) | 6.587 | 191.646 | 63.354 |
| Blue ALAN level (dn) | 11.251 | 169.717 | 36.273 |
| Panchromatic ALAN level (nW/cm2/sr) | 4.800 | 159.820 | 46.287 |
| Panchromatic ALAN (mean difference) (nW/cm2/sr) | -18.786 | 49.996 | 0.290 |
| Panchromatic ALAN (max difference) (nW/cm2/sr) | -70.390 | 107.510 | -1.815 |
| HBASE (mean) (%) | 5.979 | 100.000 | 84.457 |
| HBASE (standard deviation) | 0.000 | 33.648 | 12.047 |
| **Nashville dataset (n=5,680)** | | | |
| Red ALAN level (dn) | 4.784 | 210.018 | 31.819 |
| Green ALAN level (dn) | 4.287 | 189.544 | 24.478 |
| Blue ALAN level (dn) | 7.985 | 168.177 | 20.051 |
| Panchromatic ALAN level (nW/cm2/sr) | 0.630 | 361.520 | 21.483 |
| Panchromatic ALAN (mean difference) (nW/cm2/sr) | -20.979 | 152.166 | 0.007 |
| Panchromatic ALAN (max difference) (nW/cm2/sr) | -117.960 | 218.170 | -4.782 |
| HBASE (mean) (%) | 1.688 | 100.000 | 71.797 |
| HBASE (standard deviation) | 0.000 | 30.919 | 9.019 |
| **Tianjing dataset (n=5,851)** | | | |
| Red ALAN level (dn) | 3.446 | 228.292 | 36.674 |
| Green ALAN level (dn) | 2.433 | 132.322 | 21.654 |
| Blue ALAN level (dn) | 2.998 | 52.529 | 10.245 |
| Panchromatic ALAN level (nW/cm2/sr) | 1.300 | 215.670 | 20.605 |
| Panchromatic ALAN (mean difference) (nW/cm2/sr) | -11.671 | 79.840 | -0.005 |
| Panchromatic ALAN (max difference) (nW/cm2/sr) | -55.160 | 147.140 | -2.373 |
| HBASE (mean) (%) | 3.388 | 100.000 | 72.228 |
| HBASE (standard deviation) | 0.000 | 39.663 | 13.770 |

TABLE AIII

Comparison of Linear, Kernel and Elastic map models performance upon training (**Atlanta dataset**) and testing sets

| Approach | Quality indicator | | | | | | |
|---|---|---|---|---|---|---|---|
| | Correlation | | | WMSE | | | Contrast similarity RGB image |
| | R | G | B | R | G | B | |
| *Training set (Atlanta dataset)* | | | | | | | |
| Linear regression | 0.819 | 0.803 | 0.749 | 0.808 | 0.888 | 0.294 | 0.910 |
| Kernel regression | 0.824 | 0.813 | 0.748 | 0.650 | 0.713 | 0.209 | 0.924 |
| Random Forest Regression | 0.938 | 0.934 | 0.915 | 0.299 | 0.328 | 0.117 | 0.969 |
| Elastic map models [1] | 0.811 | 0.788 | 0.725 | 0.875 | 0.919 | 0.243 | 0.952 |
| *Testing sets (Beijing, Haifa, Khabarovsk, London, Naples, Nashville, Tianjing datasets) [2]* | | | | | | | |
| Linear regression | 0.860 | 0.870 | 0.758 | 0.348 | 0.322 | 0.242 | 0.793 |
| Kernel regression | 0.858 | 0.870 | 0.751 | 0.307 | 0.250 | 0.215 | 0.800 |
| Random Forest Regression | 0.848 | 0.852 | 0.714 | 0.238 | 0.224 | 0.308 | 0.836 |
| Elastic map models | 0.843 | 0.847 | 0.733 | 0.230 | 0.247 | 0.296 | 0.856 |

The following notes are applicable to TABLES AIII-AX:

[1] For elastic map approach, the results of the best-performing model are reported;

[2] Averaged – across seven testing sets – levels are reported.

TABLE AIV

Comparison of Linear, Kernel and Elastic map models performance upon training (**Beijing dataset**) and testing sets

| Approach | Quality indicator | | | | | | |
| | Correlation | | | WMSE | | | Contrast similarity |
| | R | G | B | R | G | B | RGB image |
| *Training set (Beijing dataset)* | | | | | | | |
| Linear regression | 0.879 | 0.885 | 0.784 | 0.220 | 0.196 | 0.109 | 0.985 |
| Kernel regression | 0.888 | 0.889 | 0.789 | 0.163 | 0.149 | 0.081 | 0.988 |
| Random Forest Regression | 0.959 | 0.960 | 0.926 | 0.068 | 0.061 | 0.036 | 0.994 |
| Elastic map models [1] | 0.857 | 0.858 | 0.759 | 0.257 | 0.215 | 0.121 | 0.990 |
| *Testing sets (Atlanta, Haifa, Khabarovsk, London, Naples, Nashville, Tianjing datasets) [2]* | | | | | | | |
| Linear regression | 0.848 | 0.856 | 0.755 | 1.149 | 0.582 | 0.246 | 0.963 |
| Kernel regression | 0.795 | 0.837 | 0.714 | 0.740 | 0.393 | 0.255 | 0.959 |
| Random Forest Regression | 0.839 | 0.842 | 0.700 | 0.992 | 0.504 | 0.252 | 0.967 |
| Elastic map models | 0.812 | 0.808 | 0.671 | 0.865 | 0.524 | 0.273 | 0.961 |

TABLE AV
Comparison of Linear, Kernel and Elastic map models performance upon training (**Haifa dataset**) and testing sets

| Approach | Quality indicator | | | | | | |
| | Correlation | | | WMSE | | | Contrast similarity |
| | R | G | B | R | G | B | R |
| *Training set (Haifa dataset)* | | | | | | | |
| Linear regression | 0.861 | 0.887 | 0.845 | 0.558 | 0.370 | 0.091 | 0.964 |
| Kernel regression | 0.879 | 0.910 | 0.871 | 0.388 | 0.252 | 0.067 | 0.978 |
| Random Forest Regression | 0.957 | 0.963 | 0.948 | 0.141 | 0.106 | 0.029 | 0.993 |
| Elastic map models [1] | 0.872 | 0.895 | 0.833 | 0.236 | 0.200 | 0.074 | 0.983 |
| *Testing sets (Atlanta, Beijing, Khabarovsk, London, Naples, Nashville, Tianjing datasets) [2]* | | | | | | | |
| Linear regression | 0.820 | 0.847 | 0.730 | 4.223 | 2.979 | 1.669 | 0.931 |
| Kernel regression | 0.851 | 0.848 | 0.719 | 2.628 | 2.992 | 2.166 | 0.949 |
| Random Forest Regression | 0.850 | 0.850 | 0.728 | 1.922 | 2.046 | 2.360 | 0.959 |
| Elastic map models | 0.826 | 0.840 | 0.721 | 1.852 | 2.281 | 2.004 | 0.952 |

TABLE AVI
Comparison of Linear, Kernel and Elastic map models performance upon training (**Khabarovsk dataset**) and testing sets

| Approach | Quality indicator | | | | | | |
| | Correlation | | | WMSE | | | Contrast similarity |
| | R | G | B | R | G | B | RGB image |
| *Training set (Khabarovsk dataset)* | | | | | | | |
| Linear regression | 0.879 | 0.882 | 0.788 | 0.223 | 0.230 | 0.093 | 0.978 |
| Kernel regression | 0.883 | 0.897 | 0.812 | 0.151 | 0.150 | 0.074 | 0.974 |
| Random Forest Regression | 0.959 | 0.962 | 0.930 | 0.067 | 0.061 | 0.027 | 0.992 |
| Elastic map models [1] | 0.881 | 0.887 | 0.804 | 0.198 | 0.214 | 0.086 | 0.986 |
| *Testing sets (Atlanta, Beijing, Haifa, London, Naples, Nashville, Tianjing datasets) [2]* | | | | | | | |
| Linear regression | 0.846 | 0.858 | 0.758 | 0.400 | 0.338 | 0.226 | 0.858 |
| Kernel regression | 0.848 | 0.831 | 0.732 | 0.359 | 0.316 | 0.235 | 0.870 |
| Random Forest Regression | 0.831 | 0.835 | 0.703 | 0.390 | 0.332 | 0.237 | 0.877 |
| Elastic map models | 0.812 | 0.828 | 0.710 | 0.424 | 0.409 | 0.245 | 0.901 |

TABLE AVII
Comparison of Linear, Kernel and Elastic map models performance upon training (**London dataset**) and testing sets

| Approach | Quality indicator | | | | | | |
| | Correlation | | | WMSE | | | Contrast similarity |
| | R | G | B | R | G | B | RGB image |
| *Training set (London dataset)* | | | | | | | |
| Linear regression | 0.857 | 0.862 | 0.778 | 0.336 | 0.279 | 0.169 | 0.956 |
| Kernel regression | 0.874 | 0.869 | 0.815 | 0.210 | 0.196 | 0.108 | 0.967 |
| Random Forest Regression | 0.955 | 0.955 | 0.939 | 0.093 | 0.082 | 0.048 | 0.982 |
| Elastic map models [1] | 0.841 | 0.841 | 0.768 | 0.305 | 0.252 | 0.109 | 0.979 |
| *Testing sets (Atlanta, Beijing, Haifa, Khabarovsk, Naples, Nashville, Tianjing datasets) [2]* | | | | | | | |
| Linear regression | 0.853 | 0.863 | 0.756 | 1.155 | 1.050 | 0.649 | 0.938 |
| Kernel regression | 0.826 | 0.860 | 0.678 | 0.575 | 0.623 | 0.761 | 0.935 |
| Random Forest Regression | 0.850 | 0.850 | 0.711 | 0.595 | 0.682 | 1.017 | 0.937 |
| Elastic map models | 0.822 | 0.822 | 0.680 | 0.844 | 0.955 | 0.877 | 0.949 |

TABLE AVIII
Comparison of Linear, Kernel and Elastic map models performance upon training (**Naples dataset**) and testing sets

| Approach | Quality indicator | | | | | | |
| | Correlation | | | WMSE | | | Contrast similarity |
| | R | G | B | R | G | B | RGB image |
| *Training set (Naples dataset)* | | | | | | | |
| Linear regression | 0.866 | 0.874 | 0.687 | 0.211 | 0.257 | 0.126 | 0.985 |
| Kernel regression | 0.895 | 0.898 | 0.741 | 0.120 | 0.146 | 0.099 | 0.992 |
| Random Forest Regression | 0.961 | 0.961 | 0.903 | 0.045 | 0.056 | 0.043 | 0.996 |
| Elastic map models [1] | 0.875 | 0.872 | 0.646 | 0.112 | 0.129 | 0.091 | 0.992 |
| *Testing sets (Atlanta, Beijing, Haifa, Khabarovsk, London, Nashville, Tianjing datasets) [2]* | | | | | | | |
| Linear regression | 0.825 | 0.824 | 0.704 | 3.740 | 4.588 | 2.391 | 0.937 |
| Kernel regression | 0.821 | 0.803 | 0.567 | 2.705 | 2.950 | 2.616 | 0.950 |
| Random Forest Regression | 0.842 | 0.848 | 0.652 | 2.869 | 2.966 | 3.103 | 0.959 |
| Elastic map models | 0.822 | 0.818 | 0.728 | 2.620 | 2.716 | 2.261 | 0.953 |

TABLE AIX
Comparison of Linear, Kernel and Elastic map models performance upon training (**Nashville dataset**) and testing sets

| Approach | Quality indicator | | | | | | |
| | Correlation | | | WMSE | | | Contrast similarity |
| | R | G | B | R | G | B | RGB image |
| *Training set (Nashville dataset)* | | | | | | | |
| Linear regression | 0.868 | 0.876 | 0.831 | 0.458 | 0.344 | 0.087 | 0.971 |
| Kernel regression | 0.900 | 0.900 | 0.833 | 0.246 | 0.224 | 0.073 | 0.982 |
| Random Forest Regression | 0.965 | 0.965 | 0.946 | 0.096 | 0.085 | 0.031 | 0.993 |
| Elastic map models [1] | 0.878 | 0.883 | 0.821 | 0.288 | 0.242 | 0.069 | 0.983 |
| *Testing sets (Atlanta, Beijing, Haifa, Khabarovsk, London, Naples, Tianjing datasets) [2]* | | | | | | | |
| Linear regression | 0.836 | 0.851 | 0.748 | 0.995 | 0.949 | 0.872 | 0.921 |
| Kernel regression | 0.856 | 0.866 | 0.746 | 0.778 | 0.838 | 0.774 | 0.946 |
| Random Forest Regression | 0.839 | 0.845 | 0.716 | 0.846 | 0.866 | 0.893 | 0.945 |
| Elastic map models | 0.816 | 0.826 | 0.717 | 0.839 | 0.886 | 0.789 | 0.951 |

TABLE AX
Comparison of Linear, Kernel and Elastic map models performance upon training (**Tianjing dataset**) and testing sets

| Approach | Quality indicator | | | | | | |
| | | R | | G | | B | |
| | R | G | B | R | G | B | RGB image |
| *Training set (Tianjing dataset)* | | | | | | | |
| Linear regression | 0.905 | 0.899 | 0.729 | 0.173 | 0.198 | 0.134 | 0.978 |
| Kernel regression | 0.911 | 0.903 | 0.768 | 0.153 | 0.155 | 0.073 | 0.981 |
| Random Forest Regression | 0.967 | 0.965 | 0.912 | 0.061 | 0.067 | 0.040 | 0.991 |
| Elastic map models [1] | 0.878 | 0.871 | 0.726 | 0.358 | 0.399 | 0.128 | 0.987 |
| *Testing sets (Atlanta, Beijing, Haifa, London, Naples, Nashville datasets) [2]* | | | | | | | |
| Linear regression | 0.847 | 0.857 | 0.751 | 1.624 | 0.726 | 0.206 | 0.967 |
| Kernel regression | 0.860 | 0.863 | 0.529 | 1.322 | 0.567 | 0.243 | 0.968 |
| Random Forest Regression | 0.844 | 0.843 | 0.651 | 1.430 | 0.663 | 0.217 | 0.967 |
| Elastic map models | 0.821 | 0.825 | 0.679 | 1.269 | 0.670 | 0.222 | 0.966 |